%% file: article.tex
\documentclass[a4paper, twocolumn]{article}
\input{setup}

\usepackage{appendix}
\usepackage[sorting=none]{biblatex}
\input{Sections/title.tex}

\begin{document}

\date{}
\maketitle

\begin{strip}
    \centering
    \begin{minipage}{.9\textwidth}
        \begin{abstract}
            \vspace{-3em}

\input{Sections/abstract.tex}
        \end{abstract}
        \hspace{2cm}
    \end{minipage}
\end{strip}


\input{Sections/introduction.tex}
\input{Sections/literature.tex}
\input{Sections/tests1.tex}
\input{Sections/tests2.tex}

\input{Sections/conclusion.tex}

\section*{Acknowledgments}
A. R. acknowledges financial support by National Funds through the Portuguese funding agency, FCT - Fundação para a Ciência e a Tecnologia, within project LA/P/0063/2020; from FCT under PhD grant 2022.12332.BD; and from the European Union’s Horizon Europe research and innovation program under EPIQUE Project GA No. 101135288. 

NW acknowledges support from NSERC and the US DOE, Office of Science, National Quantum Information Science
 Research Centers, Co-design Center for Quantum Advantage (C2QA) under Contract No. DE-SC0012704
 (Basic EnergySciences, PNNLFWP76274).
\begin{appendices}
    \renewcommand{\appendixpagename}{\large \textbf{APPENDICES}}
    \vspace{1.5em}
    \begin{center}  
    \appendixpagename
    \end{center}

\input{Apps/acronyms.tex}
\input{Apps/expdesign.tex}

\input{Apps/smc.tex}

\input{Apps/mcmc.tex}
\input{Apps/subsampling.tex}
\input{Apps/expsetup.tex}
\end{appendices}


\printbibliography

\end{document}

%% file: setup.tex
\usepackage{hyperref} 
\usepackage[dvipsnames, table]{xcolor}
\usepackage{geometry}
\usepackage{mathtools}
\usepackage{authblk}  
\usepackage{amssymb}
\usepackage{amsmath}
\usepackage{soul}
\usepackage{subcaption}
\usepackage{siunitx}
\usepackage{scalerel}
\usepackage{stackengine}
\usepackage{fbox}
\usepackage{dashbox}
\usepackage{bm}
\usepackage{enumitem}
\usepackage{cuted}
\usepackage{abstract}
\usepackage{titlesec}
\usepackage{booktabs}
\setlist{topsep=4pt,itemsep=0pt,parsep=2pt}

\titleformat{\section}  
  {\fontsize{12}{14}\bfseries} 
  {\thesection} 
  {1em} 
  {} 
  [] 

\titleformat{\subsection}  
  {\fontsize{11}{13}\sffamily} 
  {\thesubsection} 
  {2em} 
  {} 
  [] 

\newcommand{\opsub}[2]{\mathbf{\hat{#1}_#2}}

\newcommand{\ket}[1]{\left\lvert \ #1 \ \right\rangle}

\newcommand{\dd}[0]{\mathrm{d}}

\mathchardef\hyph="2D
\newcolumntype{g}{>{\columncolor{gray!15}}c}

\usepackage{float}
\floatstyle{boxed}
\newfloat{algorithm}{h}{loa}
\floatname{algorithm}{Algorithm}

%% file: Sections/title.tex
\title{Calibration of Quantum Devices \\ via Robust Statistical Methods}
\author[1,2,3]{Alexandra Ramôa \thanks{alexandra.ramoa@inl.int}} 
\author[4]{Raffaele Santagati}
\author[5,6,7]{Nathan Wiebe}

\affil[1]{International Iberian Nanotechnology Laboratory (INL), Portugal}
\affil[2]{High-Assurance Software Laboratory (HASLab), Portugal}
\affil[3]{Department of Computer Science, University of Minho, Portugal}
\affil[4]{Quantum Lab, Boehringer Ingelheim, Germany}
\affil[5]{Department of Computer Science, University of Toronto, Canada}
\affil[6]{Pacific Northwest National Laboratory, Richland WA, USA}
\affil[7]{Canadian Institute for Advanced Studies, Toronto, Canada}

\date{\today}

%% file: Sections/abstract.tex
Bayesian inference is a widely used technique for real-time characterization of quantum systems. It excels in experimental characterization in the low data regime, and when the measurements have degrees of freedom. A decisive factor for its performance is the numerical representation of the Bayesian probability distributions. In this work, we explore advanced statistical methods for this purpose, and numerically analyze their performance against the state-of-the-art in quantum parameter learning.
In particular, we consider sequential importance resampling, tempered likelihood estimation, Markov Chain Monte Carlo, random walk Metropolis (RWM), Hamiltonian Monte Carlo (HMC) and variants (stochastic gradients with and without friction, energy conserving subsampling), block pseudo-marginal Metropolis-Hastings with subsampling, hybrid HMC-RWM approaches, and Gaussian rejection filtering. We demonstrate advantages of these approaches over existing ones, namely robustness under multi-modality and high dimensionality. We apply these algorithms to the calibration of superconducting qubits from IBMQ, surpassing the standard quantum limit and achieving better results than Qiskit's default tools. In Hahn echo and Ramsey experiments, we reduce the uncertainty by factors of $10$ and $3$ respectively, without increasing the number of measurements; conversely, we match the performance of Qiskit's methods while using up to to 99.5\% less experimental data.  We additionally investigate the roles of adaptivity, dataset ordering and heuristics in quantum characterization. Our findings have applications in challenging quantum characterization tasks, namely learning the dynamics of open quantum systems.

%% file: Sections/introduction.tex
\section{Introduction}

Characterizing quantum systems is crucial for quantum science, underpinning processes such as calibration, certification, and sensing. Bayesian inference is one of the most promising approaches, as a framework capable of optimal use of the quantum resources~\cite{Higgins_2007, Ferrie_2011}, which has been applied to an extensive range of practical problems. Examples include state or process tomography, Hamiltonian learning, noise model estimation, and others \cite{Huszar_2012, Lukens_2020b, Granade_2016, Hincks_2018, Granade_2012, Gentile_2021, Gebhart2023, Obrien2022}. 

Bayesian inference gives us rational criteria for characterizing quantum systems. It systematically updates probabilistic estimates of the system's unknown parameters based on experimental measurement results, combining prior knowledge with new evidence, and progressively improving the quality of the model describing the quantum system. 


The main open problems concerning Bayesian learning for quantum systems arise on two (related) fronts: finding good experiments \cite{Ferrie_2012, Ferrie_2011, rainforth2024}, and representing the Bayesian probability distributions \cite{Betancourt_2018, Daviet_2016, Doucet_2001}. These are  optimization and integration problems respectively.

Furthermore, an accurate description of the model of the system becomes even harder when the system size increases. For example, a potentially exponentially larger number of parameters is required. Even just storing the probability distribution representing the model parameters becomes rapidly intractable.

For open quantum systems, it is often necessary to also describe a bath to which the system is coupled. This means their description needs even more intricate formalisms, and learning their underlying variables commands informative data and robust processing. Random measurements and brute-force algorithms, which succeed for simple models, are likely to fail in more complex scenarios, where problems like high dimensionality and multi-modality make exploring the parameter space more challenging \cite{Granade_2017, Betancourt_2018}.

On the integration front, Monte Carlo methods are a prominent solution \cite{Stan, Salvatier_2016, Betancourt_2018, Doucet_2001, Granade_2012}. They estimate expectations over probability distributions by probabilistically choosing samples at which to evaluate the integrand, and using them in numerical integration \cite{Doucet_2001, Del_Moral_2006, Betancourt_2018}. Methods within this family differ in how they produce samples and thus in how efficient they are. Unlike deterministic approaches such as grid-based quadrature, their required resources do not necessarily scale exponentially with the dimension \cite{Betancourt_2018}.

Recent Bayesian methods have gone beyond parameter exploration to model learning, where the specific form and number of parameters are learnt from the experimental data \cite{Gentile_2021, Flynn_2022}. However, these approaches are strongly limited by the speed of the inference process and the memory required to store the multidimensional parameter distributions. It is then of fundamental importance to compare known techniques in order to guide the design of future methods for the characterization of open quantum systems.

\subsection{Contributions}

Here, we apply advanced statistical techniques to the characterization of open quantum systems via Bayesian inference,  to further optimize the inference process and design appropriate control on the quantum systems. While Bayesian inference has often been applied to the characterization of quantum systems, these instances often target uni-dimensional estimation problems where simple statistical tools suffice.  We demonstrate the shortcomings of these tools through numerical simulations and propose more robust alternatives.

In particular, we report and compare results from different techniques, such as sequential importance resampling with the Liu-West filter and with Markov Chain Monte Carlo kernels, Hamiltonian Monte Carlo (HMC), stochastic gradient HMC with and without friction, HMC with energy conserving subsampling, random walk Metropolis-Hastings (RWM), tempered likelihood estimation, block pseudo-marginal Metropolis-Hastings with subsampling, hybrid approaches that adaptively switch between HMC and RWM, and Gaussian rejection filtering. 

We apply these techniques to several quantum problems, namely, estimation of decoherence effects (energy loss and dephasing) and frequencies in quantum hardware. We observe improved performance with respect to Qiskit's default fitters \cite{Qiskit} in the calibration of IBMQ devices, especially in low data regimes.

We propose heuristics for the experimental design, assess the impact of often neglected factors such as the dataset ordering, and explore how commonly used strategies fail in different contexts.

We supply pseudo-code for the key algorithms, and make available the code used for these experiments. 

\subsection{Document structure}

The rest of this paper is organized as follows. Section \ref{sec:background} overviews the background and existing literature on inference applied to quantum systems. Section \ref{sec:tests1} shows the performance of various methods for iterative phase estimation and oscillatory dynamics that arise frequently in quantum physics. Section \ref{sec:tests2} presents the setup and results of various experiments on quantum hardware, including Hahn echo ($T_2$ estimation), energy loss ($T_1$ estimation), and Ramsey experiments. Finally, section \ref{sec:conclusion} discusses the findings and concludes with insights into the efficiency of the proposed methods.

The appendices supply details and pseudo-code for the main algorithms used in this paper. Appendix \ref{app:acronyms} presents a list of acronyms. Appendix \ref{app:bayesian_experimental_design} overviews the problem of experimental design, and techniques for precession dynamics; \ref{sec:sequential_monte_carlo} describes the sequential Monte Carlo algorithm and tempered likelihood estimation; \ref{sec:mcmc} introduces Markov Chain Monte Carlo and Hamiltonian Monte Carlo; and \ref{app:subsampling} discusses subsampling strategies, in particular for HMC. Finally, appendix \ref{app:expsetup} details the experimental set up.

The code used to produce the results is available at \cite{repository}.

%% file: Sections/literature.tex
\section{Background}
\label{sec:background}

Bayesian inference can be used to estimate parameters governing the evolution of a quantum system, provided the following conditions are fulfilled \cite{rainforth2024, Granade_2012, qinfer, Hincks_2018, msc}:
\begin{enumerate}
    \item We can measure the system;
    \item The results of these measurements have a known probabilistic dependence on the parameter(s) to be estimated.
\end{enumerate}

Additionally, in most fruitful applications of the Bayesian framework, the following additional condition holds:
\begin{enumerate}
    \setcounter{enumi}{2}
    \item The results of the measurements have a known probabilistic dependence on experimental control(s). 
\end{enumerate}

One common example is a two-level system undergoing oscillations at an unknown frequency, i.e., a harmonic oscillator~\cite{Ferrie_2012, Wiebe_2016, Granade_2012, Wiebe_2014a}: 

\begin{equation}
    \mathbf{P}(1\mid \omega;t)=\sin^2(\omega t).
\end{equation}

The frequency $\omega$ is the parameter to be estimated, whereas the evolution time is the controllable parameter. 

In a frequentist framework, calling $N_1$ the number of times we observe $1$ as an outcome out of a total number $N_{\rm exp}$ of experiments, we could estimate $\omega$ by choosing a fixed time $t$ and doing a number of measurements:


\begin{equation}
    \omega \approx \arcsin \left( \sqrt{\frac{N_1}{N_{\rm exp}}} \right).
\end{equation}

We could improve our estimate by considering experiments at different times and performing a fit to a curve.
On the other hand, when performing Bayesian inference, we systematically apply Bayes' rule to update our beliefs based on observations. A key role in updating our belief with Bayes' rule is played by the likelihood function, defined as

\begin{equation}
    \mathbf{L}(\theta \mid D; E) = \mathbf{P}(D \mid \theta; E),
\end{equation}

where $\theta$ is a parameter, $D$ is a datum, and $E$ is an experimental control. In the example above, these correspond to the frequency, binary outcomes and evolution times respectively, and we have $\mathbf{L}(\omega \mid D; t)= \sin^2(\omega t)^D\cos^2(\omega t)^{1-D}$ for a binary outcome $D$.

Given a prior distribution $\mathbf{P}(\theta)$ describing our initial knowledge about the parameters, Bayes' rule expresses a relation between our prior probability distribution and a posterior distribution $\mathbf{P}(\theta \mid D; E)$:

\begin{equation}
    \label{eq:bayes}
    \mathbf{P}(\theta \mid D; E) = \frac{\mathbf{L}(\theta \mid D; E)\mathbf{P}(\theta)}{\mathbf{P}(D ; E)},
\end{equation}

where $\mathbf{P}(D ; E)$ can be regarded as a normalization constant.

In the harmonic oscillator example, the prior could be a uniform distribution in the domain $[0, 2\pi[$. The Bernstein–von Mises theorem proves, under fairly inclusive assumptions, that the Bayesian results are asymptotically correct from a frequentist point of view. Their dependence on the prior can be removed through data collection \cite{Vaart_1998}, i.e., observation is king; the importance of the prior in defining our final posterior decreases as the wealth of data used to update our knowledge increases. 

Multiple data points can be either batch processed or considered in sequential updates, where the posterior of one iteration serves as the prior for the next. The advantage of Bayesian inference is that we can optimize the experimental controls in a look-ahead fashion. That is, before performing $N$ experiments, we can optimize their controls; however, the cost of this optimization scales exponentially with $N$. An alternative is pursuing a locally optimal, or greedy, strategy, considering a look-ahead of a single experiment \cite{Ferrie_2011, Granade_2012}. 

Alternatively, there are other strategies, namely the $\sigma^{-1}$ and particle guess heuristics (PGH) \cite{Ferrie_2012, Wiebe_2014a}, among others. The $\sigma^{-1}$ heuristic, introduced for the precession example, chooses measurement times inversely proportional to the uncertainty, and the PGH performs similarly but removes the need to calculate the uncertainty. For details on the optimization process and alternatives, refer to appendix~\ref{app:bayesian_experimental_design}.

In addition to experimental design, another determining factor for the performance of Bayesian learning is the strategy used to encode information from equation (\ref{eq:bayes}). This equation describes a probability distribution, which we can evaluate for any parameter values we choose. Since exact probability distributions cannot be represented efficiently on a computer, approximations such as numerical integration or variational approximations must be employed. We focus on the former, which is more general in nature. In this approach, the probability distribution on the parameter space is approximated by sampling a number of discrete points in the multi-parameter space~\cite{Doucet_2001}.  The robustness of the sampling method is crucial for the cost-to-performance ratio of this approach, as it impacts both the experimental design and the accuracy of the process.

Arguably the most popular method for numerical integration in Bayesian quantum learning is via Sequential Importance Resampling (SIR) - a type of Sequential Monte Carlo (SMC)/particle filter - with the Liu-West Filter (LWF)~\cite{Liu_2001}, adopted in \cite{Granade_2012, Chase_2009, Wiebe_2014a, Wiebe_2014b, Wang_2017, Granade_2016, Hincks_2018, Flynn_2022, Gentile_2021} as well as by the available software~\cite{qinfer}. This is a relatively lightweight and straightforward strategy, especially due to LWF. Its main advantage is that it considerably reduces the number of computations of the likelihood function; the drawback is that it preserves only the first two statistical moments of the distribution. The performance of the LWF depends on a parameter, denoted $a \in [0,1]$, which governs a trade-off between distribution preservation and exploration. 

For $a=1$, the LWF does nothing; for other values, normal distributions are kept invariant, but others are only guaranteed to maintain their mean and variance. Common choices are $a=0.98$ \cite{Granade_2012}, $a=0.9$ \cite{Wiebe_2014a}, and $a=0.995$ \cite{Liu_2001}. High values retain more structure at the expense of convergence speed, while lower values are better for normal distributions. For general distributions, a delicate trade-off arises between correctness and efficiency. This may jeopardize correctness for less well-behaved distributions and lead to complete failure of the inference process in some cases; for instance, in the presence of multimodality, whether intrinsic or due to ambiguity in the dataset. Other approaches that work under the normality assumption are the Gaussian rejection filtering (GRF) of \cite{Wiebe_2016}, where for each Bayesian update a Gaussian distribution is fit according to the data, and Gaussian conjugate posteriors of \cite{Evans_2019}.

In \cite{Granade_2017}, the SMC-LWF protocol is adapted to multimodal distributions using clustering. While this provides a working solution, it addresses a symptom rather than the root cause and doesn't address other shortcomings in the representation. As an alternative, SMC can be made asymptotically correct by replacing the LWF or SMC with a more rigorous approach. 

Markov chain Monte Carlo (MCMC) is one such option \cite{South_2019}. MCMC produces successive states that converge to or preserve a target probability distribution by means of carefully crafted transitions. One simple example is random walk Metropolis, where the transitions are a random walk (Gaussian perturbation on the current state) followed by acceptance or rejection of the new state step according to the ratio of probabilities for the proposed and original states \cite{Betancourt_2018, Metropolis_1953}. For details, refer to appendix \ref{sec:mcmc}.

Simple MCMC methods have recently been applied to Bayesian learning for quantum problems. Na\"ive Metropolis-Hastings is applied to quantum tomography in \cite{Mai_2017}. \cite{Lu_2019, Williams_2017} use slice sampling, with the latter acknowledging the shortcomings of this strategy for e.g. multimodal distributions and mentioning SMC as a possible solution. It is noted that SMC can introduce inaccuracies, but this strongly depends on the implementation. While these inaccuracies might arise when using  SMC with LWF, more robust variants exist.

Further improvements on vanilla Metropolis-Hastings (MH) have been shown for pseudo-Bayesian quantum tomography by taking advantage of the preconditioned Crank-Nicolson (pCN) algorithm~\cite{Lukens_2020, Lukens_2020b, Mai_2021}.

This type of MCMC is rather recent, having been introduced in \cite{Cotter_2013}. By adapting the step size, it offers better acceptance probabilities for high-dimensional spaces compared to standard MH algorithms. However, it is not informed by the geometry of the target distribution, and so is still expected to produce diffusive transitions. Simple MCMC methods must navigate a trade-off in the exploration: proposals that are too conservative converge slowly, whereas proposals that are too bold have high rejection rates. This can, however, be overcome by geometrically informed algorithms, which can reconcile bold proposals with large acceptance rates.

One such method is Hamiltonian Monte Carlo (HMC)\footnote{The ideas behind pCN and HMC are not mutually exclusive, and preconditioned Crank-Nicolson (pCN)-like improvements can be adopted for HMC \cite{Cotter_2013,Beskos_2017,Alenlov_2019}.} \cite{Duane_1987,Neal_2011,Betancourt_2018}, which brings two added benefits: good mixing, and robustness under complex distribution features \cite{Beskos_2017}. HMC, which is the gold standard for modern inference \cite{Stan, Salvatier_2016}, draws inspiration from classical mechanics to produce an efficient Markov chain that leverages the differential structure of the target function. 

None of the above consider embedding MCMC in an SMC scheme, which has several advantages. SMC is more reliable than isolated MCMC and is suitable for online estimation, enabling adaptivity—a crucial resource for many quantum applications. If post-processing is required, a still more reliable option is tempered, or annealed, likelihood estimation (TLE) \cite{Neal2001}, a well-established method for complex statistical simulation problems where a sampler traverses a sequence of $N$ distributions $\textbf{P}_i(\theta)$ consisting of the target distribution to the power of increasing exponents, $\textbf{P}_\text{target}(\theta)^{\gamma_i}$, with $\gamma_N=1$.

Other works on this subject typically employ less general approaches or focus on aspects other than the numerical representation. In \cite{Sergeevich_2011}, an exact representation (a Fourier series) makes a simple single-parameter problem analytically tractable, but this method does not generalize well to more complex inference problems. Similarly, rough approximations such as the Gaussian mixtures of \cite{Craigie_2021} can be efficient for a few parameters, but are generally not scalable to a higher number of parameters. Some papers focus primarily on the experimental design and barely \cite{Huszar_2012, Fiderer_2021} or not at all \cite{Ferrie_2011} on the discretization of the probability distribution; \cite{Ferrie_2012} tackles the same problem but makes a normality assumption to make it at once general and analytically tractable. Likewise, some others improve or generalize other aspects, while using the standard SMC with Liu-West approach \cite{Ferrie_2014, Ferrie_2014b,Gentile_2021}.

The main aim of this work is to explore numerical methods for Bayesian inference applied to quantum systems and evaluate their efficiency, including techniques used in leading software such as PyMC3 \cite{Salvatier_2016} and Stan \cite{Gelman_2015,Stan}. We exploit this knowledge to characterize quantum devices, with a particular focus on quantum channels that capture decoherence processes or other open system effects. We also explore the interplay between experimental design and statistical representation, proposing novel strategies and extending the particle guess heuristic of \cite{Wiebe_2014a} to more general scenarios. 

%% file: Sections/tests1.tex
\section{Simulated tests}
\label{sec:tests1}

\subsection{Iterative phase estimation}
\label{sub:phase_estimation}

\begin{figure}[!ht]
  \centering
  \includegraphics[width=\columnwidth]{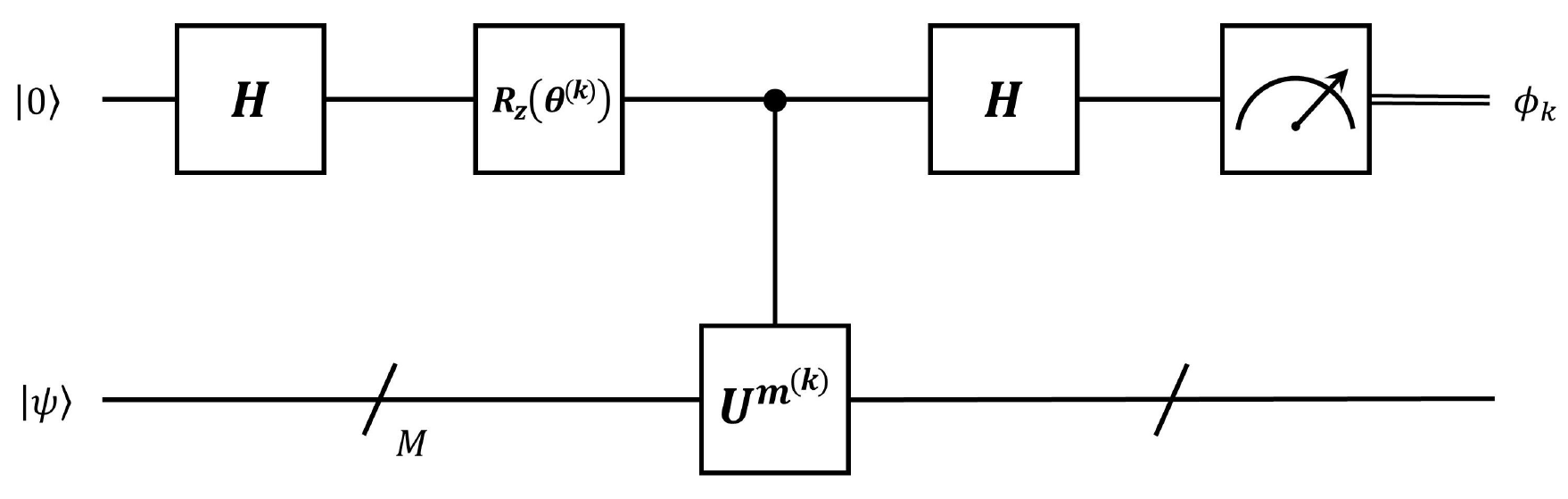}
  \caption{Iterative phase estimation circuit diagram.}
  \label{fig:phase_est_circ}
\end{figure}

Iterative phase estimation (IPE) relies on a simpler quantum circuit than the canonical quantum phase estimation algorithm, combining it with classical processing and feedback \cite{Kitaev_1996,Kitaev_2002,Svore_2014,Dob_ek_2007}. The circuit of Figure \ref{fig:phase_est_circ} can be used for this purpose. It has two main benefits: it replaces the deep circuit with multiple shallower ones, and precision increases with the number of iterations rather than that of ancilla qubits. However, it achieves this at the cost of losing the ability of coherent quantum phase estimation to project the input state into an eigenstate of the unitary~\cite{Simon2025}. 

Several routes are possible within IPE. A common approach is to deterministically obtain bits of the phase from last to first, using successive calls to the circuit in Figure \ref{fig:phase_est_circ} for a pre-determined sequence of integers $m^{(k)}$ and adaptive angles $\theta^{(k)}$ \cite{Kitaev_1996,qiskit_textbook,Kitaev_2002,Svore_2014,Dob_ek_2007}. 

Another proposed algorithm, Bayesian phase estimation, applies Bayesian inference to this problem, exploiting the two degrees of freedom ($m$ and $\theta$, dropping the iteration label $k$ to simplify notation) as controllable parameters \cite{Wiebe_2016, Paesani_2017}; this fits the framework described in section \ref{sec:background}. For representation, the authors use Gaussian rejection filtering. Here, we compare this strategy with Markov chain Monte Carlo, namely Hamiltonian Monte Carlo and random walk Metropolis. One problem with applying HMC to this type of distribution is that it has points of zero probability, which result in discontinuities that affect exploration (details are provided in Appendix \ref{sub:hmc}). We diagnose this by screening for low acceptance rates (with a threshold $0.01$), and performing an RWM step after HMC if they're found. This aids the algorithm to escape the problematic regions. 

\begin{figure*}[!htb]
\captionsetup[subfigure]{width=.9\textwidth}%
\begin{subfigure}{.5\textwidth}
  \centering
  \includegraphics[width=\textwidth]{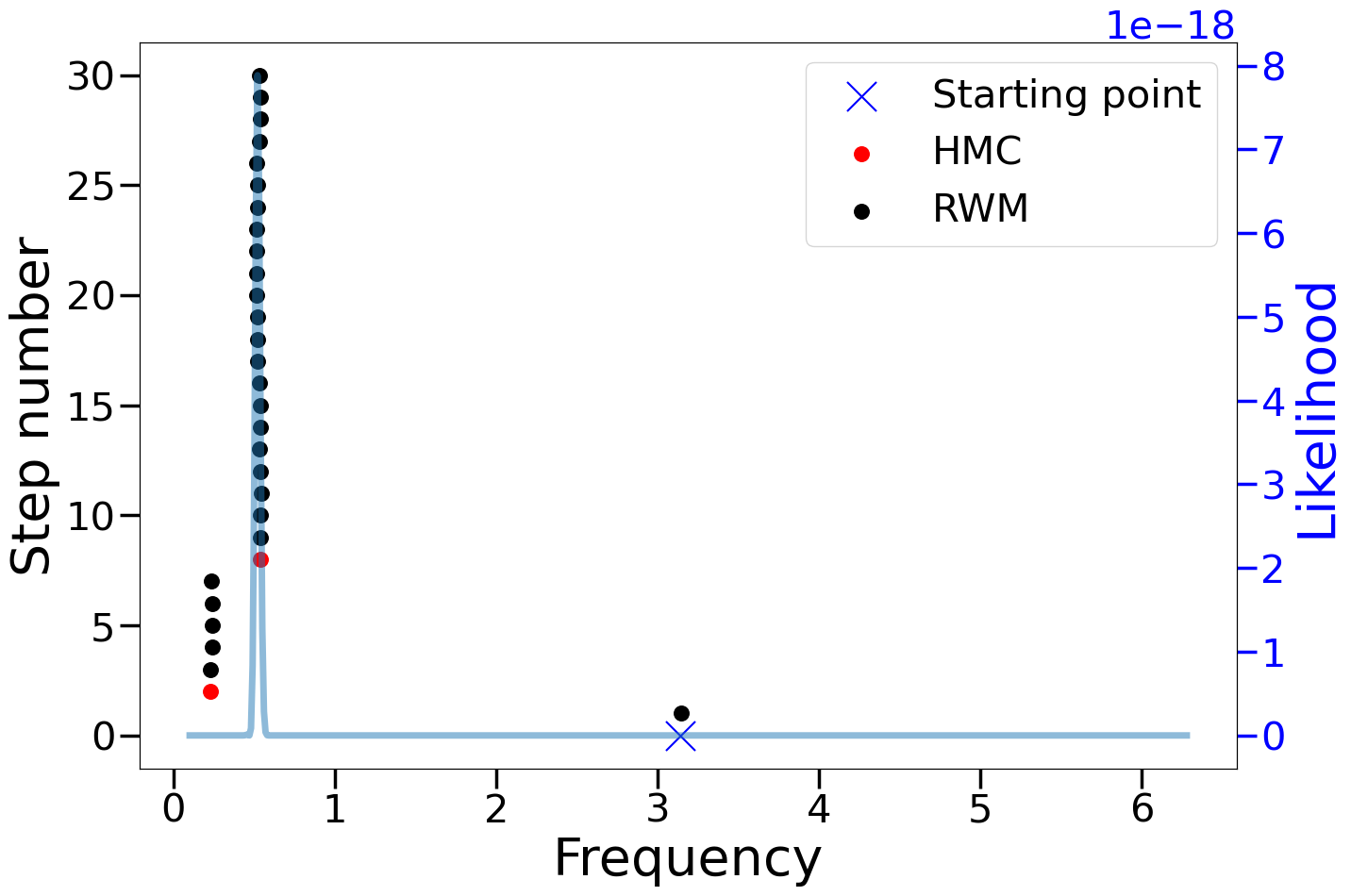}
  \caption{Successive states (upward evolution) generated by Markov chain Monte Carlo. The experiment controls were chosen in advance.}
  \label{fig:ipe_hmc}
\end{subfigure}%
\begin{subfigure}{.5\textwidth}
  \centering
  \includegraphics[width=\textwidth]{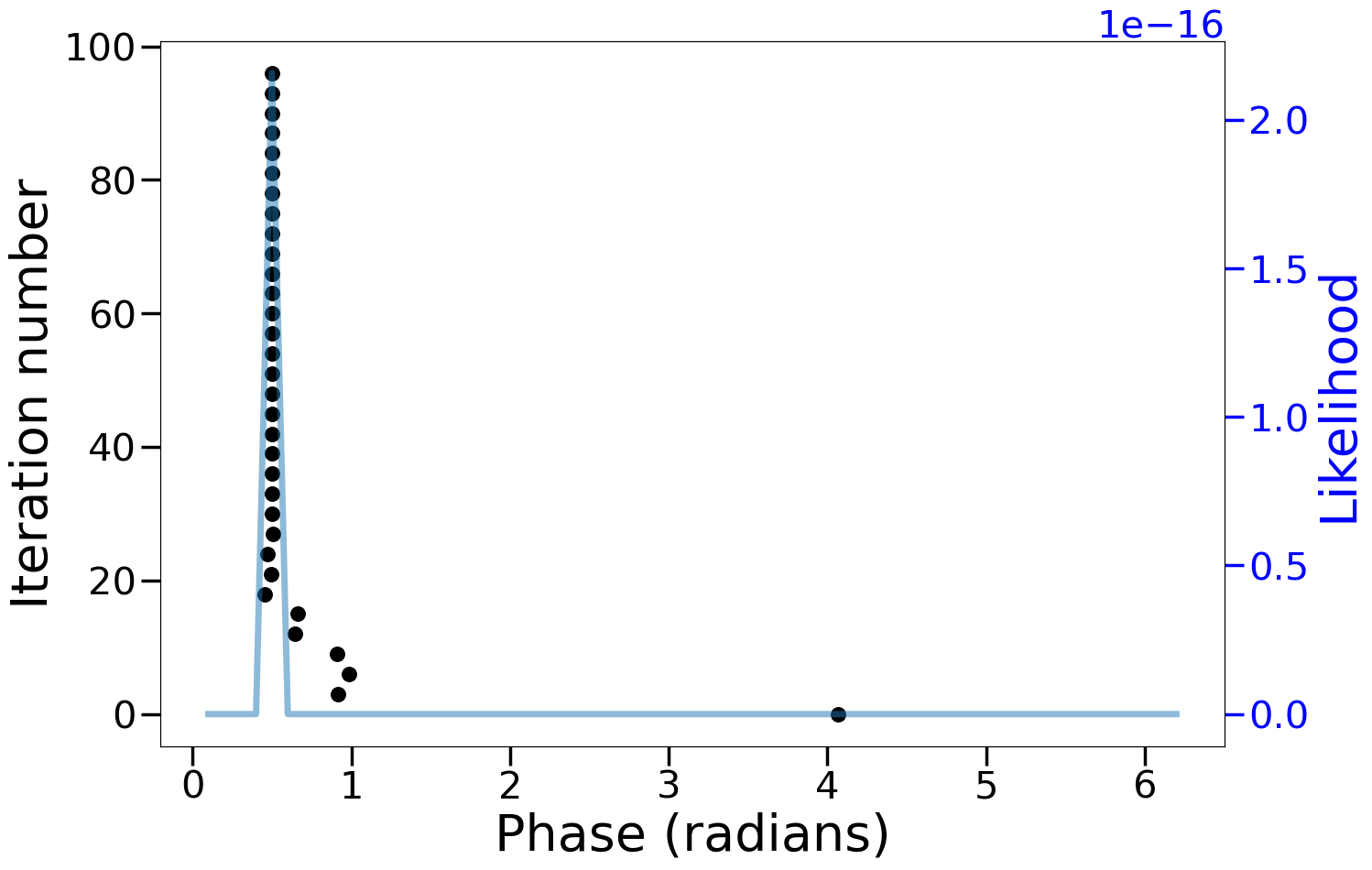}
  \caption{Evolution of the distribution means through the iterations of Gaussian rejection filtering. The experiment controls were chosen adaptively. }
  \label{fig:ipe_grf}
\end{subfigure}%
\caption{Evolution of the sampling process through the iterations/steps for Bayesian phase estimation, using Markov chain Monte Carlo (random walk Metropolis and Hamiltonian Monte Carlo) (\ref{fig:ipe_hmc}) and Gaussian rejection sampling (\ref{fig:ipe_grf}), juxtaposed with the target distribution. $100$ experiments are considered. HMC can rarely be applied due to particularities of the likelihood function, but when it is, it convergences faster than any other method.}
\label{fig:ipe_graphs}
\end{figure*}

Figure \ref{fig:ipe_graphs} shows the results of the comparison. We learn a phase set at $\phi_\text{real}=0.5$ using $100$ measurements and a flat prior on $[0,2\pi[$. One important difference between these methods is that GRF is sequential (as is sequential Monte Carlo), whereas MCMC is not. While GRF uses incremental datasets, MCMC uses the complete dataset from the start, as it targets a stationary distribution rather than a sequence of distributions. This makes MCMC incompatible with adaptive techniques.

For GRF, the experimental controls were chosen adaptively according to the strategy of \cite{Wiebe_2016}, which is a discrete version of the $\sigma^{-1}$ heuristic; the data were considered one at a time (per Bayesian update), which in this case matches the number of steps. For MCMC, since adaptivity is not possible, all combinations of $m \in [1..10]$ and $\theta \in [0..9] \cdot \pi/5$ were used, and $30$ transitions were realized. Note that even though this corresponds to fewer iterations, each of them considers all the data, in contrast with GRF where a single datum is contemplated per step (along with the previous results). This makes GRF more lightweight at the expense of correctness. It also enables adaptivity, bringing more informative data: after roughly $20$ measurements, the results are already quite decisive. We found that with datasets that small, MCMC fails to converge on most executions. 

The RWM-after-HMC fix resulted in only $3.3\%$ HMC steps, which had a very high acceptance probability (99\% on average). This illustrates the point made in section \ref{sec:background}, about how HMC is capable of combining distant proposals with very high acceptance rates, resulting in fast exploration and convergence.

In Figure \ref{fig:ipe_hmc}, it is clear that the HMC transitions are the driving force behind the bold transitions, demonstrating its potential - even in this case, where the likelihood model complicates its application. This encourages the use of HMC for applications where the likelihood function does not suffer from the null-points problem, or the use of efficient work arounds to deal with this issue, such as presented in \cite{Nishimura_2020}.

\subsection{Precession dynamics}
\label{sub:sampling_precession}

In this section, we target the study of a sinusoidal likelihood as arises in Rabi, Larmor, and Ramsey oscillations \cite{Zwiebach_2013, Vitanov_2015, Frimmer_2014, Granade_2017}, with applications in superconducting and photonic quantum computing,  magnetic field sensing, among others \cite{ramoa24, pezze2014, Craigie_2021, Santagati_2019, msc, Joas2021_single}. 
For example, in Rabi oscillations, we observe the periodic exchange of population between two quantum states driven by a near-resonant oscillating field, which represents one of the most fundamental operations performed on a qubit.
These dynamics are also similar to those of iterative phase estimation as described in the previous section, and to quantum amplitude estimation \cite{ramoa24}, although the latter cases typically have a discrete domain.

We then consider a two-level quantum system evolving under the Hamiltonian $H = \frac{\omega}{2}\sigma_x$. The  outcome distribution for a measurement in the $z$ basis is then $\mathbf{P}(+\mid \omega;t)=\left(\sin^2(\omega t) \right)^x\left(\cos^2(\omega t) \right)^{1-x}$. Using this generative model, Bayesian inference can be used to learn the frequency $\omega$.

\begin{figure*}[!htb]
\captionsetup[subfigure]{width=.9\textwidth}%
\begin{subfigure}[t]{.5\textwidth}
  \centering
  \includegraphics[width=\textwidth]{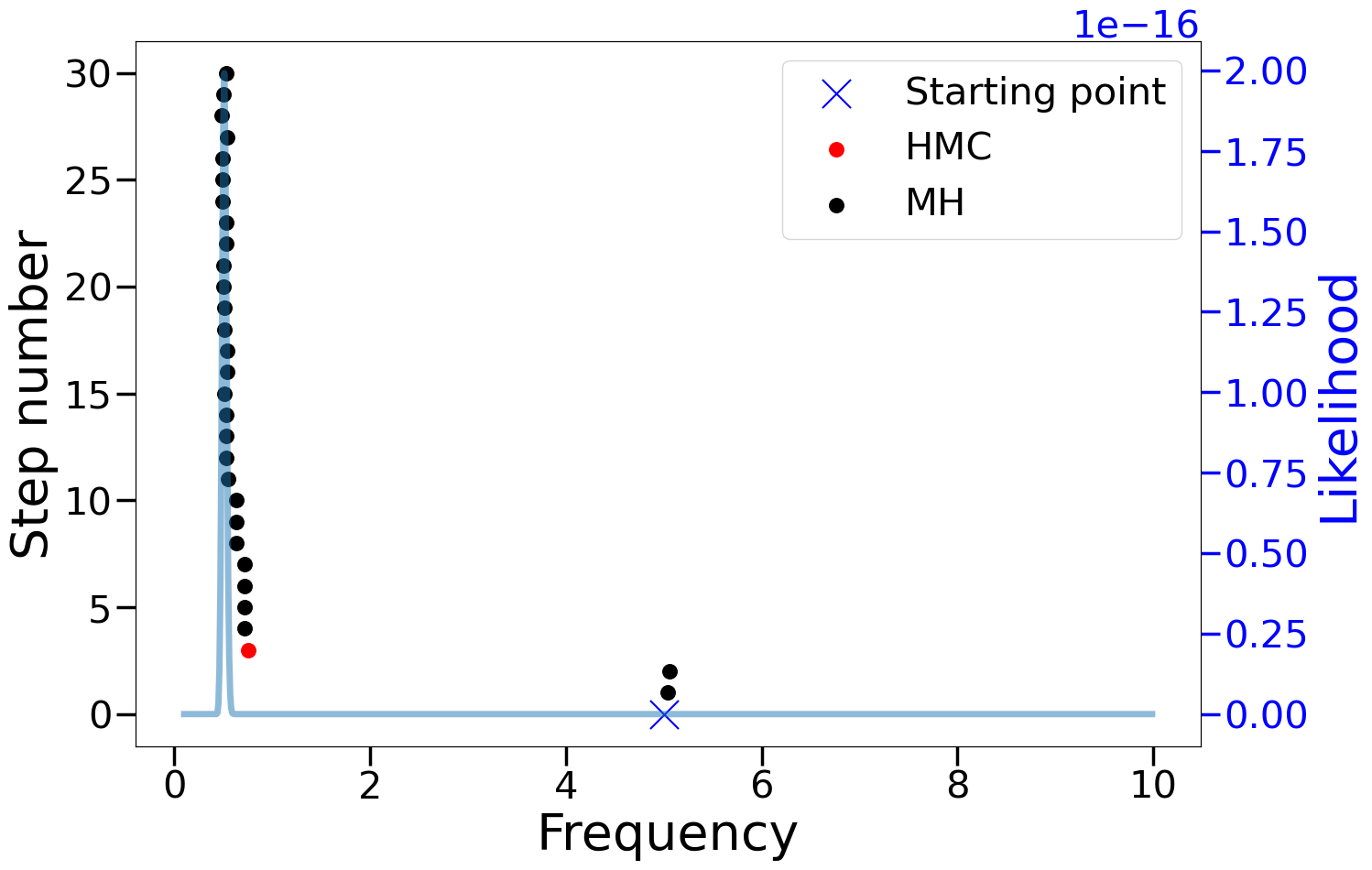}
  \caption{Successive states (upward evolution) generated by Markov chain Monte Carlo. The experiment controls were chosen in advance; all $100$ data were used at each step.}
  \label{fig:precession_hmc_rwm}
\end{subfigure}%
\begin{subfigure}[t]{.5\textwidth}
  \centering
  \includegraphics[width=\textwidth]{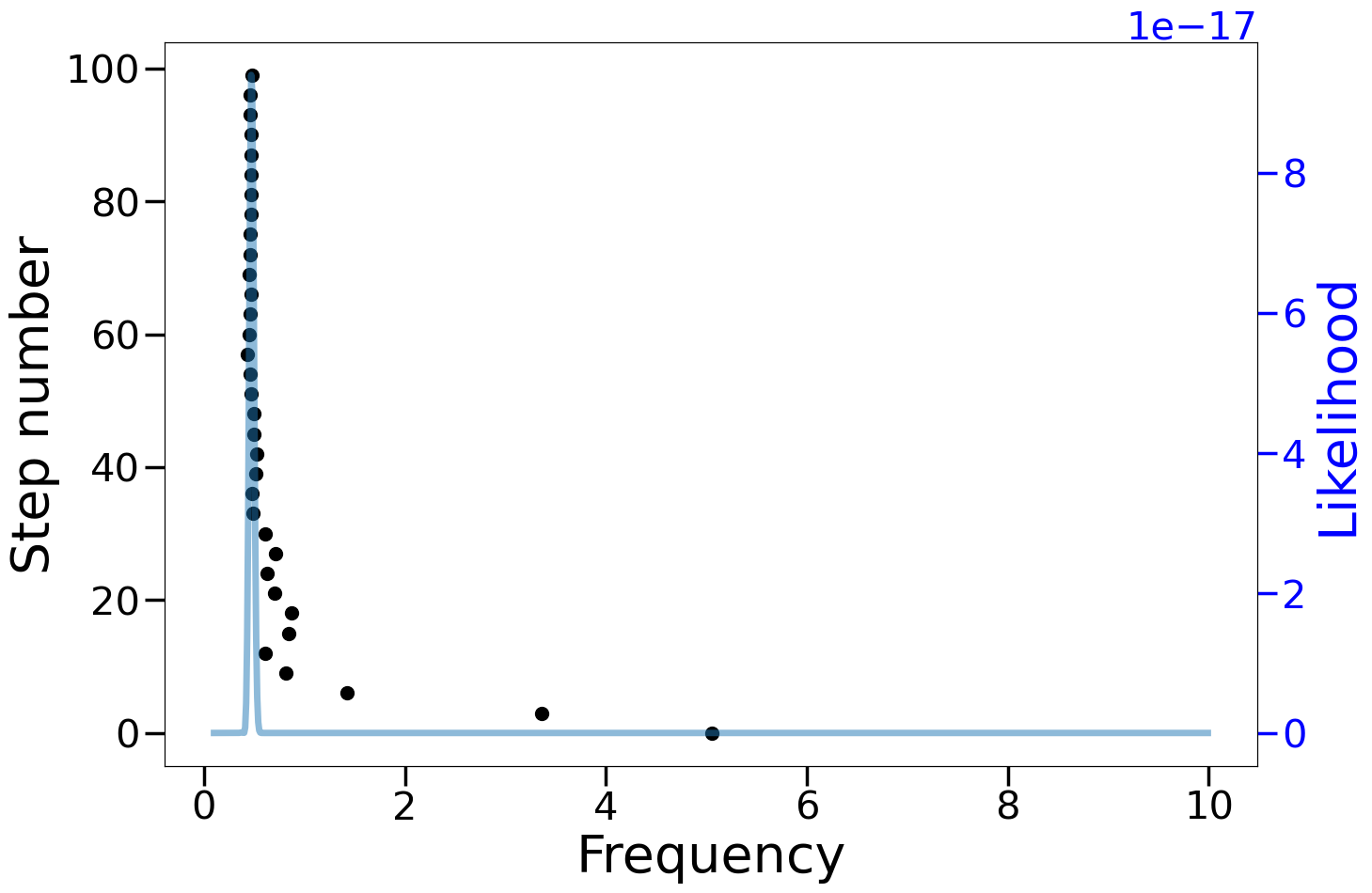}
  \caption{Evolution of the distribution means through the iterations of sequential Monte Carlo with the Liu West filter. Each iteration $i$ uses only the $i$th datum and the $(i-1)$th iteration's results.}
  \label{fig:precession_smc_lw}
\end{subfigure}%
\caption{Evolution of the sampling process through the iterations/steps for precession frequency estimation, using Markov chain Monte Carlo (random walk Metropolis and Hamiltonian Monte Carlo) (\ref{fig:precession_hmc_rwm}) and Gaussian rejection sampling (\ref{fig:precession_smc_lw}), juxtaposed with the target distribution. $100$ experiments are considered. HMC can rarely be applied due to particularities of the likelihood function, but when it is, it convergences faster than any other method.}
\label{fig:precession_graphs_dots}
\end{figure*}

Figure \ref{fig:precession_graphs_dots} shows the results of using sequential Monte Carlo with the Liu-West filter and MCMC to learn a frequency $\omega_\text{real}=0.5$ using $100$ measurements and a flat prior on $\omega \in ]0,10]$. The times were chosen in increments of $0.08$, decided with a parameter sweep. Note that what matters is not the units, but only the relative scale between times and frequencies. They're always found paired up in the form $\omega \cdot t$. 

MCMC was implemented in the same conditions as section \ref{sub:phase_estimation}, resulting in $6.7\%$ HMC steps, which had an average acceptance rate of $100\%$ . Both used 100 experiments; MCMC used 30 steps, and SMC was implemented as in Algorithm \ref{alg:sir}, with 100 particles and the LWF parameter $a=0.98$  as suggested in \cite{Granade_2012} and according to the discussion in section \ref{sec:background}. 

Despite using more steps and particles, it actually uses fewer likelihood evaluations than MCMC. SMC uses $T=100 \cdot 100 = 10^4$ evaluations for the $100$ reweightings of $100$ particles, because each reweighting only uses the latest experimental datum, exploiting its sequential nature. MCMC, which rigorously converges to and preserves the distribution based on all the experimental data, has the caveat of needing access to the full dataset for each iteration. 

If it ran for 100 iterations, RWM would use the same $T$ evaluations for $100$ steps in which it uses all $100$ data, whereas HMC would add to that $LT$ gradient evaluations, where $L$ is a hyperparameter used in the generation of HMC proposals (Algorithm \ref{alg:hmc} in the appendix). Naturally, performing both at some iterations accumulates their costs, which end up higher even for $30$ steps. Also, MCMC by itself is less robust, especially for higher frequencies (or equivalently longer evolution times) and/or broader priors, as well as incompatible with online processing (unlike SMC).

As discussed in Appendix \ref{sub:mcmc_smc}, Markov chains can be made more robust by exploiting the coupling from the past method. As compared to pure MCMC, MCMC within an SMC scheme is more fit to handle multi-modality and sequential estimation. As compared to SMC with simpler resampling kernels such as LW, it is more scalable, and correct in more challenging scenarios where normality does not hold. This can be made obvious by extending the domain to encompass negative frequencies as in \cite{Granade_2017}, which evidences one of the downfalls of the LWF or other methods operating under the assumption of normality and/or unimodality. Due to the cosine function's symmetry, the domain extension gives rise to multimodality, which these methods are incapable of dealing with. More generally, complex problems where multiple explanations are possible for the same dataset translate into multi-modal distributions. Hence, in section \ref{sec:tests2}, we adopt SMC with MCMC for learning the parameters of quantum devices.

Finally, we present results for tempered likelihood estimation (TLE) (appendix \ref{sub:tle}), where we sample from a sequence of increasing powers of the posterior with the goal of traversing progressively more challenging distributions to arrive at the target (exponent $1$).  We test TLE-SHMC, both based on the full data and with energy-conserving subsampling and control variates (appendix \ref{app:subsampling}). In the latter case, the likelihood evaluations were based on a subset of the data for both the reweightings and HMC. The subsampling indices for each iteration and particle were chosen in block pseudo-marginal Metropolis-Hastings steps, whose goal is to avoid a deadlock by introducing correlation between consecutive sets of subsampling indices. Here we use $3$ blocks, inducing roughly $67\%$ correlation. Refer to the appendix for details.

As discussed in appendix \ref{app:subsampling}, subsampling introduces noise. To offset this, we apply control variates: auxiliary random variables that do not change the expected value, but can decrease the variance in the estimate if well chosen. One option is to choose approximations of the target; we use a Taylor expansion. Due to its  gradient-based nature, they do not work well as is due to high target curvature. This can be controlled in several ways. One option would be to keep the evolution times short, but this tends to be inefficient. Another one would be to narrow down the prior. 

Such a narrowing down can be done by a warm-up phase on the first observations (short evolutions) if prior knowledge is insufficient to justify it. Section \ref{sec:tests2} shows the warming-up alternative; for these tests, a less general prior was chosen. The real frequency was set at $\omega = 0.8$. 

 The $400$ measurement times were chosen randomly up to $t_\text{max}=100$, and when subsampling $50/400$ were evaluated. Apart from that, all conditions were matched for the two methods. The tempering coefficients were $10$ and evenly spaced. The resulting standard deviations were $\sigma_\text{FD} = 8.2 \times 10^{-4}$ and $\sigma_\text{subs} = 8.6 \times 10^{-4}$ for the full data and subsampling cases respectively. The subsampled distribution thus has an increase of less than 5\% in the uncertainty while decreasing the data usage by $88\%$. 
 
\subsection{Multivariate function (multiparameter estimation)}
\label{sub:multi_cos}

This section considers a multi-parameter generalization of the two previous examples, with the purpose of testing methods in more challenging settings.
This can be achieved by using a vector $\Vec{\omega}$ containing the parameters ${\rm dim}$ and $\omega_d$, that we use to define a binomial likelihood as a function of the observation $D \in \{0,1\}$:
\begin{equation}
\label{eq:multimodallikelihood}
    L(\Vec{\omega} \mid D) = \frac{1}{"} \sum_{d=1}^{{\rm dim}} \cos^2 (\frac{\omega_d t}{2})
\end{equation}

We use this toy example as a test case, as this type of sum-of-cosines arises in e.g. quantum multiphase estimation \cite{Gebhart_2021}.

Clearly, this introduces redundancy, and thus multimodality of the likelihood function. The number of modes reflects how the parameters affect the likelihood function equally. The number of ways the frequencies can be distributed across the multiple dimensions is the number of permutations of ${\rm dim}$ items, i.e. ${\rm dim}!$. For instance, for ${\rm dim}=2$, there are $2$ modes, because having $\omega_1 = A$ and $\omega_2 = B$ is indistinguishable from having the opposite, as both dimensions contribute identically to the likelihood function. 

With this, we can also subsample in a more general setting, namely, stochastic gradient HMC. Note that subsampling introduces noise in the gradients used to simulate the fictitious dynamics that underpin the efficient exploration of HMC. This noise is sufficient to make the trajectory diverge, leading to incorrect results. One possible remedy is introducing friction in the equations of motion. The intuition, supported by numerical results, is that this friction helps control divergence. Details are provided in appendix \ref{sub:sg_hmc}.

\begin{figure*}[!htb]
\captionsetup[subfigure]{width=.9\textwidth}%
\begin{subfigure}[t]{\textwidth/2}
  \centering
  \includegraphics[width=.7\textwidth]{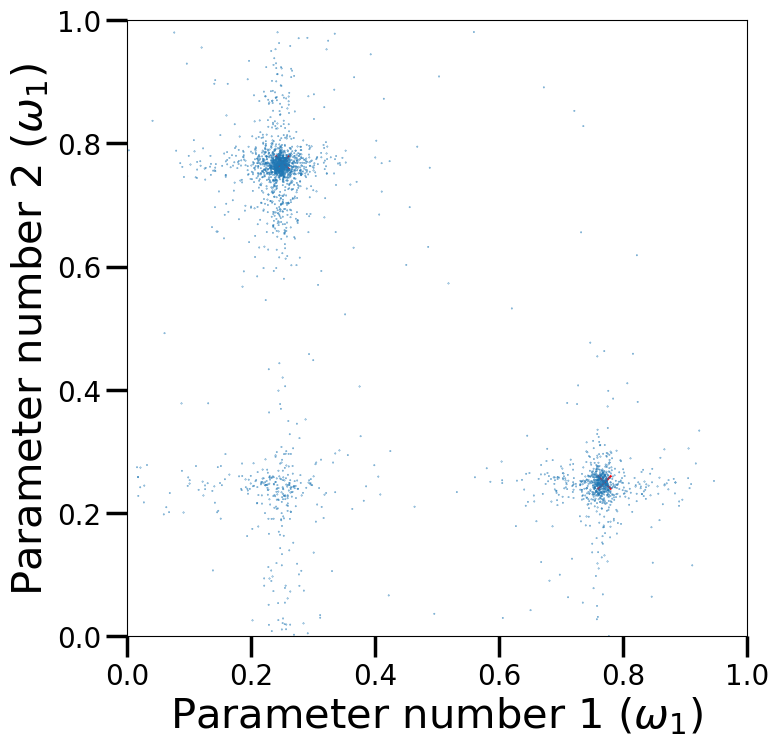}
  \caption{}
  \label{fig:sg_hmc}
\end{subfigure}%
\begin{subfigure}[t]{\textwidth/2}
  \centering
  \includegraphics[width=.7\textwidth]{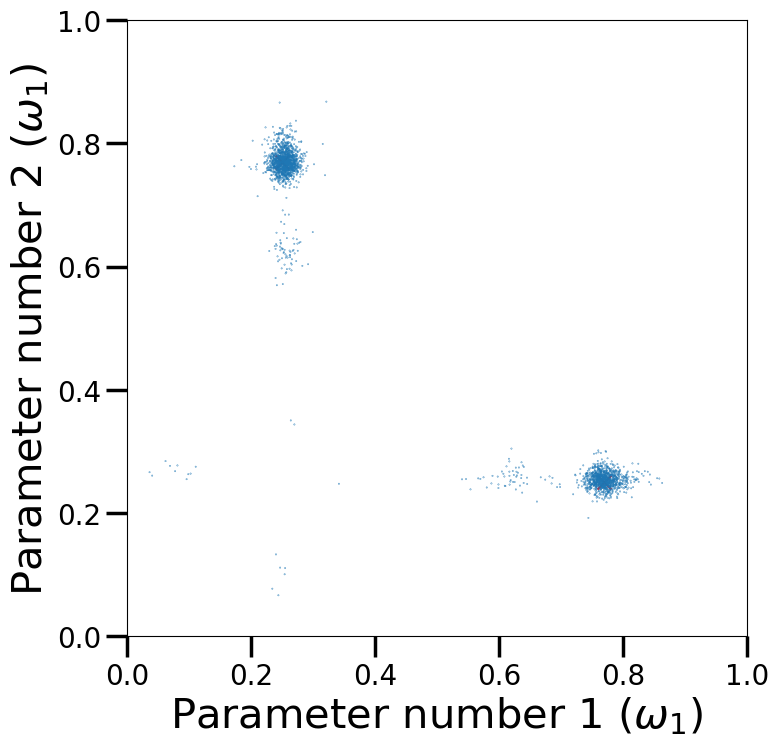}
  \caption{}
  \label{fig:sg_hmc_friction}
  \end{subfigure}%
\caption{Results of multi-parameter (2-dimensional) inference for a multimodal sum-of-sinusoids likelihood (Equation \ref{eq:multimodallikelihood}), using stochastic gradients in Hamiltonian Monte Carlo within sequential Monte Carlo while introducing friction (\ref{fig:sg_hmc_friction}) and not (\ref{fig:sg_hmc}). Friction is a construction meant to aid convergence in the fabricated dynamics HMC relies on. The modes are marked with red 'x' markers, but are covered by the particles.}
\label{fig:sg_hmc_both}
\end{figure*}

The effect of introducing this friction term is shown in Figure \ref{fig:sg_hmc_both}. It does help control the dispersion: we can see that the particles gather more neatly around the modes. While friction would slow down exploration in the ideal case, its benefits outweigh this concern when subsampling. However, the effect is limited.

We also test multi-dimensional generalizations of SMC-powered adaptivity in a unit hyper-cube domain. Often, the adaptive  $\sigma^{-1}$ heuristic and particle-guess heuristic (PGH) are employed (appendix \ref{app:precession_heuristics}). The former chooses evolution times $t_i \propto \sigma_{i-1}^{-1}$, where $\sigma_{j}$ is the standard deviation of the Bayesian distribution in a given iteration, whereas the latter replaces the standard deviation by a cheaper proxy. This results in exponentially increasing evolution times. 

However, the intuition behind it does not hold under multimodality, as in the case of likelihood model \ref{eq:multimodallikelihood}, or in practical scenarios where a problem or dataset admits redundant explanations, which often happens for likelihood functions with many parameters. Alternatively, one may cluster the particles and consider the average variance, but clustering is a challenging task, especially when the number of modes is unknown.  As an alternative to the $\sigma^{-1}$ heuristic in such cases, we propose a space occupation-based metric as a straightforward proxy for the uncertainty:

\begin{equation}\label{eq:heuristic_t}
  \scriptstyle
  t_k \propto \big(\texttt{occupationrate}_{k-1} \cdot \widehat{\text{ESS}}_{k-1}/\texttt{nparticles} \big)^{-1}.
\end{equation}

The occupation rate is a proxy for the standard deviation, as it similarly relates to the uncertainty or spread in the distribution: the more the particles are concentrated in a small region of the parameter space, the more peaked the distribution they represent is, and lower the uncertainty. However, in SMC, it is often the case that the particles do not move in all iterations, but only when a resampling step is triggered due to low effective sample size (ESS). As such, in most iterations, the particles remain in their original locations, the occupation rate remains unchanged, and the evolution time $t_k$ stalls. To curb this, we add a second factor: the ratio of the ESS to the total number of particles (which upper bounds the ESS; the lower bound is zero), resulting in Equation \ref{eq:heuristic_t}. In SMC, this factor is by construction $1$ immediately after resampling occurs, and then decreases exponentially with each iteration until the next resampling step. Thus, it ensures that the evolution times increase in all iterations as intended, producing similar behavior to the original heuristic. 

To benchmark this approach, we compare it against two other methods: random times $t \in [0,100]$, and increasing times $t^\text{(max)}_{k}= C_1 \cdot \big(\texttt{floor}(k/C_2) +1\big)$, with constants $C_1$, $C_2$ optimized in a parameter sweep. Results are shown in Table \ref{tb:2d_exps}.

\begin{table*}[!htb]
\centering
\renewcommand{\arraystretch}{0.8}
\begin{tabular}{@{}l c c c@{}}
\toprule
 & \textbf{Median} $\boldsymbol{\sigma}$ & \textbf{Mean} $\boldsymbol{\sigma}$ & \textbf{Success rate} \\
\midrule
\textbf{Random}         & $2.0 \times 10^{-4}$ & $2.1 \times 10^{-3}$ & 65\% \\
\textbf{Adaptive}       & $7.3 \times 10^{-5}$ & $1.6 \times 10^{-3}$ & 85\% \\
\textbf{Inc. random}    & $1.8 \times 10^{-4}$ & $1.7 \times 10^{-3}$ & 53\% \\
\bottomrule
\end{tabular}
\caption{Results of multi-parameter (2-dimensional) inference for a multimodal sum-of-sinusoids likelihood (Equation \ref{eq:multimodallikelihood}), using different choices of measurement times: random ($t_i \sim \texttt{uniform}[0,100]$), increasing random ($t_i \sim \texttt{uniform}[0, C_1 \cdot \big(\texttt{floor}(k/C_2) +1\big)]$, and adaptive (Equation \ref{eq:heuristic_t}). Our adaptive strategyperforms best. The metrics are explained in the text. $100$ experiments were performed for each.}
\label{tb:2d_exps}
\end{table*}

For defining \textit{success}, assessing performance, and averaging standard deviations (over multiple runs), the mode locations were used \textcolor{red}. Success was considered as the conjunction of accuracy, precision, correctness, and mode coverage. For evaluating these points, thresholds were considered for the average distance to a mode, the variance around each mode, the difference between real and estimated error, and the discrepancy between the number of particles attributed to each mode. The considered metric was the weighted average of standard deviations associated with each mode, each particle being assigned to its closest one. This is what is meant by \textit{standard deviation}.

The adaptive method shows significant improvement in the median standard deviation. Although this advantage shrinks if the mean is considered instead, suggesting a degree of instability in the performance, it still outperforms the other methods across all metrics. 

We note another notable limitation of the PGH: its susceptibility to decoherence. After a point, it will choose evolutions beyond the coherence time, leading to uninformative data and thus a learning plateau, even if the noise is modeled perfectly. In this case, it may be beneficial to cap the evolution within the coherence time of the system, or to switch the strategy entirely. Some works have used this heuristic for open quantum systems, but the learning saturates \cite{fioroni2025}. 

\begin{figure}[!htb]
\captionsetup[subfigure]{width=.9\textwidth}%
\begin{subfigure}[t]{\textwidth/4}
  \centering
  \includegraphics[width=.95\textwidth]{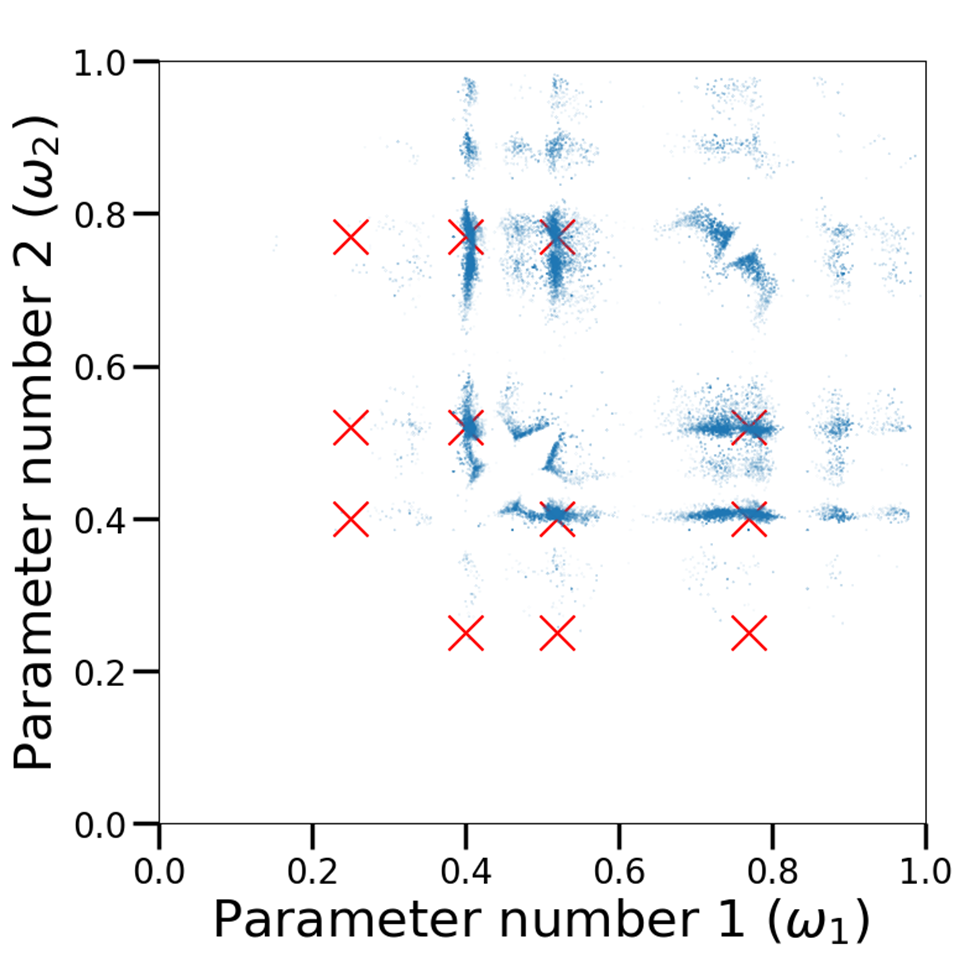}
  \caption{}
  \label{fig:4d_sir_a}
\end{subfigure}%
\begin{subfigure}[t]{\textwidth/4}
  \centering
  \includegraphics[width=.95\textwidth]{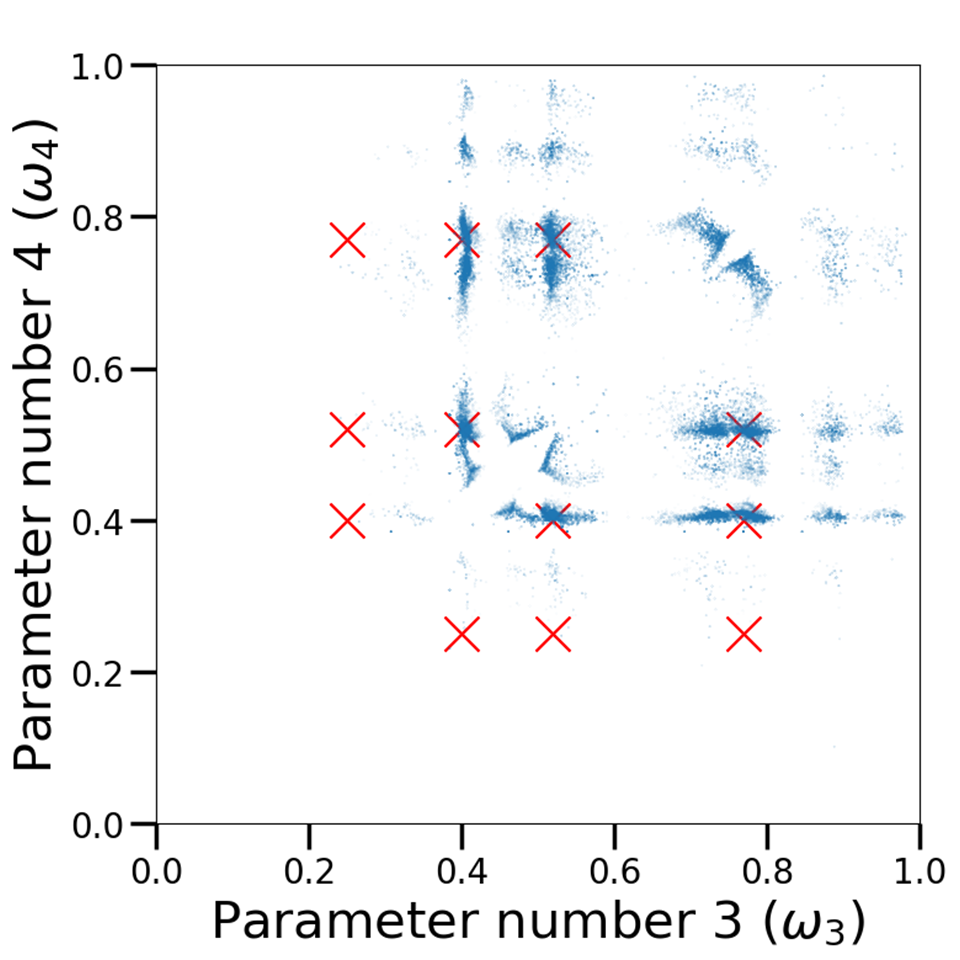}
  \caption{}
  \label{fig:4d_sir_b}
  \end{subfigure}%
  
\begin{subfigure}[t]{\textwidth/4}
\centering
\includegraphics[width=.95\textwidth]{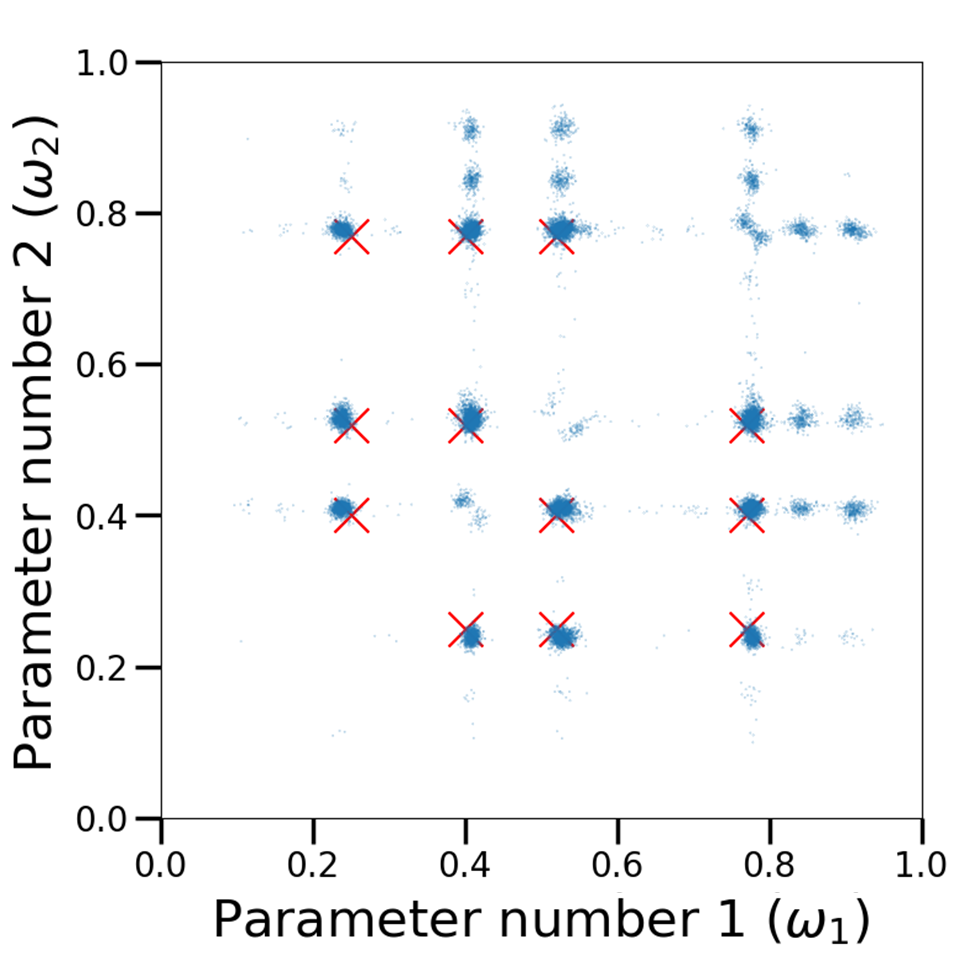}
\caption{}
\label{fig:4d_tle_a}
\end{subfigure}%
\begin{subfigure}[t]{\textwidth/4}
\centering
\includegraphics[width=.95\textwidth]{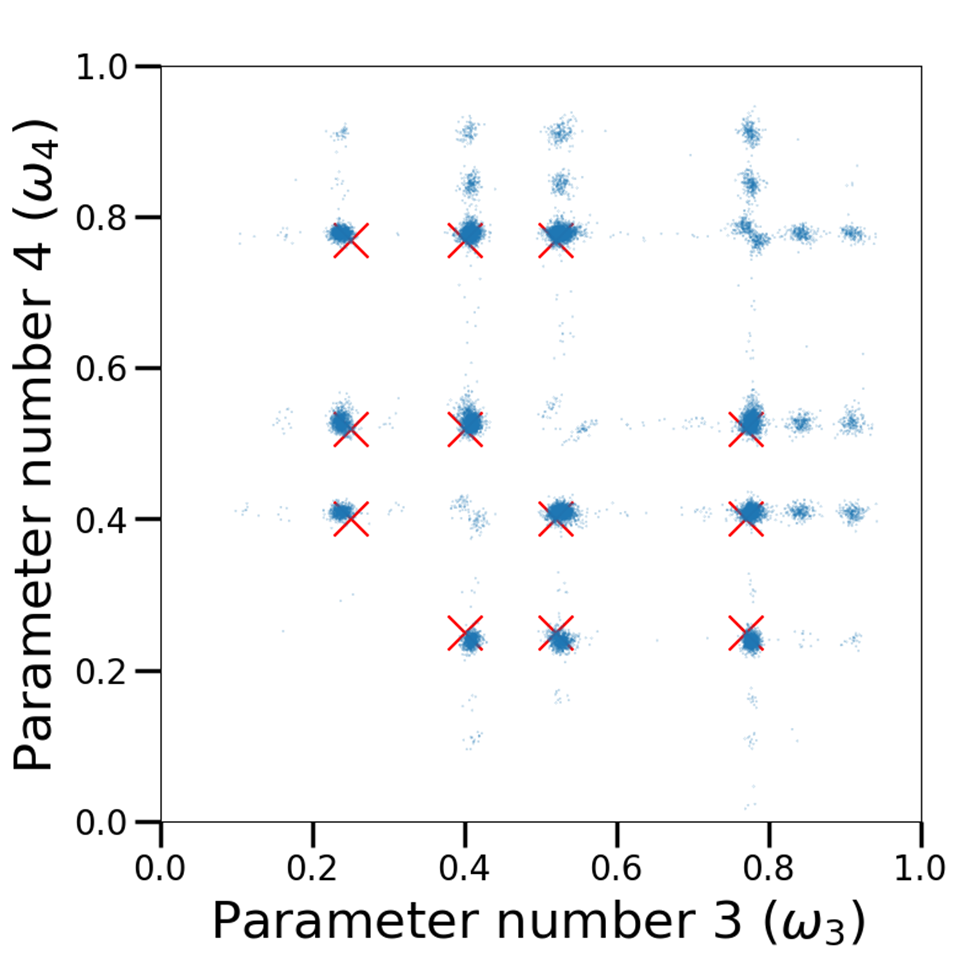}
\caption{}
\label{fig:4d_tle_b}
\end{subfigure}%
\caption{Results of multi-parameter (4-dimensional) inference for a multimodal sum-of-sinusoids likelihood (Equation \ref{eq:multimodallikelihood})  for two variations of sequential Monte Carlo: with the prior as importance function (\ref{fig:4d_sir_a} and \ref{fig:4d_sir_b}), and tempered likelihood estimation (\ref{fig:4d_tle_a} and \ref{fig:4d_tle_b}).
The modes are marked with red 'x' markers, with dimensions plotted pairwise. Tempered estimation performs better, providing increased mode coverage.}
\label{fig:4d_est}
\end{figure}

Finally, Figure \ref{fig:4d_est} shows the results of 4-dimensional estimation ($24$ modes) using the SMC of Algorithm \ref{alg:sir} and TLE-SMC, both with nearly $100\%$ HMC moves. Both used $12^4$ particles, $250$ data, and random times on $]0,100]$. For TLE, five coefficients were used. The added robustness of TLE is clearly evident, as is the increased complexity resulting from the extra dimensions.

%% file: Sections/tests2.tex
\section{Experiments on quantum hardware}
\label{sec:tests2}

This section presents the results of applying the previously explored techniques to the characterization of IBMQ qubits. The experiments were performed using IBM quantum services, Qiskit and Qiskit-Pulse \cite{Alexander_2020}. For details, refer to appendix \ref{app:expsetup}. 

Subsections \ref{sub:exp_t2} and \ref{sub:exp_t1} present the results of coherence time estimation ($T_2$ and $T_1$ respectively - the characteristic times for dephasing and amplitude damping). Subsections \ref{sub:exp_ramsey_2d} and \ref{sub:exp_ramsey_1d} concern Ramsey estimation \cite{Zwiebach_2013,Gross_2013,Frimmer_2014,Vitanov_2015,Deutsch_2005} without and with refocusing pulses.

Except where stated otherwise, we plot median results across $100$ independent executions; measurement times are chosen non-adaptively on a grid; the posterior distribution is discretized via SMC with the prior as importance function and MCMC move steps (appendices \ref{sec:sequential_monte_carlo} and \ref{sec:mcmc}); and the chosen Markov kernel is RWM with a proposal variance proportional to the current SMC estimate of the variance. This proportionality was tuned to obtain a stable particle acceptance rate around $65\%$.

For each set of results, the backend estimates at the time of data collection are presented where relevant and applicable. 

\subsection{Hahn echo experiment}
\label{sub:exp_t2}

\begin{figure}[!ht]
    \centering
    \includegraphics[width=\columnwidth]{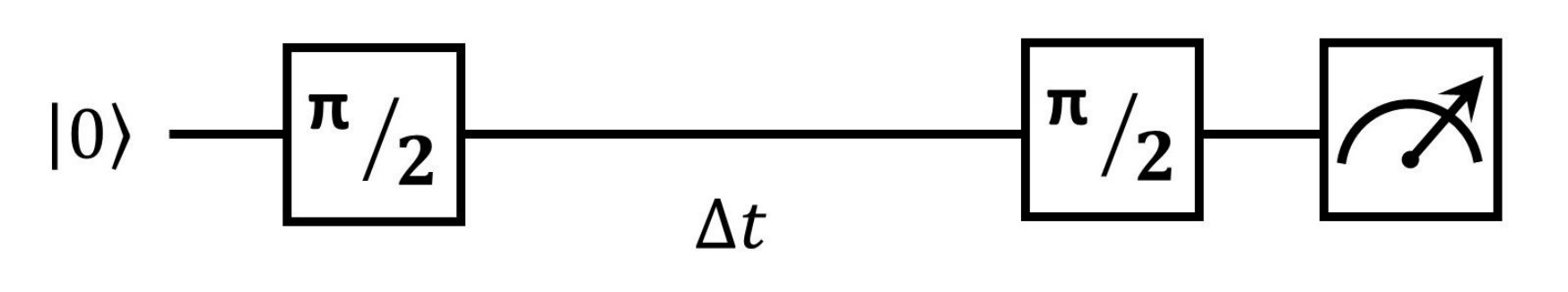}
    \caption{Pulse schedule for measuring $T_2$. }
    \label{fig:t2_circ}
\end{figure}

In this experiment, the parameter of interest is the characteristic time constant $T_2$. Dephasing and energy relaxation combine to produce a $T_2$ which can be estimated using the circuit \ref{fig:t2_circ} for several waiting times $\Delta t$ between half-pi pulses. 

The purpose of the half-pi pulses is to take the Bloch vector to the equatorial plane, where the $T_2^*$ is most relevant, and then back for the computational basis measurement. In the $x$ basis, the likelihood is: 
\begin{equation}
    \label{eq:t2}
    \mathbf{P}(+ \mid \Delta t) = \frac{1}{2} + \frac{e^{- \Delta t/T_2}}{2}
\end{equation}

A Hadamard gate directly translates this to $\mathbf{P}(0 \mid \Delta t)$, but any rotation that takes the Bloch vector to the $x$-$y$ plane will do, such as rotations of $\pi/2$ radians around the $x$ or $y$ axis.

\begin{figure*}[!htb]
\captionsetup[subfigure]{width=.9\textwidth}%
\begin{subfigure}{.5\textwidth}
  \centering
  \includegraphics[width=\linewidth]{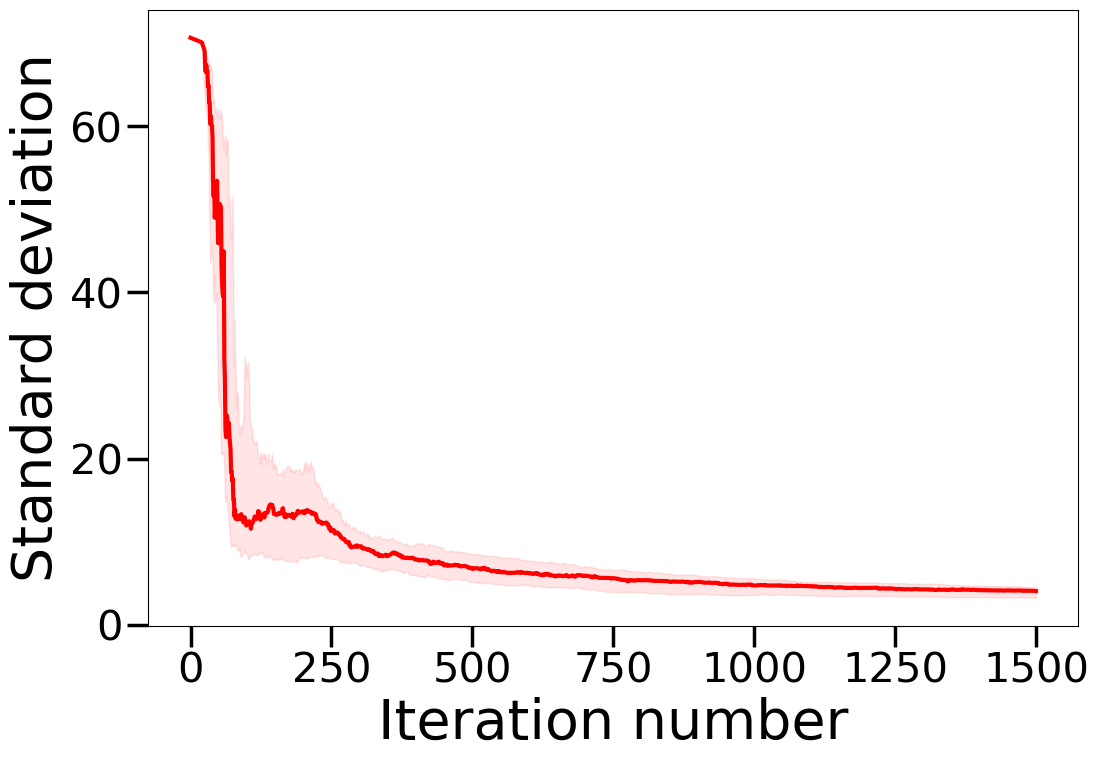}
  \caption{Data added by order of ascending evolution time. }
  \label{fig:T2_non_reversed}
\end{subfigure}
\begin{subfigure}{.5\textwidth}
  \centering
  \includegraphics[width=\linewidth]{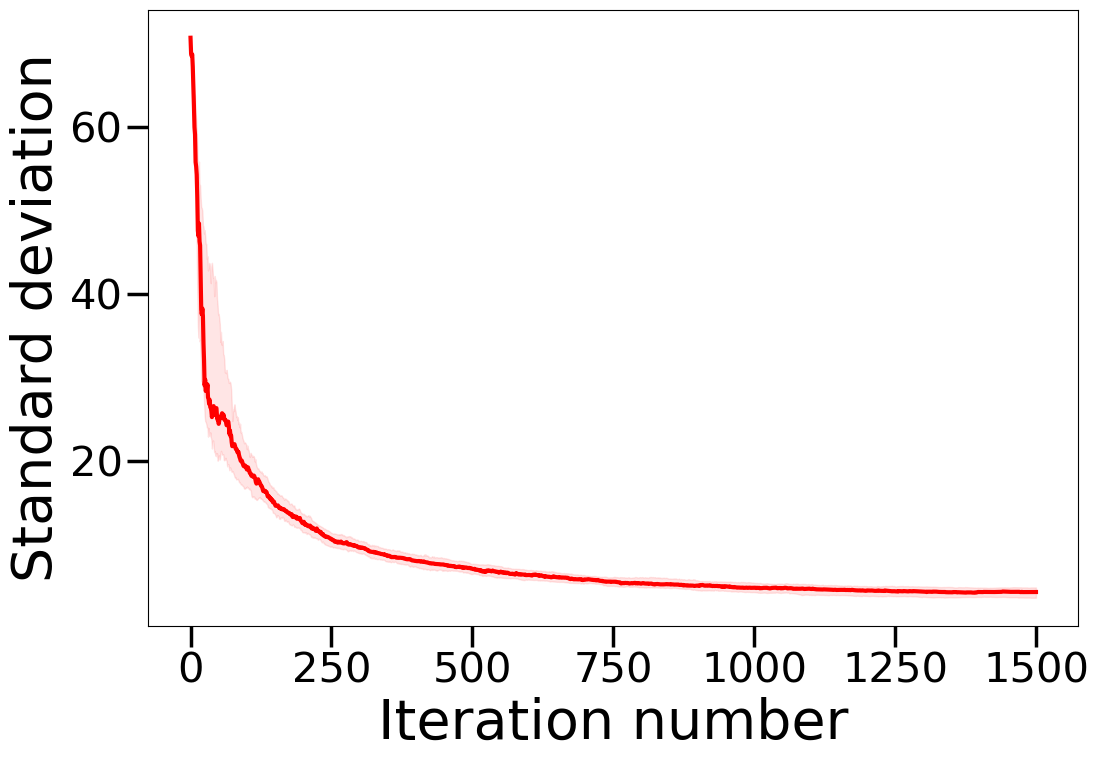}
  \caption{Data added by order of descending evolution time. }
  \label{fig:T2_reversed}
\end{subfigure}
\caption{Evolution of the standard deviation of the dephasing time $T_2$ during the Bayesian inference process, for the IBMQ device \texttt{ibmq\_rome} and two different orders of data addition. Both results were based on the same dataset containing $1500$ observations. The shaded region represents the interquartile range. The ascending order results in a less smooth and less effective process.}
\label{fig:T2_evolution}
\end{figure*}

In the case of exponential decay models such as this one, the sampler benefits from long evolution times being considered sooner, because the information is more challenging near the origin. This stands in contrast with the precession frequency example, where the evolution times typically grow exponentially and are tacitly considered in increasing order.
To illustrate this, in Figure \ref{fig:T2_evolution} we report the standard deviation at each iteration. The only difference between Figure~\ref{fig:T2_evolution}a and b is the order in which the data are fed to the sampler. In the case of figure \ref{fig:T2_non_reversed}, the shortest times were considered first, whereas in that of \ref{fig:T2_reversed}, the opposite was done. The sharp drop in the uncertainty, followed by a plateau, and the larger inter-run variability in the first case are signs of a sharp drop in effective sample size and undue particle concentration (owing to particle collapse). Conversely, the second graph shows a steadier learning process. Thus, the longer measurement times were included earliest in the sequence of SMC target distributions.

\begin{figure}[!htb]
    \centering
    \includegraphics[width=\columnwidth]{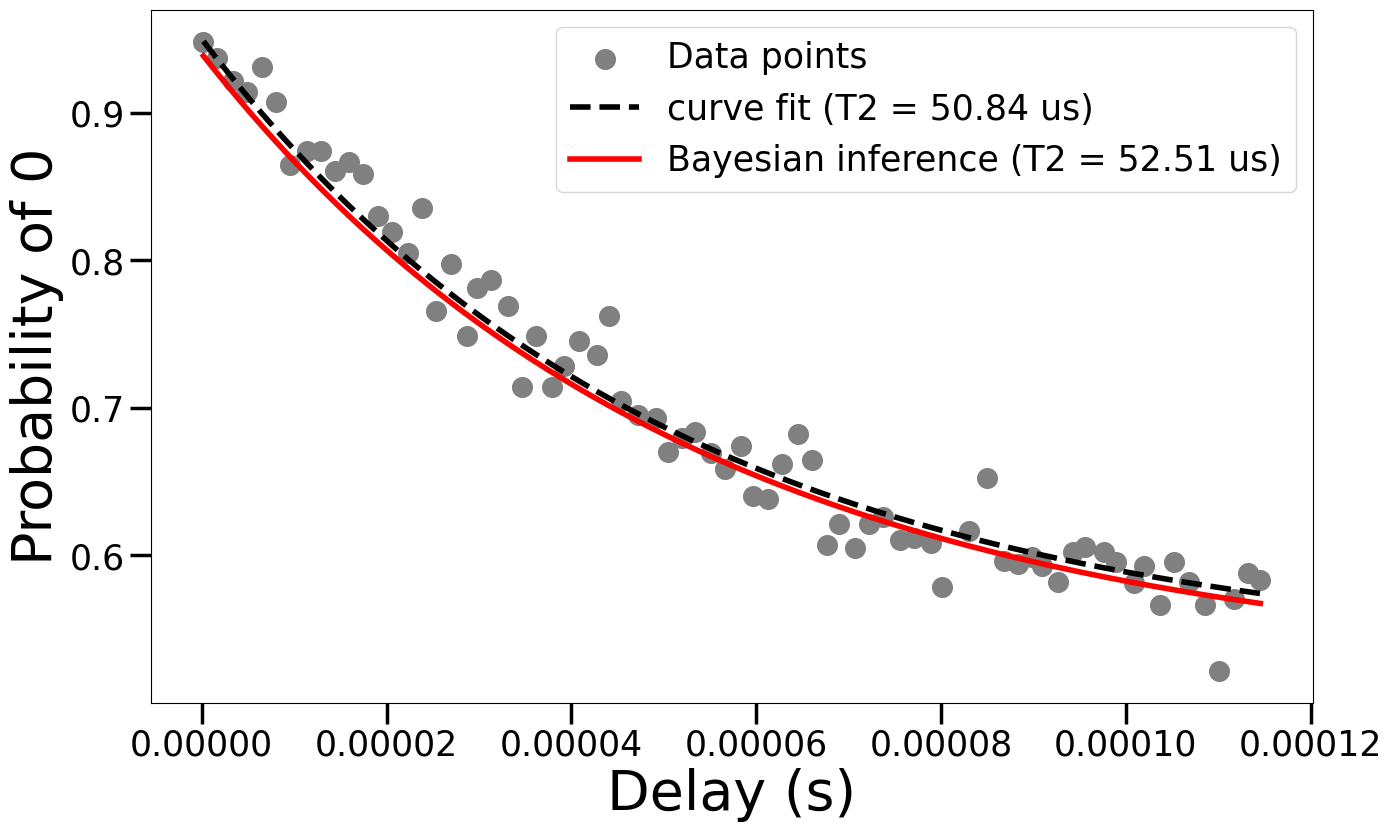}
    \caption{Curves produced by $T_2$ estimates for the IBMQ device \texttt{ibmq\_casablanca}. The curve fit relied on $512$ shots per point; for the inference, only $4.3\%$ of these data were used, even though the achieved uncertainties are similar.}
    \label{fig:T2_bayes_adj}
\end{figure}

Due to systematic errors, the coefficient and constant of equation \ref{eq:t2} may not be exactly $1/2$. For this reason, we change equation \ref{eq:t2} into $\mathbf{P}(1 \mid \Delta t)=A\exp(-t/T_2)+B$, as is done by default in the Qiskit fits. We then include a rough estimate of two constants $A$ and $B$ in the inference model. This can be achieved with a higher shot count for $t=0$ and some large $t \rightarrow \infty$ (implying a constant measurement overhead). Results are shown in figure \ref{fig:T2_bayes_adj} for the \texttt{ibmq\_casablanca} backend. For Bayesian inference, we used $100$ shots (roughly $20\%$ of the amount used by the Qiskit fitter) for the calibration of each $A$ and $B$.

The conditions for Bayesian inference were as follows. A $50$ particle SMC approximation was used, with a flat prior on $]0,250]\mu s$. A single Markov (RWM) move was used per particle per step. The threshold effective sample size for resampling was set at $\widehat{\text{ESS}}=0.8$. A total of $1500$ different measurement times were used for $75$ different measurement times; the repetition of measuring times was due to the restrictions of Qiskit, but also reflects the difficulty of single-shot readout for some systems, e.g. nuclear spins (although it is feasible \cite{Joas2021_single}). There were thus 1500 data/steps, plus $2 \cdot 512$ additional shots to customize $A$ and $B$. Of the total steps, 33 (2\%) triggered a resampling stage on average. The measurement times were chosen in constant increments within $t \in ]0,115] \mu s$. The results were obtained by taking the medians over $100$ runs, split equally across $10$ different datasets. 

Quantitative results are shown in Table \ref{tb:t2}, where they are also compared to those of the Qiskit fitter. The latter was applied in two settings. First, using $512$ shots per step. In this case, both the obtained values and their associated uncertainties were close to our Bayesian approach, despite the $256\%$ increase in shots. Second, allowing the curve fitter roughly as many total measurements as were used for our Bayesian inference algorithm. In this case, the achieved error was over 3 times larger than that obtained by Bayesian inference.

\begin{table*}[!htb]
\centering
\renewcommand{\arraystretch}{0.8}
\begin{tabular}{@{}l c c r@{}}
\toprule
 & $\mathbf{T_2}$ ($\mu s$) & $\boldsymbol{\sigma}$ ($\mu s$) & \textbf{No. of measurements} \\
\midrule
\textbf{Bayesian inference} & 52.51 & 4.4 & $2\,524$ \\
\textbf{Qiskit fitter}      & 52.39 & 3.7 & $38\,400$ \\
\textbf{Qiskit fitter}      & 56.70 & 13.9 & $2\,535$ \\
\bottomrule
\end{tabular}
\caption{Dephasing time estimation results for a qubit of the IBMQ device \texttt{ibmq\_casablanca}. The registered backend estimate was at the time $T_2 = 57.58\mu s$ for the two first lines, and $48.31\mu s$ for the last one. Our inference algorithm performs better: it achieves similar uncertainty to Qiskit's fitter using 11 times less data, and an uncertainty over 3 times smaller using the same number of data.}
\label{tb:t2}
\end{table*}

Note that when using the Qiskit fitter in the low-data regime, we do not use the same data as were used for our Bayesian inference approach, instead controlling for the timespan and shot number. Providing the fitter the same  data yielded an error in the order of the thousands of microseconds. To improve its performance, we redistribute the shots by fewer measurement times, optimizing this set of times. For the inference, no such optimization was performed; the measurement times were spaced out uniformly up to a fixed maximum time, which was the same for both strategies.

These comparative tests demonstrate that Bayesian inference excels at extracting information from small datasets, as compared to standard methods such as curve fitting. This is expected to be enhanced if using Bayesian experimental design, which we do not due to implementation constraints. 

\subsection{T1 estimation}
\label{sub:exp_t1}

\begin{figure}[!ht]
    \centering
    \includegraphics[height=1.75cm]{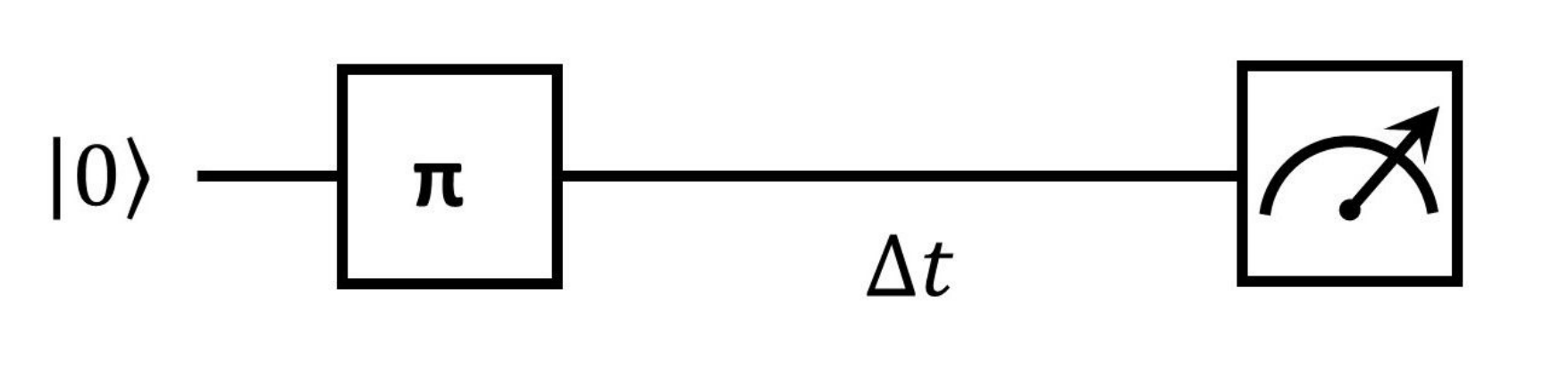}
    \caption{Pulse schedule for measuring $T_1$.}
    \label{fig:t1_circ}
\end{figure}

In this experiment, the parameter of interest is the energy relaxation time $T_1$. This parameter can be estimated using the circuit of figure \ref{fig:t1_circ} for several waiting times $\Delta t$ since the initialization at $\ket{1}$. The pi pulse realizes a rotation of $\pi$ radians in the Bloch sphere, taking the $\ket{0}$ state to the $\ket{1}$ state. It can be a $\opsub{\sigma}{x}$ gate, in which case the rotation is around the $x$ axis.

The resulting likelihood takes the form: 
\begin{equation}
\label{eq:t1}
  \mathbf{P}(1 \mid \Delta t) = e^{- \Delta t/T_1}
\end{equation}

\begin{figure}[!htb]
    \centering
    \includegraphics[width=\columnwidth]{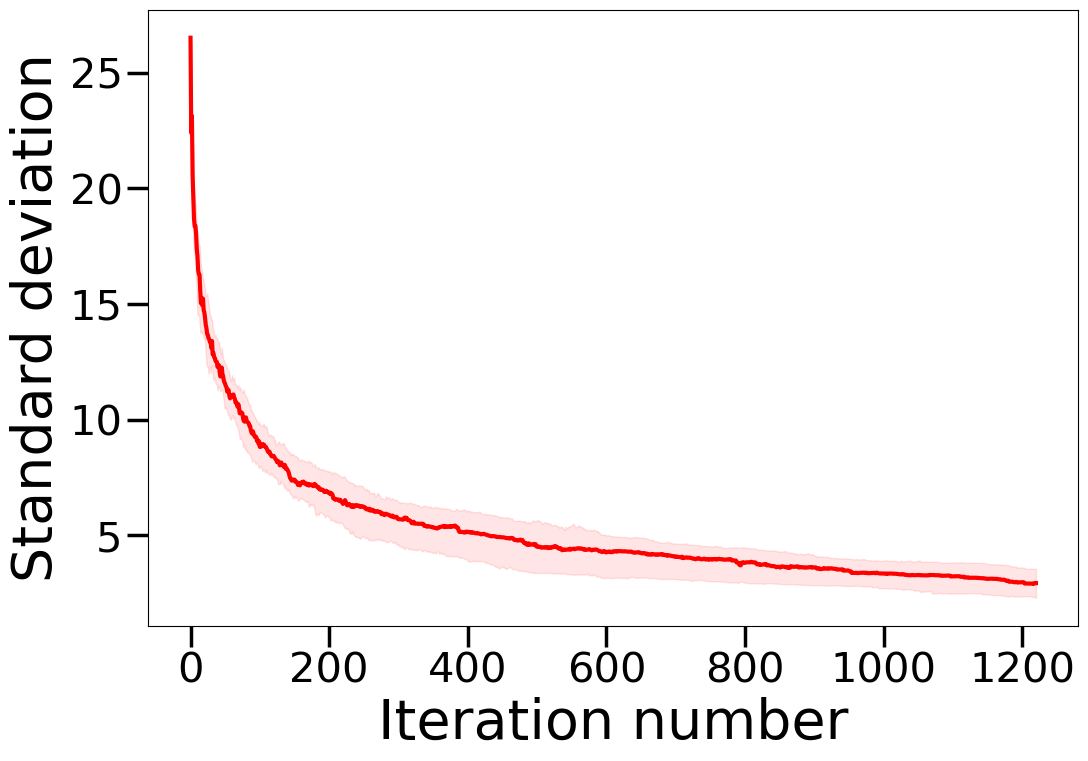}
    \caption{Evolution of the standard deviation during the estimation of the energy relaxation time constant $T_1$ of IBMQ device \texttt{ibmq\_guadalupe}. The dataset contemplated $1500$ observations. The shaded region represents the interquartile range. As expected, the behavior is quite similar to that of Figure \ref{fig:T2_reversed}.}
    \label{fig:t1_evolution}
\end{figure}

Figure \ref{fig:t1_evolution} shows the evolution of the uncertainty as the inference proceeds. The final standard deviation was $2.9$

The conditions were similar to section \ref{sub:exp_t2}, apart from two points. The prior was set on $]0,100]\mu s$ (this doesn't have much of an impact after the first few iterations), and the measurement times on $]0,50]\mu s$. A standard deviation of $\sigma = 2.9 \mu s$ for an estimate of $T_1=62.45\mu s$ was achieved.

\subsection{Ramsey experiment}
\label{sub:exp_ramsey_2d}

This subsection presents the results of performing Ramsey experiments on IBM quantum devices. The experimental scheme for Ramsey interferometry is based on two $\pi/2$ pulses of frequency $\omega$ detuned by $\delta$ with respect to an estimate of the qubit's resonant frequency $\widetilde{\Omega}_0$. Their application is separated by free evolution for a time $\Delta t$ during which the only action on the system is the parallel field $\opsub{H}{0}$. During this period, with no driving, the Bloch vector oscillates at a frequency $\delta = \Omega_0-\omega$ around the $z$ axis, $\Omega_0$ being the resonant frequency and the quantity of interest.

\begin{figure}[!ht]
    \centering
    \includegraphics[width=\columnwidth]{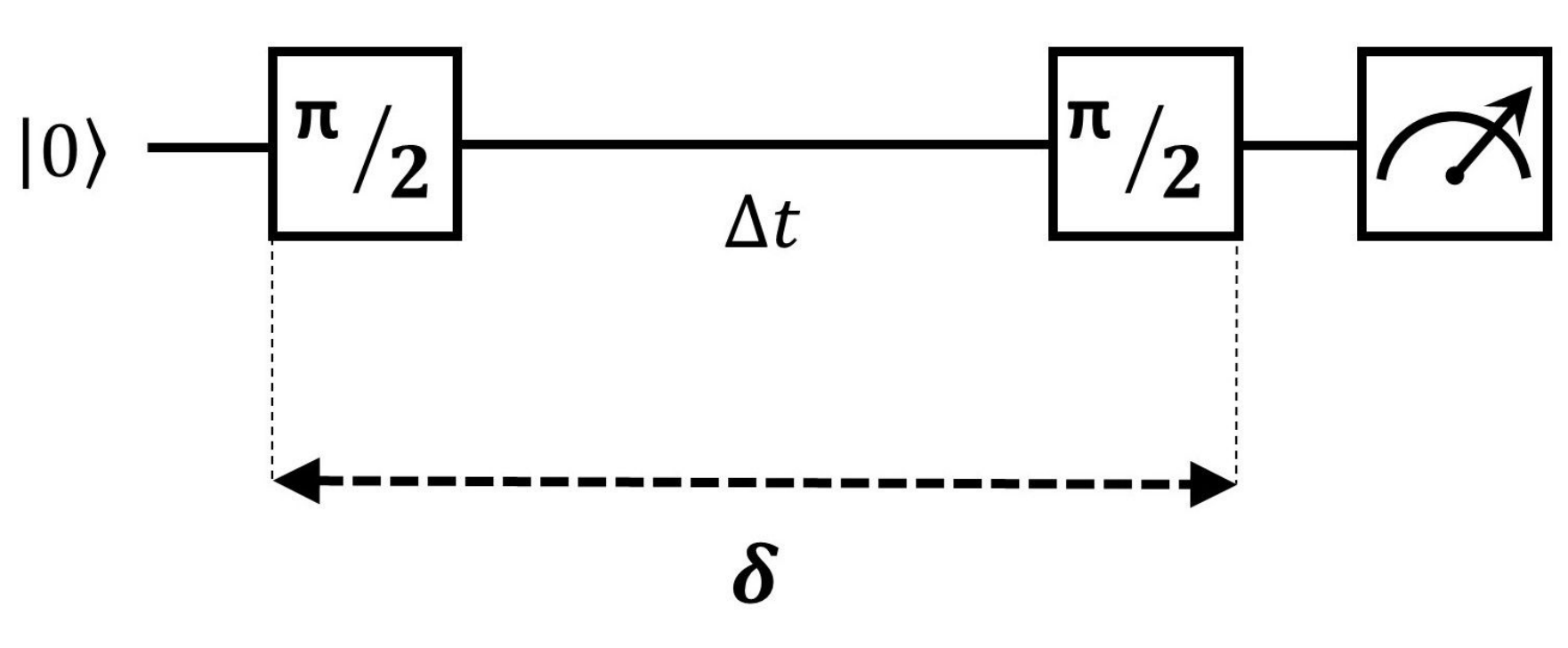}
    \caption{Pulse schedule for estimating the detuning $\delta_\text{meas}$ in a Ramsey experiment. This is known as a \textit{Ramsey sequence}. $\boldsymbol{\delta}$ is an experimental control determining the pulse frequency. The frequency estimate can then be corrected to higher precision as $\Omega_0 = \widetilde{\Omega}_0+\delta_\text{meas}$}..
    \label{fig:ramsey_circ}
\end{figure}

This technique is central for high precision metrology, including applications as sensitive as atomic clocks \cite{Vitanov_2015}. The basic sequence to be used for the Ramsey experiments is depicted in figure \ref{fig:ramsey_circ}. 

Measuring in the $z$ basis gives (assuming initialization at $\ket{0}$):
\begin{equation}
    \label{eq:ramsey_p1}
    \mathbf{P}(1 \mid t) = \cos^2
    \left( \frac{\delta}{2}\cdot \Delta t \right)
\end{equation}

In practice, the oscillation pattern is unlikely to exactly recreate equation \ref{eq:ramsey_p1}, due to experimental error and environment effects. The protocol suffers from decoherence, as well as other sources of error such as fluctuations in the detuning. The result is that the sinusoidal wave suffers from damping, which can be modeled by an exponential decay envelope:
\begin{equation}
    \label{eq:ramsey_p1_decay}
    \mathbf{P}(1 \mid t) = e^{- t/T_2^*}\cos^2
    \left( \frac{\delta}{2}\cdot \Delta t \right)
    + \frac{1-e^{-t/T_2^*}}{2}
\end{equation}

This imposes a symmetric effect on the basis states, since decoherence is expected to affect $\ket{+}$ and $\ket{-}$ equally (the $\pi/2$ pulses translate these results into the computational basis). This expression can also be derived by considering Lorentzian noise in the real parameter, and replacing it with a pair of hyperparameters \cite{Granade_2012}.

We use the circuit from figure \ref{fig:ramsey_circ} to estimate the detuning $\delta$ and coherence time $T_2^*$ from equation \ref{eq:ramsey_p1_decay}. Instead of targeting the latter directly, the factor $\gamma_2^*=1/T_2^*$ is considered, so that the two parameters are matched in units (though not necessarily scale) as suggested in \cite{Granade_2012}. 

For the isolated characteristic time estimation of subsections \ref{sub:exp_t1} and \ref{sub:exp_t2}, the data were considered by order of increasing evolution times. For isolated frequency estimation, we have the opposite case. Here, both parameters are combined, creating a more delicate interplay. We found numerically that increasing times worked best, but this depends on the magnitude of the parameters and the sampling details.

\begin{figure}[!htb]
    \centering
    \includegraphics[width=8cm]{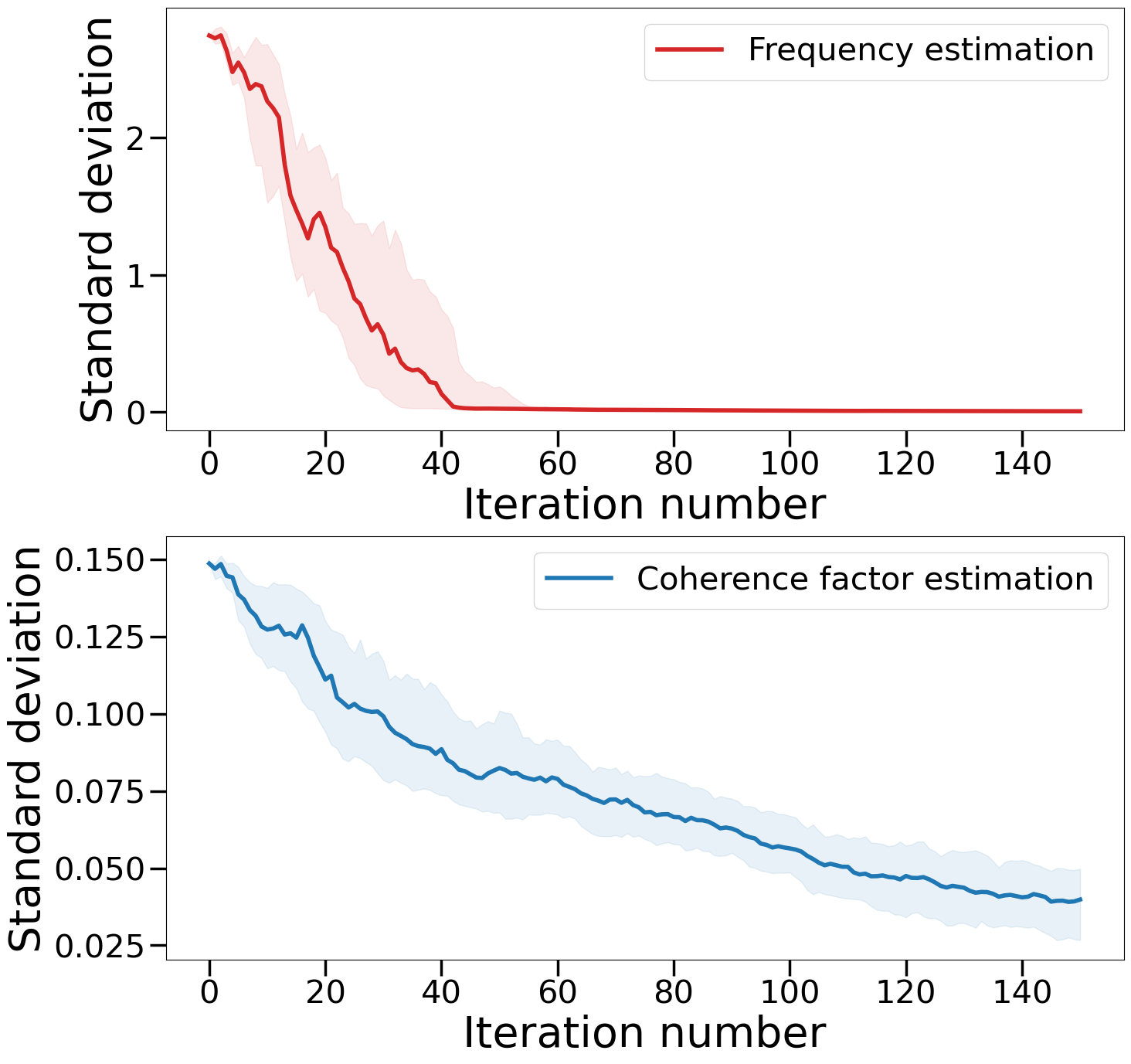}
    \caption{Evolution of the standard deviations during the estimation of the detuning frequency $\delta$ and coherence factor $\gamma_2^*$ of IBMQ device \texttt{ibmq\_armonk}. The dataset contemplated $150$ observations. The solid curves are the median results for 300 runs split by 100 datasets. The shaded regions represent interquartile ranges. The algorithm learns both parameters, although at different rates due to how the likelihood depends on each of them.}
    \label{fig:ramsey_evol}
\end{figure}

An example of the evolution of standard deviations through the iterations is shown in Figure \ref{fig:ramsey_evol}, where the difference between parameter learning rates is evident.

\begin{figure}[!htb]
    \centering
    \includegraphics[width=8cm]{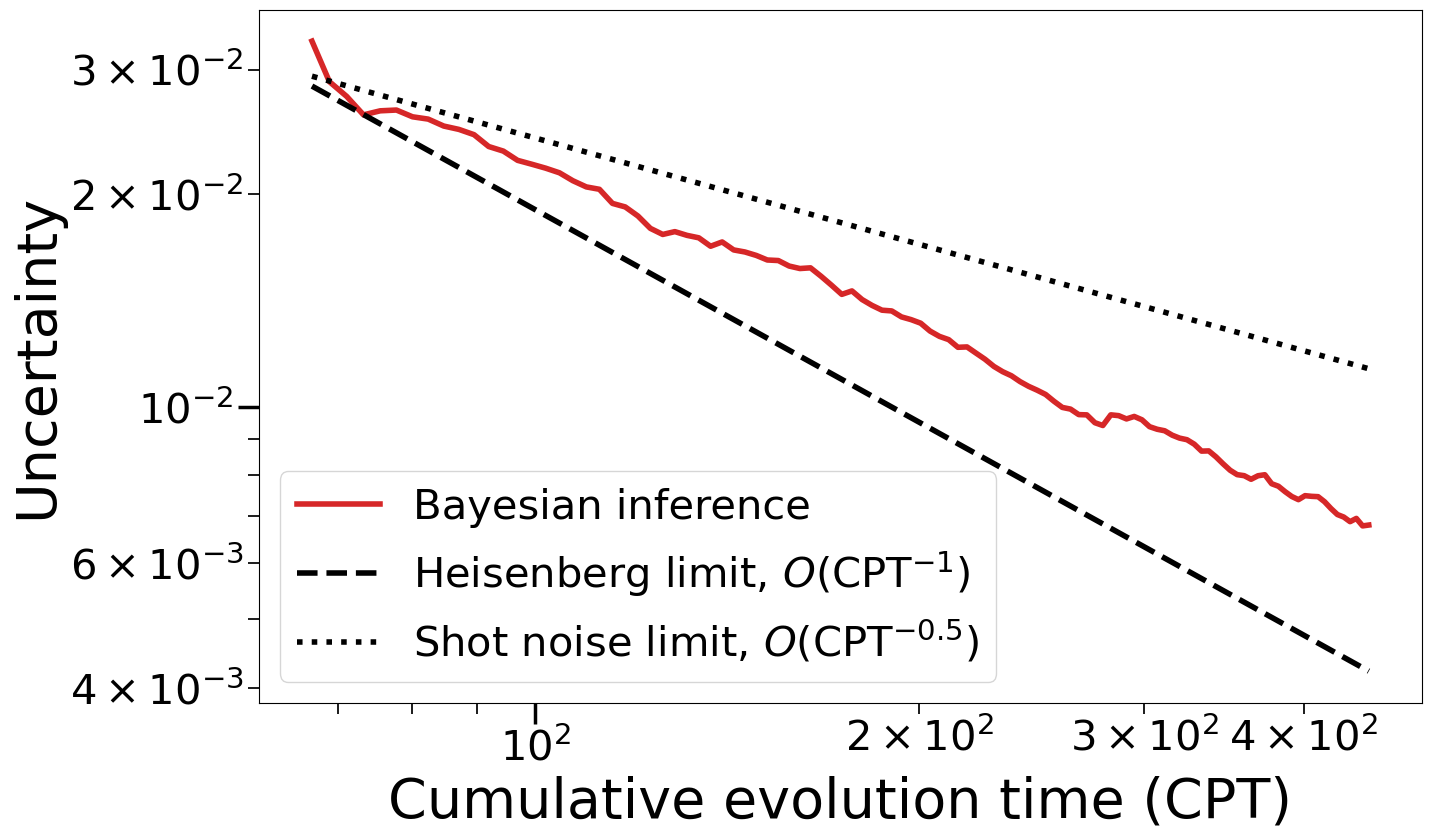}
    \caption{Evolution of the standard deviation during the estimation of the detuning frequency $\delta$ of IBMQ device \texttt{ibmq\_armonk}, as a function of the cumulative evolution time. Although it fails to saturate the Heisenberg limit, the Bayesian estimation process achieves sub-standard quantum limit performance despite the noise.}
    \label{fig:vs_HL}
\end{figure}

We can additionally compare the algorithm performance with the fundamental classical and quantum limits of metrology, Figure~\ref{fig:vs_HL}. In spite of the strong decoherence (the estimated coherence time was $11\mu s$), we observe quantum-enhanced estimation. The learning rate falls between the classical and quantum limits, roughly $\mathcal{O} (\text{CPT}^{-0.75})$.  

\begin{figure}[!htb]
\centering
\includegraphics[width=\linewidth]{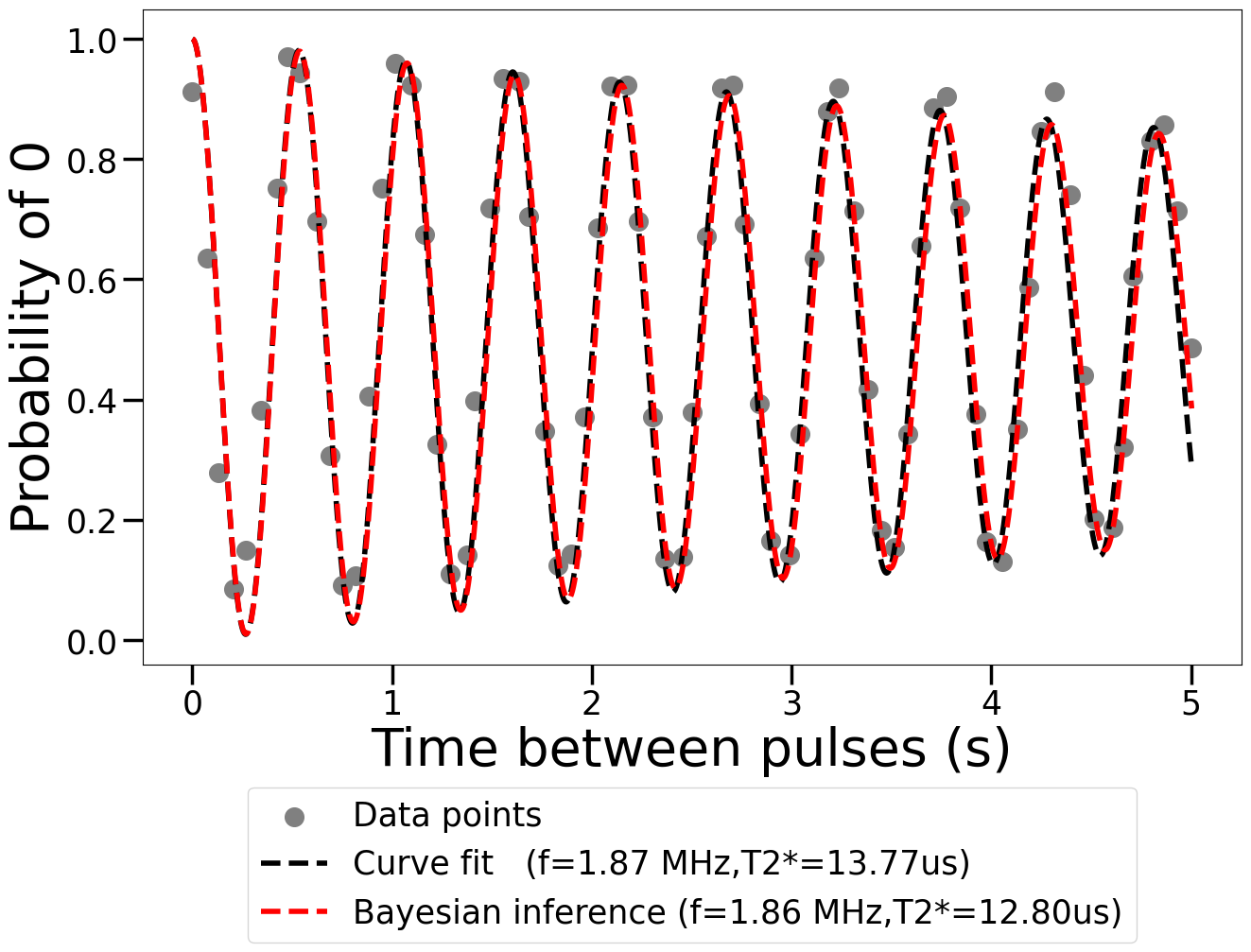}
\caption{Results for detuning frequency $\delta$ and coherence factor $\gamma_2^*$ estimation using Bayesian inference and a curve fit, superposed on the data points used for the latter, which used 512 shots each. For Bayesian inference, only $0.4\%$ (2) of the data were used, even though the results are nearly indistinguishable.}
\label{fig:ramsey_curves}
\end{figure}

 A visual comparison of the predictive power of the results obtained by inference and by curve fitting are presented in figure \ref{fig:ramsey_curves}. Both methods seem to fit the data well, and are nearly undistinguishable despite the mismatch in resource usage.

$T_2^*$ is susceptible to large variations, so comparisons are suceptible to time fluctuations. Qiskit does not make estimates of this constant available (despite still supplying a built-in fitter for it), likely due to its volatility.

The default Qiskit fitter for is geared towards coherence time estimation, and not frequency calibration. We thus focus on the estimates for this quantity, though the frequency still brings added complexity. Table \ref{tb:ramsey} presents quantitative final results for $T_2^*$ estimation. Our algorithm achieves a ten-fold reduction in uncertainty while using the same number of data. For reference, the uncertainties in the estimation of the induced oscillation frequency were $0.4\%$ for inference, and $15\%$ for curve fitting with the same number of shots. 

\renewcommand{\arraystretch}{0.8}
\begin{table}[!htb]
\centering
\begin{tabular}{@{}l c r@{}}
\toprule
 & $\sigma$ ($\mu s$) & \textbf{No. of measurements} \\
\midrule
\textbf{Bayesian} & 3 & $150$ \\
\textbf{Curve fit}      & 30 & $150$ \\
\bottomrule
\end{tabular}
\caption{Unechoed dephasing time estimation results for the IBMQ device \texttt{ibmq\_armonk}. Bayesian inference achieves a standard deviation $10$ times lower than the Qiskit fitter for the same No. of measurements.}
\label{tb:ramsey}
\end{table}

For the inference, a $225$ particle SMC approximation was used, with a flat prior on $f \in ]0.0,5.0]MHz$, $T_2^* \in [3.00,25.00]\mu s$. Three Markov moves were used per particle per step. The threshold effective sample size for resampling was set at $\widehat{\text{ESS}}=0.8$. A total of $75$ different measurement times was used for 2 shots each, signifying 150 data/steps. Of these, 28 (19\%) triggered a resampling stage on average. The measurement times were chosen in constant increments within $t \in ]2.0,5] \mu s$.

For a detuning of $\delta = 1.83MHz$, the inference final estimates were $\delta = 1.863 MHz$ and $T_2^*=9\mu s$, and their associated uncertainties $\sigma = 0.007 MHz$ and $3\mu s$. 

\subsection{Echoed Ramsey experiment}
\label{sub:exp_ramsey_1d}

We now perform \textit{Hahn-Ramsey} experiments on IBM quantum devices, testing adaptive experimental design (appendix \ref{app:bayesian_experimental_design}) on \texttt{ibmq\_armonk}. These modified experiments, detailed in appendix \ref{app:exp_ramsey_1d}, alleviate the effect of decoherence, allowing us to consider single-parameter estimation for conservative time evolutions. 

\begin{figure}[!htb]
    \centering
    \includegraphics[width=\columnwidth]{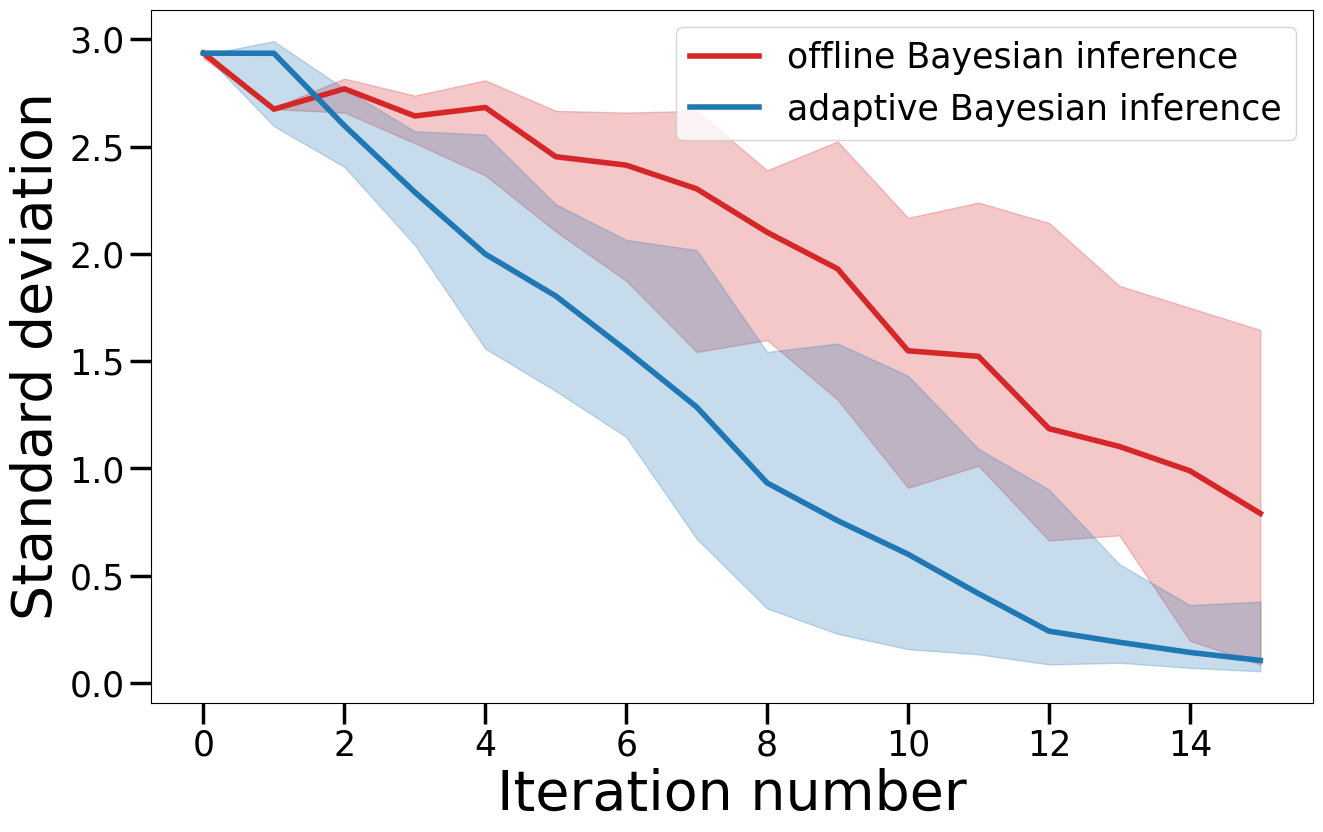}
    \caption{Evolution of the standard deviation during the estimation of a detuning frequency in the IBM device \texttt{ibmq\_armonk}, using adaptive and offline Bayesian inference. The shaded regions represent interquartile ranges. The adaptive approach learns faster.}
    \label{fig:adaptive}
\end{figure}

The strategy was compared to the non-adaptive (offline) estimation of before. Adaptivity serializes the data collection process; due to limited usage of the IBMQ devices, only $15$ data/steps were used for both methods, and $100$ runs. Still, the difference was clear (figure \ref{fig:adaptive}).

A $100$ particle SMC approximation was used, with a flat prior on $\delta \in ]0.0,10.0]MHz$. One Markov move was used per particle per step. The threshold effective sample size for resampling was set at $\widehat{\text{ESS}}=0.5$. A total of $15$ single-shot measurements for different times were used, signifying $15$ steps. Of these, $7$ ($47\%$) and $9$ ($60\%$) triggered a resampling stage on average for the adaptive and offline methods, respectively. The offline measurement times were chosen in constant increments within $t \in ]0.2,2] \mu s$, and the adaptive ones were chosen from $20$ guesses around $1/\sigma_\text{curr}$ to have the lowest variance (as suggested in appendix \ref{app:precession_heuristics}). The results were obtained by taking medians over $100$ runs split equally by $10$ different datasets. The quantitative results are presented in Table \ref{tb:ramsey_adaptive}. The adaptive approach achieves lower uncertainty.

\begin{table*}[!htb]
\centering
\renewcommand{\arraystretch}{0.8}
\begin{tabular}{@{}l c c c c@{}}
\toprule
 & $\boldsymbol{\delta}$ (MHz) &  $\boldsymbol{\sigma}$ (MHz) & \textbf{No. of measurements} & \textbf{Precision} ($\sigma^2 \text{CPT}$) \\
\midrule
\textbf{Adaptive} & 1.8 & 0.1 & 15 & 0.38 \\
\textbf{Offline} & 2.3 & 0.8 & 15 & 9.3 \\
\bottomrule
\end{tabular}
\caption{Results of using adaptive and offline inference to learn a detuning frequency on the IBMQ device \texttt{ibmq\_armonk}. The detuning relative to the prevailing backend estimate was $1.83 MHz$. The adaptive strategy accelerates the learning process, achieving a lower uncertainty and a better precision.}
\label{tb:ramsey_adaptive}
\end{table*}

An additional quantity was considered: the \textit{precision}, defined as in \cite{Santagati_2019} ($\sigma^2\Delta t_\text{acc}$, where $\Delta t_\text{acc}$ is the cumulative evolution time). This is meant to penalize long measurement times, and judge whether the adaptive method unfairly benefits from lengthier evolutions. According to this metric, adaptivity still outperforms the offline method. Note that penalties of this type can be incorporated into the utility function, if curtailing simulation runtime is critical.

\begin{figure}[!htb]
  \centering
  \includegraphics[height=5cm]{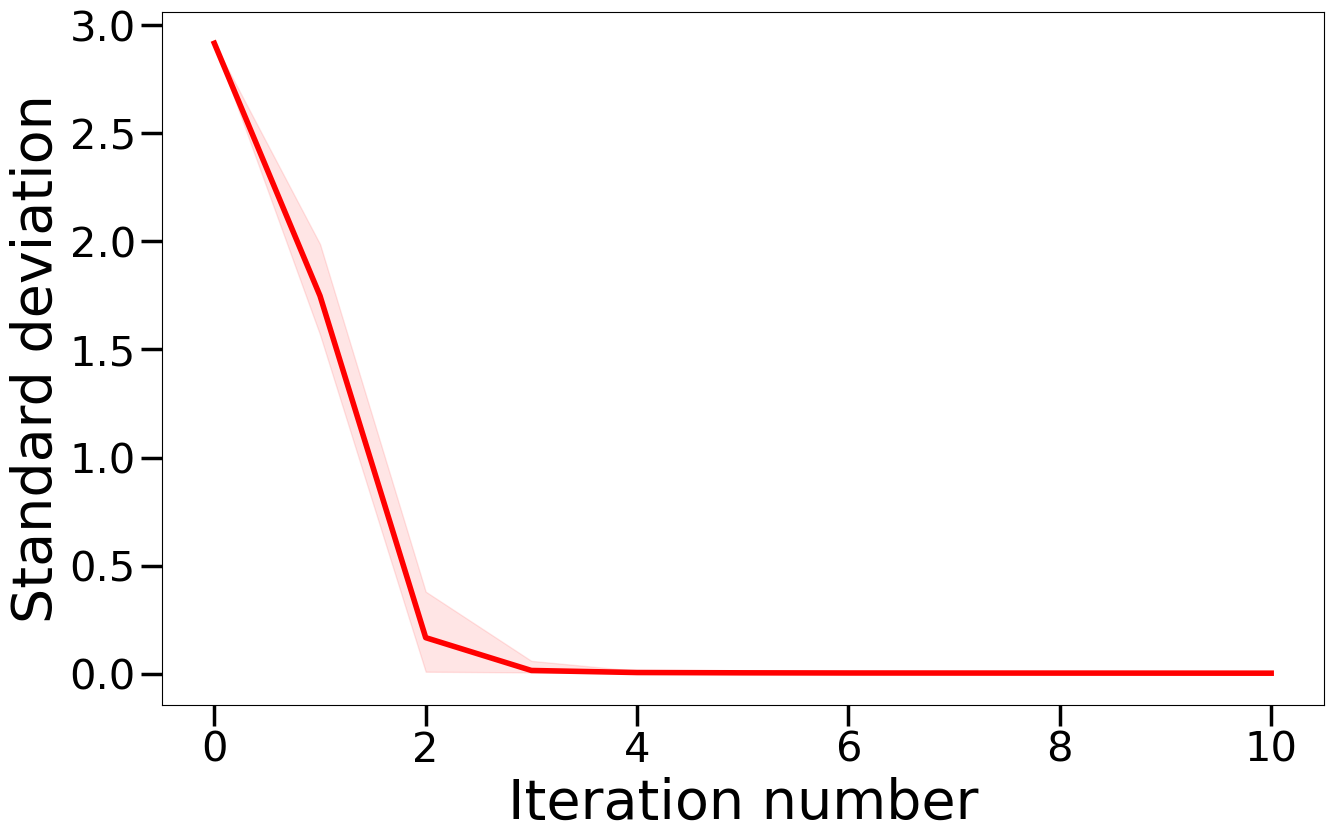}
\caption{Evolution of the standard deviation during the estimation of a detuning frequency on the IBMQ device \texttt{ibmq\_armonk}, using Bayesian inference with tempered likelihood estimation with Hamiltonian Monte Carlo. The dataset contemplated $75$ observations, and ten tempering coefficients were used.}
\label{fig:tempered_hmc_all}
\end{figure}

Next, TLE with HMC move steps was tested (figure \ref{fig:tempered_hmc_all}) for frequency estimation in the device \texttt{ibmq\_armonk}. A $100$ particle SMC approximation was used, with a flat prior on $f \in ]0.0,10.0]MHz$. One Markov (HMC) move was used per particle per step. The HMC hyperparameters were set to $M=\/\sigma_\text{curr}^2$, $L=20$, $\epsilon = 0.001$, producing an acceptance rate of $68\%$. The threshold effective sample size for resampling was set at $\widehat{\text{ESS}}=0.5$. A total of $75$ different measurement times was used for a single shot each. The tempering coefficients were spaced on a grid for a total of $10$, signifying $10$ steps. Of these, $3$ ($30\%$) triggered a resampling stage on average. The measurement times were chosen in constant increments within $t \in ]0.2,10] \mu s$. The results were obtained by taking medians over $100$ runs split equally by $10$ different datasets. The quantitative results are presented in table \ref{tb:ramsey_1d}. The algorithm correctly identifies the detuning frequency, with an uncertainty of the order of the nanoseconds. 

\begin{table*}[!htb]
\centering
\renewcommand{\arraystretch}{0.8}
\begin{tabular}{@{}l c c c@{}}
\toprule
 & $\boldsymbol{\delta}$ (MHz) & $\boldsymbol{\sigma}$ (MHz) & \textbf{No. of measurements} \\
\midrule
\textbf{Bayesian inference} & 1.830 & $6 \times 10^{-3}$ & 75 \\
\bottomrule
\end{tabular}
\caption{Hahn-Ramsey frequency estimation results for the IBMQ device \texttt{ibmq\_armonk}, using tempered likelihood estimation with Hamiltonian Monte Carlo resampling. The detuning relative to the prevailing backend estimate was $1.83 MHz$, and it was correctly identified by the algorithm. }
\label{tb:ramsey_1d}
\end{table*}

\begin{figure*}[!htb]
    \centering
\captionsetup[subfigure]{width=.9\textwidth}%
\begin{subfigure}[t]{.45\textwidth}
  \centering
  \includegraphics[width=\linewidth]{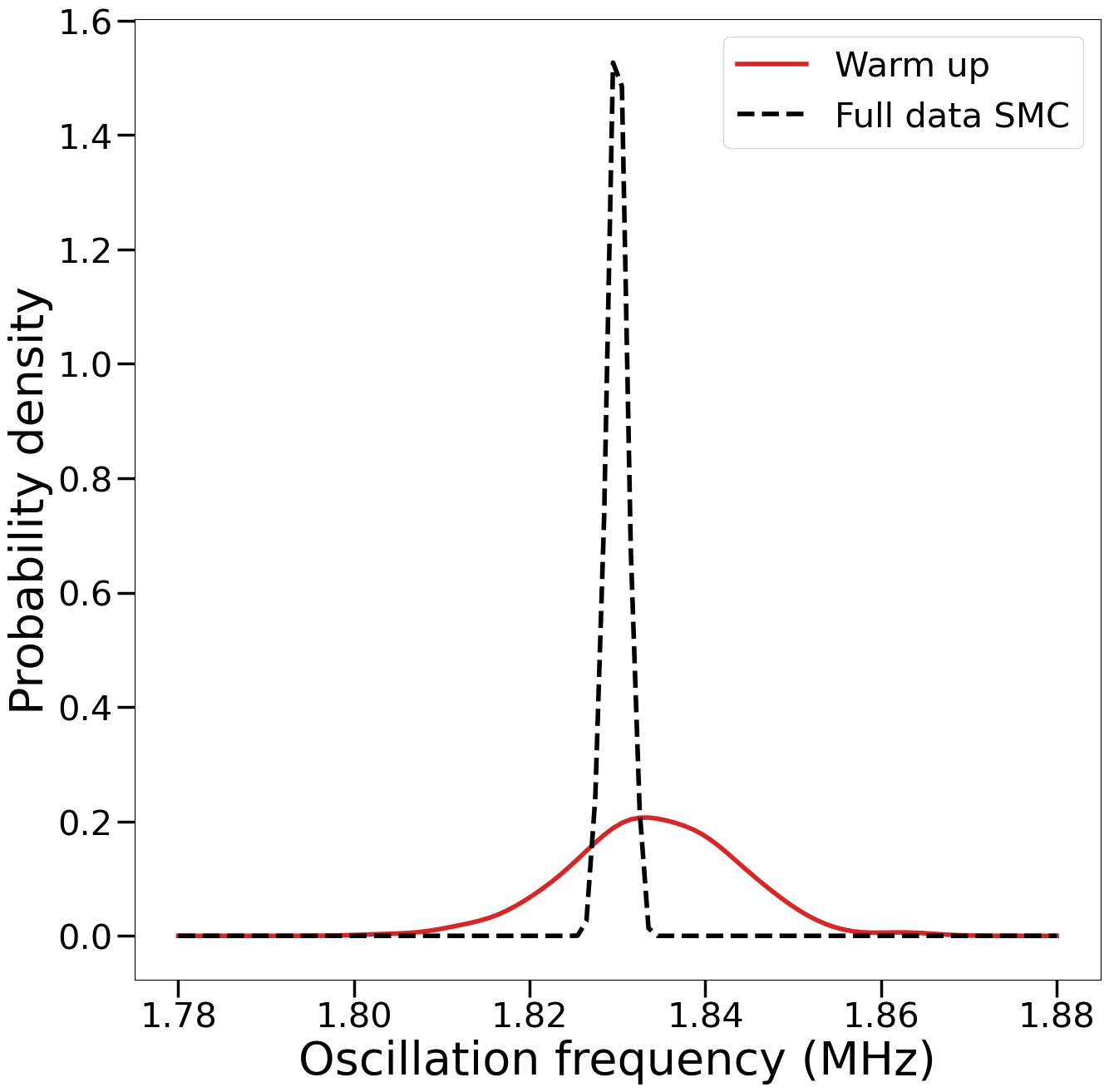}
  \caption{Kernel density estimates of the full posterior (375 data) and of the warmed up distribution for subsampling (150 data).}
  \label{fig:subs_warmup}
\end{subfigure}
\begin{subfigure}[t]{.45\textwidth}
  \centering
    \includegraphics[width=\linewidth]{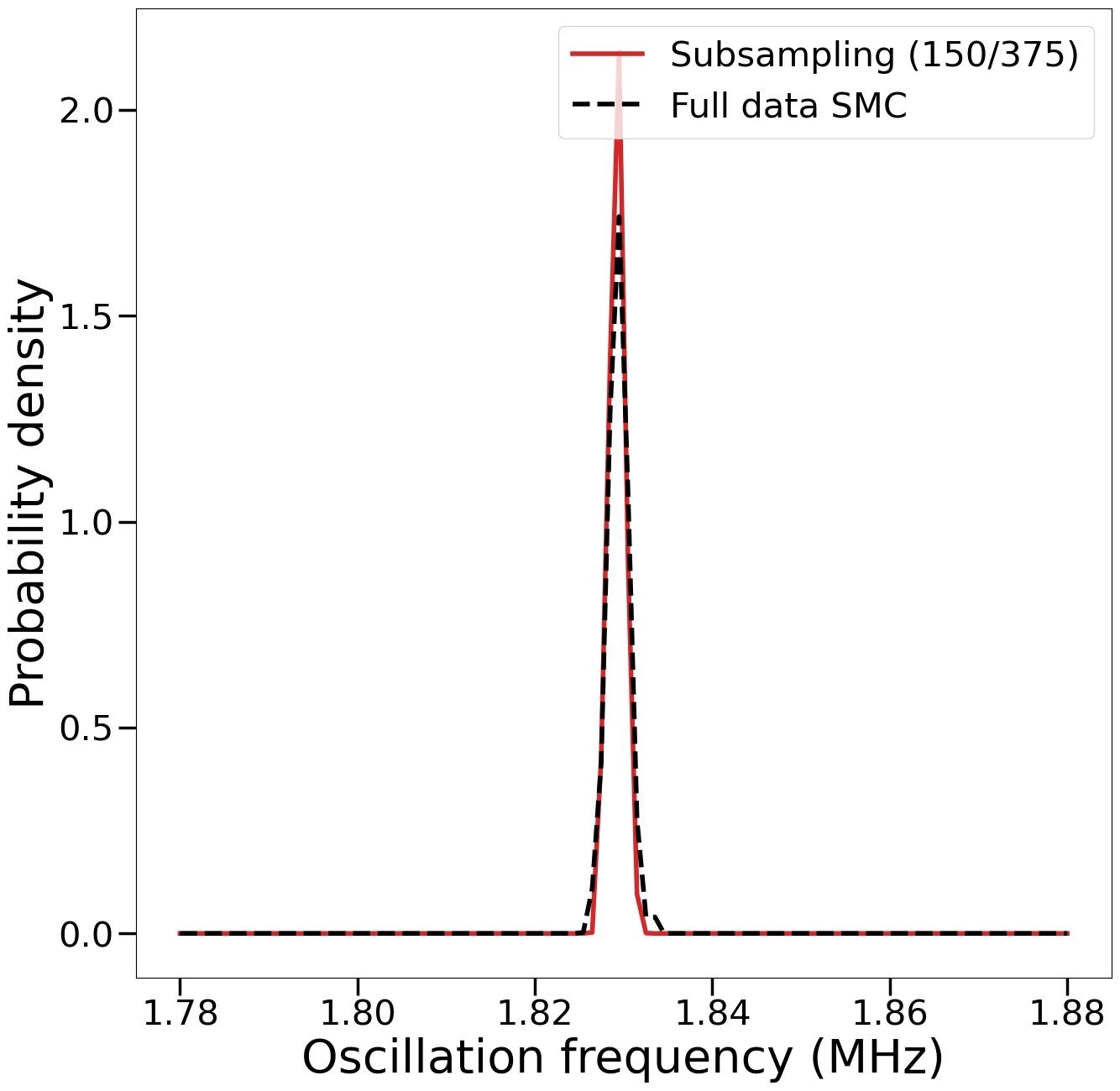}
  \caption{Kernel density estimates of the full posterior (375 data) and of the subsampling posterior (150 observations).}
  \label{fig:subs_final}
\end{subfigure}
\caption{Results of using tempered likelihood estimation with random walk Metropolis resampling and data subsampling (at 40\%) with control variates to learn a detuning frequency on the IBMQ device \texttt{ibmq\_armonk}. The prior was flat in $]0,10]MHz$. $375$ observations and $10$ tempering coefficients were used, $3$ of which were employed in the warm up when subsampling. The subsampled distribution achieves similar results to the original one, while cutting likelihood evaluations by over half.}
\label{fig:subs_tle}
\end{figure*}

Finally, we tested frequency estimation via subsampling TLE with RWM move steps with control variates (appendix \ref{app:subsampling}). For practical reasons, the prior is narrowed down via a warm up on the lower times before employing control variates. The results are presented in figure \ref{fig:subs_tle}. 

The experiment was performed in similar conditions to the previous one, except for a few aspects. The number of particles was $200$, and they were resampled at each step. A total of $75$ different measurement times was used for $5$ shots each, totaling $375$ data.  

The quantitative results are presented in table \ref{tb:ramsey_1d_subs}. The difference in the estimator between the full data and subsampling cases was negligible. The standard deviation was also close between methods.

\begin{table*}[!htb]
\centering
\renewcommand{\arraystretch}{0.8}
\begin{tabular}{@{}l c c c@{}}
\toprule
 & $\boldsymbol{\delta}$ (MHz) & $\boldsymbol{\sigma}$ (MHz) & \textbf{No. of measurements} \\
\midrule
\textbf{Full data TLE}    & 1.831 & $1.2 \times 10^{-3}$ & 375 \\
\textbf{Subsampling TLE}  & 1.832 & $1.0 \times 10^{-3}$ & 150 \\
\textcolor{black!60}{\textbf{Warm-up}} & \textcolor{black!60}{1.83} & \textcolor{black!60}{$1.3 \times 10^{-2}$} & \textcolor{black!60}{150} \\
\bottomrule
\end{tabular}
\caption{Hahn-Ramsey frequency estimation results for the IBMQ device \texttt{ibmq\_armonk}, using TLE with HMC move steps based on the full data or subsampling. The detuning relative to the prevailing backend estimate was $1.83 MHz$. }
\label{tb:ramsey_1d_subs}
\end{table*}

%% file: Sections/conclusion.tex
\section{Discussion}
\label{sec:conclusion}

Bayesian inference can be fruitfully applied to the characterization of key processes in quantum computing, with applications in sensing, phase estimation, and device calibration, among others. The performance of characterization procedures hinges on two crucial factors: the experimental design and the quality of the numerical representation of Bayesian statistics. The importance of these factors becomes particularly evident in challenging scenarios, such as learning the dynamics of open quantum systems, where a high number of parameters and the presence of redundant or degenerate explanations often leads to multimodality: multiple distinct parameter configurations can yield similar likelihoods.

Collecting experimental data from quantum systems often presents greater challenges than performing classical computations. This is why, when characterizing them, we need to use advanced statistical techniques that can deliver reliable performance, even in complex scenarios. The algorithms explored in this work, for example, excel in problems where experimental data collection is costly, real-time estimation is desired, or likelihood functions are difficult to sample. We demonstrate how these methods enable a trade-off between data efficiency and classical computational efficiency, leveraging classical processing to minimize the experimental data needed for a given uncertainty.

Using quantum systems and circuits as test cases, we extensively analyze and test these strategies, comparing them with simpler methods that have been widely adopted for quantum characterization.  Sequential Monte Carlo (SMC) with the prior as importance function and Liu-West resampling (LW), a popular choice in quantum metrology, is an interesting lightweight choice for simple cases but is unfit to handle multimodality or complex features. One possible solution is to replace the Liu-West filter with Markov Chain Monte Carlo, which unlike the former does not assume normality. 

Switching to a tempered likelihood estimation framework further increases both cost and robustness in high-dimensional, multi-modal scenarios. For a model with $4$ dimensions and $16$ modes, it increased the fraction of correctly identified modes from $50\%$ to $100\%$. However, this method is not well-suited to sequential estimation: it samples from a static distribution. The same concept applies to pure Markov Chain Monte Carlo and Hamiltonian Monte Carlo. The latter was shown to enable a fast exploration under certain conditions, finding a target mode in $1/9$ of the iterations as compared to SMC-LW. To reap these benefits while recovering the sequential nature, one may embed it in a sequential Monte Carlo framework.

We also demonstrate how subsampling strategies can reduce the processing costs, namely the likelihood evaluations. For Ramsey experiments, subsampling reduced the number of likelihood evaluations by $60\%$ without increasing the uncertainty. 

We then show how Bayesian experimental design 
 combined with heuristics (such as the $\sigma^{-1}$ and particle guess heuristics) or with other techniques (such as standard dataset ordering) can fail under some circumstances. We propose and test alternatives for some of these cases, observing performance improvements. Using a space occupation based heuristic, we halve the median uncertainty and increase the success rate by $20\%$ as compared with reference strategies. 

Lastly, we use the algorithms to calibrate IBMQ superconducting quantum hardware, learning qubit resonance frequencies and coherence times. Our experiments show that the algorithms presented here can surpass the standard quantum limit and learn faster than standard methods. For Hahn echo and Ramsey experiments, we obtain uncertainties respectively $10$ and $3$ times smaller than the default methods, while using the same number of experiments; conversely, we achieve similar results using $93.4\%$ and $99.6\%$ less experimental data. We additionally test an adaptive heuristic for Hahn-Ramsey experiments, observing a reduction of the uncertainty by a factor of $8$ for the same number of measurements.

An interesting direction for future work is to use these methods to characterize more complex quantum systems, for instance, learning the coefficients for the unitary and dissipative terms governing the dynamics of an open quantum system in the Lindblad formalism.  
Reliable statistical techniques are bound to be especially important in this scenario, given the complex, high-dimensional posteriors that are likely to arise.  

Further research could also explore variational Bayesian inference as compared to the numerical approaches considered here. Another topic of relevance would be Bayesian experimental design for arbitrary likelihood models, as most heuristics specifically target a simple oscillator model.

%% file: Apps/acronyms.tex
\section{Table of acronyms}
\label{app:acronyms}

\begin{table}[ht]
\centering
\caption{List of Acronyms}
\begin{tabular}{ll}
\toprule
\textbf{Acronym} & \textbf{Definition} \\
\midrule
CLT & central limit theorem \\
GRF & Gaussian rejection filtering \\
ESS & effective sample size \\
HMC & Hamiltonian Monte Carlo \\
IPE & iterative phase estimation \\
LWF & Liu-West filter \\
MCMC & Markov Chain Monte Carlo \\
MH & Metropolis-Hastings \\
pCN & preconditioned Crank-Nicolson \\
PM & pseudo-marginal \\
RWM & random walk Metropolis \\
SHMC & sequential Hamiltonian Monte Carlo \\
SIR & sequential importance resampling \\
SMC & Sequential Monte Carlo \\
TLE & Tempered likelihood estimation \\
\bottomrule
\end{tabular}
\end{table}

%% file: Apps/expdesign.tex
\section{Bayesian experimental design}
\label{app:bayesian_experimental_design}

When performing inference, not all measurements are made equal. The structure of the likelihood function itself impacts the achievable utility of each experiment: if it varies more slowly with the parameter(s), it will produce less precise parameter estimates. Having chosen a model, the shape of the likelihood still changes depending on the controls. 

In the precession likelihoods, for a fixed parameter value, longer times mean faster oscillations, and thus more informative data. However, too fast oscillations introduce ambiguity. As the likelihoods associated to the data oscillate faster, their periodicity becomes more densely packed in parameter space. This gives rise to equivocal posteriors, where we are likely to observe multi-modality. 

We can optimize the best experiment controls by minimizing the Bayes risk (or maximizing the utility, taken to be its symmetric), given a loss function:
\begin{equation}
    \label{eq:bayes_risk_exp}
    r(\vec{E}) = 
    \mathbb{E}_{\mathbf{P}(\vec{D};\vec{E}),\mathbf{P}(\theta \mid \vec{D}; \vec{E})}
    \left[ \mathcal{L}(\theta,\hat{\theta}_X(\vec{D};\vec{E}))  
    \right]
\end{equation}

The loss can be linear, quadratic, etc. In practice one may optimize the expected variance, maximize the information gain, etc \cite{rainforth2024}.

\begin{figure*}[!htbp]
    \centering
    \includegraphics[width=\textwidth]{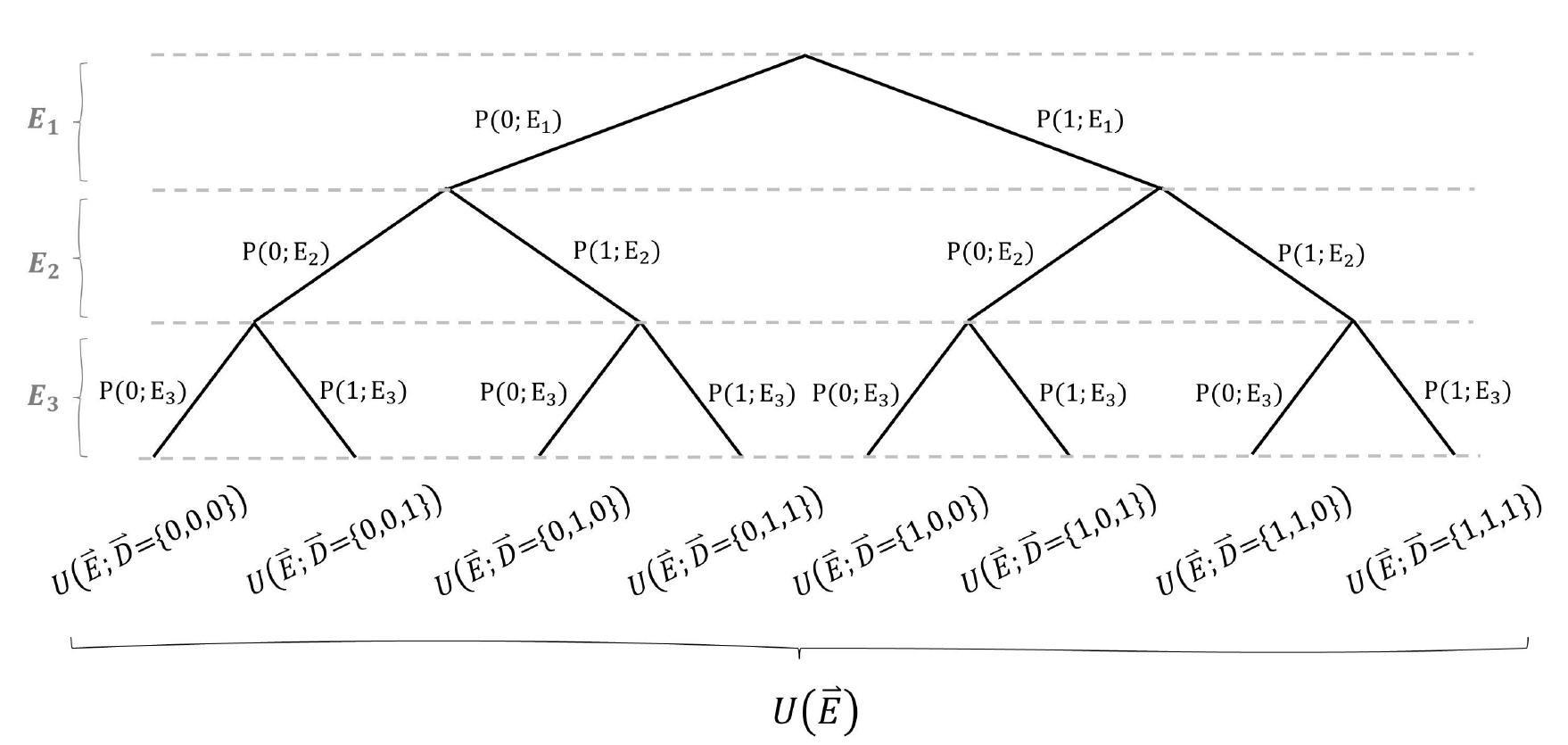}
    \caption{Binary tree illustrating the computation of the expected utility for a set of experiments $\vec{E}=\{E_1,E_2,E_3\}$. To each layer of branches corresponds one of the individual experiments, in chronological order from top to bottom. Each branch represents the occurrence of a specific datum, and is tagged with its likelihood. The leaves are marked with the utility of the data sequence traced out by the path of branches from the root to themselves.}
    \label{fig:utility_tree}
\end{figure*}

Figure \ref{fig:utility_tree} shows the operations required to compute the utility of a set of experiment controls. To make an optimal choice, this must evaluated for multiple possibilities of $\vec{E}$.  For further details on the calculations and references, refer to \cite{msc}.

In addition to computing the marginal probabilities of all $2^N$ data-vectors, we must compute their average posterior loss (e.g. the final variance). Each of these computations requires as much processing as the entire inference process did, for each possible set of experiments.

This exponential scaling with the total number of experiments is prohibitive, imposing a strong trade-off between the number of experiments and the processing cost. To alleviate this, we can exploit the parallel between the central Bayes' updates and the accessory ones by building the tree as we go, effectively intertwining its construction with the mandatory processing.

Such a coupling can be achieved by processing the data real-time, and forgoing the quest for a global maximum in favour of local optimality in what is called a \textit{greedy} algorithm \cite{Ferrie_2011}. Equipped with adaptivity, we can move along the linear path that the data conduct us through. 

While locally optimal strategies do not necessarily perform close to the global optimum, they have been shown to perform well \cite{Ferrie_2011} and achieve Heisenberg-limited estimation \cite{Higgins_2007, ramoa24}. Not only do they remove the exponential dependence on $N$, they also exploit all available information at each timestep. In contrast, for global optimization over $N$ experiments, the data from the first to the (i-1)th experiment is not used to inform the ith control. 

One may also adopt an in-between strategy consisting of a look-ahead of length $k$. Optimization costs and parallelism in the quantum component both increase with $k$. Finally, there are alternatives to this optimization process, such as machine learning or heuristics, often trading performance robustness for reduced computational costs.

\subsection{Empirical strategies for a precessing qubit}
\label{app:precession_heuristics}

Problem tailored heuristics may reduce the optimization costs as compared to general estimation strategies \cite{Ferrie_2012,Wiebe_2016,Granade_2012,Wiebe_2014a}. Adaptive heuristics can be a proxy for full-on optimization or more invested approaches, posing a less demanding alternative. 

For the precession likelihoods, \cite{Ferrie_2012} analyses the behaviour of the risk under a normality assumption for the prior. Such an assumption enables the analytical treatment of the problem, which may assist in designing practical strategies for improving inference outcomes. A flat prior can be made approximately Gaussian by an update based on a small initial set of sensible predetermined measurements. 

Assuming the warm up phase has been realized, we seek optimality through adaptivity; the analytically tractable prior can be used to arrive at an asymptotic expression for the Bayes risk (again for quadratic loss) as a function of the experiment control. This control is the measurement time, a continuous variable. 

The authors find that this quantity oscillates between a constant maximum and a lower envelope, with a frequency that increases as the inference advances (i.e. with more measurements, as the spread of the distribution is reduced).

We consider a counter $k \in \{1..N\}$ starting out at the first adaptive measurement, after the prior has been warmed up to
\begin{equation}
    \mathbf{P}(\omega) = \mathcal{N}(\mu_0,\sigma_0)
\end{equation}

At any stage, the minimum of the envelope is given in terms of the latest posterior's variance  as
\begin{equation}
    r_k^{(opt)} \equiv r(t_k^{(opt)}) = (1-e^{-1})\sigma_{k-1}^2
\end{equation}

\noindent, and occurs for
\begin{equation}
    t_k = 1/\sigma_{k-1}
\end{equation}

\noindent, $\sigma_k$ being the standard deviation of the $k$th iteration posterior,$(k+1)$th iteration prior.

As the distribution becomes sharper, the minimum of the risk approaches this lower bound location, due to the oscillations' rapid variation with respect to the bottom of the envelope. Thus, the optimal scaling of the uncertainty with the iteration number is asymptotically described by the successive envelope minima: 
\begin{gather}
    \sigma_k = \left( \sqrt{1-e^{-1}} \right)^{k}  \sigma_{0}
\end{gather}

This gives the rate at which we expect the uncertainty to shrink. These minima are associated with the times
\begin{equation}
    \label{eq:envelope_tmin}
    t_k = \frac{1}{\left( \sqrt{1-e^{-1}} \right)^{k-1}  
    \sigma_{0}}
    \approx \frac{1.26^{k-1}}{\sigma_0}
    , k \in \{1,\dots,N\}
\end{equation}

\noindent, i.e. $t_1=1/\sigma_0, t_2 = 1.26/\sigma_0,\dots$

This result is however only valid asymptotically, and does not hold for a finite number of experiments. Nonetheless, it may still be a useful insight; for instance, we could concentrate the search for an optimal solution around this lower bound. 

We can even drop full-blown optimization altogether, and instead just browse around a vicinity of the envelope's minimum by comparing the utility of a few possible $t$s. We expect that the risk's minimum will be nearby, an expectation that only becomes stronger as we move forward.

In light of this, a possible solution is to compute the current standard deviation to obtain the time given by \ref{eq:envelope_tmin}, and then perturb its location for a number of attempts (e.g. by sampling from a Gaussian centered at it) to get a few tentative $t_j$. Then we could determine the expected utility of these selected $t_j$,  and choose the one among them which yielded the maximal value.

But we don't necessarily need to calculate the current standard deviation. We could pick pairs of values $(\omega_j,\omega_j')$ at random by sampling from the current distribution:
\begin{equation}
    \omega_j,\omega_j' \sim \mathbf{P}(\cdot \mid \vec{D}_{curr})
\end{equation}

\noindent, after which we take the reciprocal of their distance $t_j=1/|\omega_j-\omega_j'|$. This has been termed the particle guess heuristic (PGH), and replaces $1/\sigma_\text{curr}$ while organically introducing variability. Also, the intuition of \ref{eq:envelope_tmin} can be generalized to higher dimensions \cite{Wiebe_2014a}, where this surrogate heuristic conveniently eschews the inversion of the covariance matrix.

This type of heuristic based guesses have been used in e.g. \cite{Granade_2012,qinfer}. With this, we may significantly reduce the complexity of choosing good experimental controls, opting for an in-between solution where we neither choose them arbitrarily nor extensively optimize them. What is more, concessions can easily be made, as experimental design can tolerate rougher calculations than the nucleus of the inference process \cite{Granade_2012}.

As a matter of fact, even settling on some apt guesses without ever considering the utility explicitly may bring better results than random measurements, for little to no added processing cost. By picking the asymptotic optimum \ref{eq:envelope_tmin}, we can hope to land in proximity to the actual optimum more than if we used indiscriminate measurements. Since as mentioned before the upper bound for the risk is flat, we never increase the maximum risk.

For these reasons, times inversely proportional to the standard deviation have been experimentally applied to the precession example in e.g. \cite{Wang_2017}, and also to phase estimation in \cite{Paesani_2017} after being proposed in \cite{Wiebe_2016}.

As an alternative, two different strategies are presented in \cite{Ferrie_2012}. One is based on locating the risk minimum with a correction to the asymptotic solution, while at the same time extending the Gaussian assumption to the successive posteriors (rather than restricting it to utility related considerations). This reduces the Bayes' update to a pair of rules that produce the mean and variance directly, thereby lessening the online processing load. This type of strategy is sometimes called \textit{offline adaptive} \cite{Lumino_2018}.

The second one is to choose the times to be exponentially spaced out (loosely emulating those of \ref{eq:envelope_tmin}, aside from a multiplying constant), i.e.
\begin{equation}
    t_k = C^k
\end{equation}

\noindent for some $C$ chosen beforehand (\textit{offline}, without making use of the data). The article proposes $C=9/8$ (for a prior domain normalized to $\omega \in ]0,1[$).

%% file: Apps/smc.tex
\section{Sequential Monte Carlo}
\label{sec:sequential_monte_carlo}

This appendix overviews SMC. For a more detailed overview, see \cite{msc}.

The purpose of SMC is to sample from a sequence of distributions, such as successive posteriors.  One possible application for sequential simulation is real-time parameter learning. However, these methods are as often used \textit{offline}, purely to aid at the post-processing stage. In that case, the goal is generally to travel across a sequence of slowly changing, increasingly complex distributions and arrive at the intended one, when it would be hard to do so directly.

The expectation of a function $f$ with respect to the k-th distribution can be calculated as:
\begin{equation}
    \label{eq:sequential_expectation}
    \mathbb{E}_{\mathbf{P}_k(\theta)} \left[f(\theta)\right]
    = \int f(\theta)\mathbf{P}_k(\theta)\mathrm{d}\theta
    \approx \sum_{i=1}^{M} f(\theta_i) \cdot w_i^{(k)}
\end{equation}

, where the $\{\theta_i^{(k)},w_i^{(k)}\}_{i=1}^M$ are produced by the SMC algorithm. 

Algorithm \ref{alg:sir} lays out the steps of an SMC algorithm, which also goes by the name of sequential importance sampling (SIS) or sequential importance sampling and resampling (SISR). The key idea is to develop an adaptive grid that relocates the grid points, often called "particles", when necessary, in order to appropriately cover the regions of relevant probability density.

\begin{algorithm*}[!ht]
\caption{Sequential importance resampling (SIR) algorithm (with the prior distribution as importance function). A ${(k)}$ superscript signals dependence of the marked quantity on the first $k$ data and no more.}
\textbf{Inputs}: data vector $\vec{D}$ (to be used incrementally: $\{D_1\}$, $\{D_1,D_2\}$,...,$\{D_1,D_2,\cdots,D_N\}$), number of particles $M$, resampling parameter $a$, threshold effective sample size $M_\text{thr}$.\\
\textbf{Computes}: successive lists of pairs $\{\theta_i^{(k)},w_i^{(k)}\}_{i=1}^M$ for (by order) $k \in \{1,\dots,N\}$ to be used in \ref{eq:sequential_expectation}. \\
\textbf{Assumes}:  ability to sample from the prior and to evaluate the likelihoods, access to a resampling routine.\\
\textbf{Assures}: The $k$-th list is computed in the $k$-th iteration, using only the datum $D_k$ and the particle history (namely the last generation of particles).
\begin{enumerate}
    \item Sample $\{\theta_i^{(0)}\}_{i=1}^M$ from the prior distribution and set $w_i^{(0)} = 1/M$ for all $i \in \{1,\dots,M\}$. Set $\{\theta_i^{(1)}\}_{i=1}^M=\{\theta_i^{(0)}\}_{i=1}^M$.
    
    \item For $k \in \{1,\cdots,N\}$:
    \begin{enumerate}
        \item For each particle $\theta_i^{(k)} \in \{\theta_i^{(k)}\}_{i=1}^M$, update the weight using the latest datum $D_k$.
        \begin{equation}
           W_i^{(k)} = w_i^{(k-1)} \cdot \mathbf{L}(\theta_i^{(k)} \mid D_k)
        \end{equation}
        Accumulate these quantities to get the normalization constant $C^{(k)}= \sum_{i=1}^M W_i^{(k)}$.

        \item For each $\theta_i^{(k)} \in \{\theta_i^{(k)}\}_{i=1}^M$, compute the normalized weights $w_i^{(k)}=W_i^{(k)}/C^{(k)}$.
        \item Calculate the approximated effective sample size
        \begin{equation}
            \widehat{\text{ESS}}_k = \left( \sum_{i=1}^M \left[ w_i^{(k)}
            \right]^2\right)^{-1}
        \end{equation}
        
        \item If $\widehat{\text{ESS}}_k < M_\text{thr}$, resample the particles $\{\theta_i^{(k)},w_i^{(k)}\}_{i=1}^M$ to get a new set $\{\theta_i^{(k)},1/M\}_{i=1}^M$.
        
        \item Set $\{\theta_i^{(k+1)}\}_{i=1}^M=\{\theta_i^{(k)}\}_{i=1}^M$. 

    \end{enumerate}

\end{enumerate}
\label{alg:sir}
\end{algorithm*}

In general, the update rule for SMC involves "canceling out" the importance distribution, which is the distribution resulting from the previous iteration:

\begin{equation}
    \label{eq:transition_weights}
    w_i^{(k+1)} \propto w_i^{(k)} 
    \frac{\mathbf{L}(\theta_i^{(k+1)} \mid D_{k+1}) \cdot \mathbf{P}(\theta_i^{(k+1)} \mid \theta_i^{(k)})}
    {\pi(\theta_i^{(k+1)} \mid \cdots)}
\end{equation}

Hence the usage of the last datum only in the case where the sequences of distributions are posteriors based on cumulative datasets.  Even still, alternatives to this reweighting can be found in the literature; for instance, \cite{Daviet_2016} suggests using a leave-one-out kernel density estimate as the transition density for the weight updates, despite not using kernel smoothing when refreshing the particle locations. If the particle cloud is a poor representation of the previous distribution (due to failure to approach equilibrium), this may allow it to recover for the subsequent iterations. This effect is also discussed and tested in \cite{msc}.

Some sources suggest a value of $\epsilon=0.5$ for the resampling threshold \cite{Doucet_2009,Granade_2012}, though sometimes more conservative values are imposed. Alternative resampling criteria may be used, e.g. the entropy of the weights \cite{Doucet_2009}.

It should be noted that the ESS is not always controlled through adaptive resampling. There are alternative ways of keeping it around or above a target value, such as choosing the sequential distributions adaptively based on a \textit{look-ahead} into prospective posteriors \cite{Gunawan_2020}. Such alternatives do however tend to be more resource-intensive, whereas this one barely adds to the overall cost of the algorithm. 

The SIR algorithm systematizes approximate inference in a flexible manner, without overt application-dependent restrictions (namely, without assumptions about the distributions, excluding those imposed by the employed resampler). By bringing the state of knowledge up to speed as soon as possible, it at once enables adaptive experimental design - or other \textit{online} assessments - and conducts the discretization so as to preserve its aptness, effectively revising it according to the turns of events.

As previously mentioned, SMC is essentially an adaptive grid, which refocuses on the regions of interest of the parameter space according to the successive cumulative datasets. This refocusing is performed by the resampling kernel, and is crucial: without it, SMC reduces to numerical integration with uniform samples, which scales exponentially with the dimension and limits the attainable precision. 

\begin{figure*}[!htb]
\captionsetup[subfigure]{width=.9\textwidth}%
\begin{subfigure}[t]{\textwidth/3}
  \centering
  \includegraphics[width=.95\textwidth]{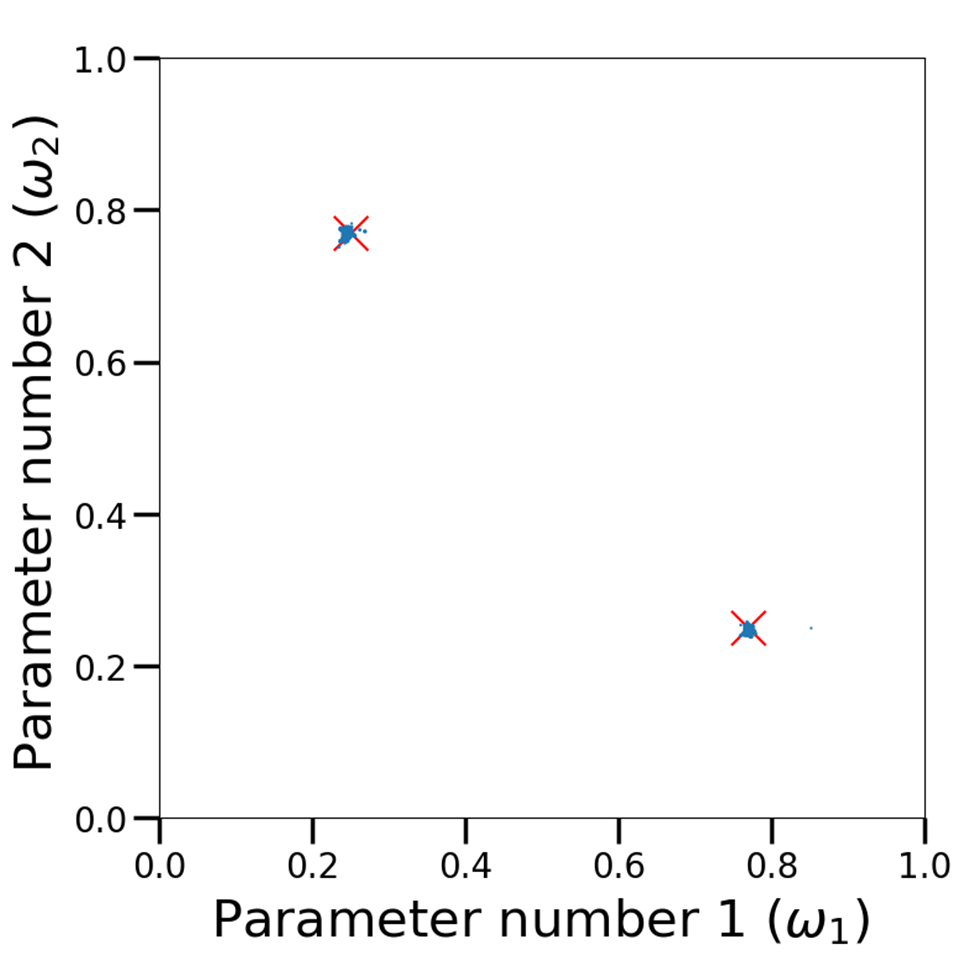}
  \caption{}
  \label{fig:2d_resampling}
\end{subfigure}%
\begin{subfigure}[t]{\textwidth/3}
  \centering
  \includegraphics[width=.95\textwidth]{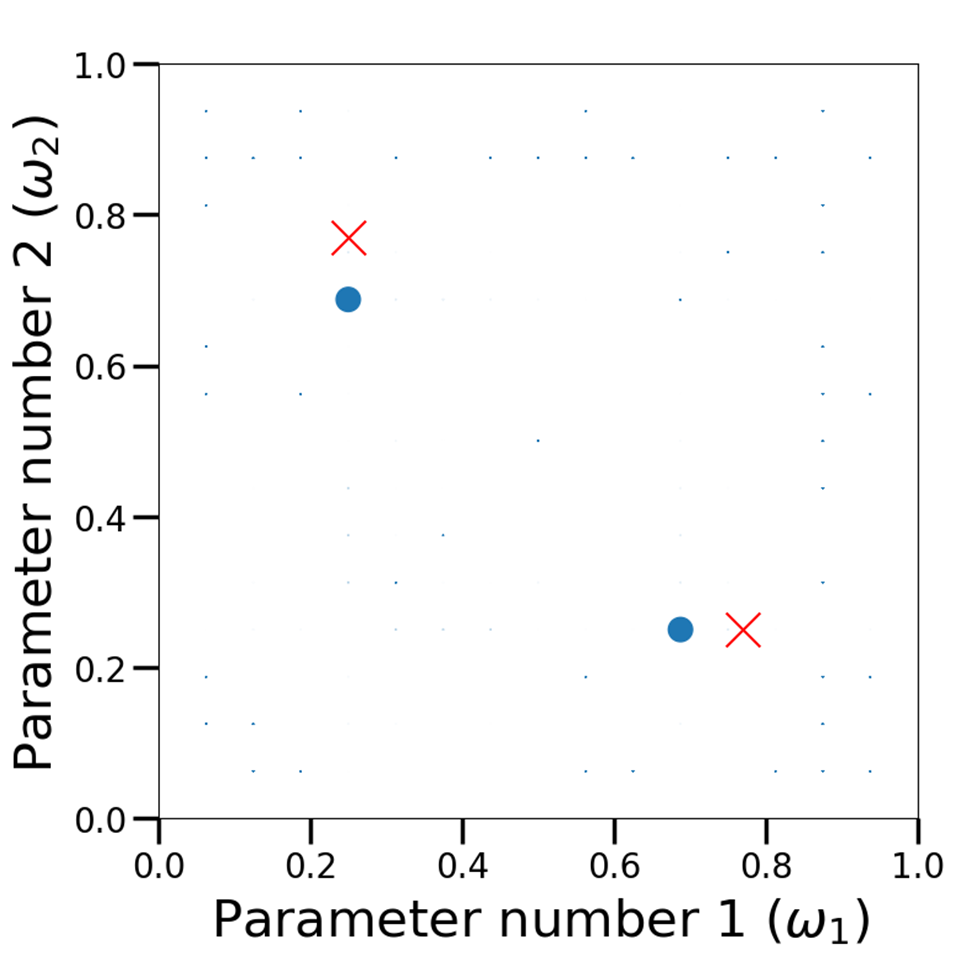}
  \caption{}
  \label{fig:2d_grid_matched}
  \end{subfigure}%
\begin{subfigure}[t]{\textwidth/3}
  \centering
  \includegraphics[width=.95\textwidth]{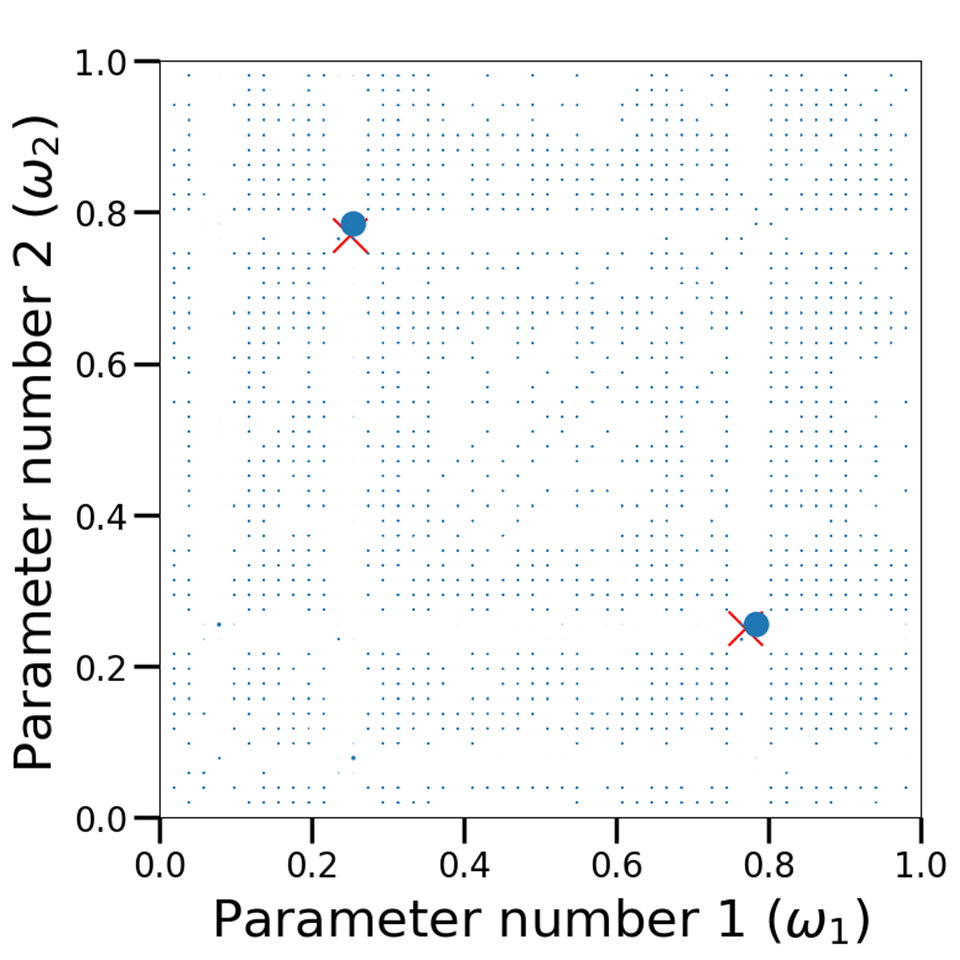}
  \caption{}
  \label{fig:2d_grid_denser}
  \end{subfigure}%
\caption{Results of multi-parameter inference for a multimodal sum-of-sinusoids likelihood (equation \ref{eq:multimodallikelihood}) using sequential Monte Carlo with random walk Metropolis propagation for $15^2=225$ particles (\ref{fig:2d_resampling}), a grid (no resampling) for $15^2=225$ particles (\ref{fig:2d_grid_matched}), and a grid (no resampling) for $50^2 = 2500$ particles (\ref{fig:2d_grid_denser}). 100 experiments are considered, and the correct modes are marked with red 'x' markers. As compared to sequential Monte Carlo, grids without resampling fail completely when using the same number of resources (particles), and produce poor results even when using $11$ times more resources. This underlines the importance of good resampling kernels..}
\label{fig:2d_cos}
\end{figure*}

Figure \ref{fig:2d_cos} illustrates this, using the multi-parameter likelihood of equation \ref{eq:multimodallikelihood}. It shows how an inference process with a limited number of particles can still converge to the correct values with the help of resampling, but fails without it. An exponentially (wrt $dim$) big increase in particle density can help with convergence (as shown in Figure \ref{fig:2d_grid_denser}), but the per-particle returns are minute - and this is for $2$ dimensions only. The final particles' proximity to the modes is at best dictated by how close to them the luckiest initial grid points happened to fall, and luck is undependable for high-dimensional spaces.

The choice of a resampler is left unspecified, as it can take several forms \cite{Doucet_2009,Granade_2012,Gunawan_2020}. Sections \ref{sub:liu_west} and \ref{sub:mcmc_smc} propose several possible choices.

\subsection{The Liu-West filter}
\label{sub:liu_west}

The Liu-West filter is one possible resampler to refresh particle locations \cite{Liu_2001}. The method is summarized in algorithm \ref{alg:liu_west}.  

\begin{algorithm}[!ht]
\caption{Liu-West resampling algorithm. The calculations assume $\theta$ to be one-dimensional, but the generalization is straightforward.}
\textbf{Inputs}: weighted sample $\{\theta_i,w_i\}_{i=1}^M$, resampling parameter $a$.\\
\textbf{Computes}: new equally weighted positions.\\
\textbf{Assumes}: Ability to sample from multinomial and Gaussian distributions.\\
\textbf{Assures}: The new sample has the same mean and variance as the original one.

\begin{enumerate}
        \item Compute the current mean and variance of the particle cloud.
        \noindent
        \begin{equation}
            \mu_k = \sum_{i=0}^{M} w_i \theta_i\\
        \end{equation} 
        
        \begin{equation}
            \sigma^2 = \sum_{i=0}^{M} w_i
            \theta_i^2-\mu^2
        \end{equation}
        
        Define
        \begin{equation}
            s = \sqrt{1-a^2} \cdot \sigma
        \end{equation}
    \item For $i \in \{1,\dots,M\}$, get a sample $\theta_i'$:
    
        \begin{enumerate}
            \item Sample a particle $\theta_s$ from the original set $\{\theta_i\}_{i=1}^M$ with probabilities given by their weights $\{w_i\}_{i=1}^M$ (i.e. a multinomial distribution), and calculate
            \begin{equation}
                \label{eq:lw_mean}
                m_i = a\theta_s + (1-a)\mu_k
            \end{equation}
            \item Sample $\theta_i'$ from a gaussian centered at $m_i$
            \begin{equation}
                \theta_i' \sim \mathcal{N}(m_i, s^2)
            \end{equation}
        \end{enumerate}     
    \end{enumerate}
\label{alg:liu_west}
\end{algorithm}

The idea is to perturb the particles about their sites, using Gaussian noise \footnote{We could also use other distributions to transform our particles into a smooth density. These distributions are commonly called kernels, and the method is called kernel smoothing .} To counteract the effect of adding noise, the particles are pulled towards the collective distribution mean, and the variance of the perturbations is chosen so as to preserve the distribution variance. This is called kernel shrinkage. The more the particles are stirred, the more we pull them towards the center to compensate.  The mean and variance are preserved, and there is a tunable parameter controlling a trade-off between exploration and structure preservation.

SMC samplers making use of this resampling strategy have often been adopted for quantum characterization examples. The method is presented in detail in \cite{Granade_2012}, and applied and/or briefly discussed in many others (e.g. \cite{Wiebe_2014b,Granade_2017}). It has also been used in experimental implementations \cite{Wang_2017}. A strong point in favour of the Liu West resampler is that it doesn't require likelihood evaluations, which is rarely the case for other methods.

On the flip-side, it only preserves the first two moments of the distribution (mean and variance, in our case representing the estimate and uncertainty respectively), which is not always sufficient. In particular, it is utterly unfit for multimodal posteriors, except for the uninteresting case where $a=0$. This also means it may converge to wrong values in the case of redundancy, which it won't be able to recover from. Furthermore, it can't be expected to yield very accurate expectations. These problems may be relieved by using more particles and resampling less often, but this is a precarious solution; not only does it defeat the purpose of using such a lightweight method, it is also bound to fail in more complex scenarious. A more sophisticated resampling strategy will be mentioned in subsection \ref{sub:mcmc_smc}.

The behaviour with respect to other moments depends on the hyperparameter $a$, which should be adapted to the application, namely the characteristics of the likelihood function. For $a=1$, we are back to the bootstrap filter, which draws random samples (with replacement) from the original particle locations. At the other extreme, choosing $a=0$ would amount to making a more drastic normality assumption: we would ignore the individual particle locations and sample from a Gaussian with the same mean and variance as the particle ensemble\footnote{It's worth noting that under such an assumption more lightweight strategies may be suitable, eliminating the need to store the entire particle cloud. For instance, \cite{Wiebe_2016} proposes a GRF approach to phase estimation, where samples are produced as for $a=0$ and the reweightings (Bayes' updates) are replaced by a acceptance-rejection step. In that case, the particles can be treated sequentially and discarded after their contributions to quantities of interest have been accumulated, greatly easing the memory requirements. The mean and variance can be calculated from 4 common counters instead of $M$ particles. Results of the application of this strategy can be found in subsection \ref{sub:phase_estimation}.}.

High values of $a$ are customarily picked so as to mostly preserve the structure of the particle cloud. Most references above choose $a=0.98$, with the original authors applying an even higher value ($0.995$), which is fitting for models with pronounced non-Gaussianity. For others, a lower value may improve convergence; $0.9$ is suggested in \cite{Wiebe_2014b} for that reason. 

One more thing to observe is that the filtering interrupts the parallelism of the SMC algorithm. While the particles can be separately reweighted (up to a constant factor) and propagated, the probabilities for the multinomial sampling step, which must be determined between these two tasks, depend on the whole particle group. 

Nevertheless, just as often the \textit{resampling} and \textit{particle propagation} or \textit{move} steps are split up in SMC algorithms, the former referring exclusively to bootstrapping. In other words, \textit{resampling} can refer to the multinomial sampling alone \cite{South_2019,Gunawan_2020}, or it can integrate the variety-introducing element as well \cite{Granade_2012}. The first case commonly occurs when the fusion isn't as natural as it is in the case of our mixture density assumption (subsection \ref{sub:mcmc_smc}), which for convenience we sample from in two steps.

 \subsection{Tempered likelihood estimation}
\label{sub:tle}

In SIR, we traverse a sequence of distributions corresponding to cumulative datasets, but other sequences can be considered. An example is \textit{tempered} or \textit{annealed} likelihood estimation (TLE), presented in detail in \cite{Neal2001} and briefly described in e.g. \cite{Del_Moral_2006,South_2019}. 

In this approach, we consider a sequence of $S$ target distributions that are powers of the target distribution:
\begin{equation}
    \mathbf{P}_s(\theta) \propto \mathbf{L}(\theta \mid \vec{D})^{\gamma_s}\mathbf{P}(\theta)
    \quad , \ 0 \leq \gamma_s \leq 1
\end{equation}

With this, the distribution starts out flattened, and is gradually made less so; by slowly increasing the exponent to $1$, we gradually lift the posterior up to its final form. 

This scheme allows us to alter the distribution continuously, which can be helpful for challenging distributions. The effect is similar to that of cumulative datasets, but more regularizing. However, it requires more likelihood evaluations for the reweightings: $N \cdot M \cdot S$ for some number of iterations $S$ instead of $N \cdot M$ ($N$ is the number of data, and $M$ that of particles).

This may be relaxed by making use of subsampling strategies, which can take advantage of the increased regularity of the sequential distributions; this will be considered in subsection \ref{app:subsampling}. Unbiased estimators coupled with variance reduction techniques may reduce the number of necessary likelihood evaluations while largely preserving accuracy, making the method at once more efficient and more parallelizable \cite{Gunawan_2020}.

SMC schemes can be used to obtain the marginal posterior distribution, or model evidence, at barely any extra cost: 
\begin{equation}
    \label{eq:marginal_smc}
    \mathbf{P}(\vec{D}) 
    \approx \prod_{t=1}^{T} \left( \sum_{i=1}^{M}  W_i^{(t)} \right)
\end{equation}

These figures can be used for model comparison.

%% file: Apps/mcmc.tex
\section{Markov Chain Monte Carlo}
\label{sec:mcmc}

This appendix contains a practical introduction to Markov Chain Monte Carlo (MCMC) methods. Reference \cite{msc} provides a more in-depth overview.

MCMC comprises a family of methods that rely on Markov chains to produce samples from a probability distribution. In its simplest form, the construction of such a chain rests only upon the ability to evaluate the probability at specific points; more advanced methods may require gradient information as well  \cite{Andrieu2003,Speagle_2020,Betancourt_2018}.

We can obtain expected values by averaging over the samples (states of the Markov chain).
\begin{equation}
    \label{eq:expectation_mcmc}
    \mathbb{E}_{\mathbf{P}( \theta \mid \vec{D})}\left[f(\theta)\right]
    \approx \frac{1}{M} \sum_{i=1}^{M} f(\theta^{(i)})
    \quad , \ \theta^{(i)} \in \mathbf{t}_\text{\tiny MC}
\end{equation}

A sufficient condition for having some distribution $\mathbf{P}_X(\theta)$ as stationary distribution of the Markov chain is that of \textit{detailed balance}, or reversibility:
\begin{equation}
    \label{eq:detailed_balance}
    T(\theta' \mid \theta) \mathbf{P}_X(\theta)
    = T(\theta \mid \theta') \mathbf{P}_X(\theta')
\end{equation}

Essentially, this says that the probability of observing a transition from state $\theta$ to state $\theta'$ or vice-versa should be the same. In our case $\mathbf{P}_X(\theta) = \mathbf{P}(\theta \mid \vec{D})$. 

The condition \ref{eq:detailed_balance} can easily be achieved using rejection sampling. We divide our transition distribution into a proposal $q$, and an acceptance/rejection step $a$:
\begin{equation}
    T(\theta' \mid \theta)  
    = q(\theta' \mid \theta) \cdot a(\theta',\theta)
\end{equation}

We can preserve the freedom of choosing the proposal distribution by tailoring the acceptance probability to it, effectively performing a \textit{ correction}. 
\begin{equation}
    \label{eq:acceptance_probII}
    a(\theta',\theta) = \min \left(1,
    \frac{q(\theta \mid \theta') \mathbf{P}_X(\theta')}
    {q(\theta' \mid \theta) \mathbf{P}_X(\theta)}
    \right)
\end{equation}

With this, it is easy to see that equation \ref{eq:detailed_balance} holds. This solution is called the MH algorithm \cite{Metropolis_1953,Hastings_1970}. An ensemble of parallel chains can ne used to extract statistics.

The procedure is described in algorithm \ref{alg:mcmc}. 

\begin{algorithm*}[!ht]
\caption{Metropolis-Hastings algorithm for posterior sampling in Bayesian inference.}
\textbf{Inputs}: starting state $\theta^{(0)}$.\\
\textbf{Computes}: sequence of states
$\mathbf{t}_\text{MC} = \{\theta^{(i)}\}_{i=1}^M$ to be used in \ref{eq:expectation_mcmc}.\\
\textbf{Assumes}: Ability to evaluate the prior, the likelihoods and a predefined proposal density $q(\theta' \mid \theta)$, and additionally to sample from the latter and from a binomial distribution.\\
\textbf{Assures}: The samples are distributed according to the posterior distribution as  $M \rightarrow \infty$.
\begin{enumerate}
    \item For $i \in \{1, \cdots, M\}$:
    
    \begin{enumerate}
        \item Sample a proposal
        \begin{equation}
            \theta^* \sim q(\cdot \mid \theta^{(i-1)})
        \end{equation}
        \item Compute the acceptance probability
        \begin{equation}
            a(\theta^*,\theta^{(i-1)}) = \min \left(1,
            \frac{q(\theta^{(i-1)} \mid \theta^*) 
            \mathbf{L}(\theta^* \mid \vec{D}) \mathbf{P}(\theta^*)}
            {q(\theta^* \mid \theta^{(i-1)}) 
            \mathbf{L}(\theta^{(i-1)} \mid \vec{D}) \mathbf{P}(\theta^{(i-1)})}
            \right)
        \end{equation}
        \item With probability $a(\theta^*,\theta^{(i-1)})$, set
        \begin{equation}
            \theta^{(i)} = \theta^*
        \end{equation}
        Otherwise, keep $\theta^{(i)} = \theta^{(i-1)}$.
    \end{enumerate}
\end{enumerate}

\label{alg:mcmc}
\end{algorithm*}

A number of initial samples is to be discarded, because the chain won't have reached equilibrium yet. This is called a burn-in period. We may want to keep only every $L$th sample, discarding the ones in between - a process called thinning. The point is that consecutive samples will be correlated due to the localized proposals; this correlation can be made negligible at or after lag $L$ (for large enough $L$). Correlated samples may bias calculations. This is particularly relevant if a \textit{compact} set of samples is desired, due to e.g. memory or post-processing constraints \cite{Link_2011}.

Both of these points are greatly impacted by the proposal distribution. The time required to reach the stationary regime (or get close enough), called the mixing time, greatly depends on it, as it determines how fast the chain explores the space. For the same reason, it influences the serial correlation between samples. Both too conservative and too bold proposals can lead to slow mixing, the latter due to high rejection rates. The exception is if the proposals are well founded and can systematically find regions of high probability, but that how to generate such proposals is an open problem.

The following two sections we will go over two well-known approaches: random walk Metropolis, and Hamiltonian Monte Carlo. Both are mostly based on \cite{Betancourt_2018}, one of the main references listed for this section.

\subsection{Random walk Metropolis}
\label{sub:rwm}

A common simplification of the Metropolis-Hastings algorithm is called simply the \textit{Metropolis} algorithm. It describes the original method, which Wilfred Hastings later generalized it to the general case. The only difference is that the proposal distribution is chosen to be symmetric, i.e. 
\begin{equation}
    q(\theta' \mid \theta)  = q(\theta \mid \theta')
\end{equation}

With this, the acceptance ratio from \ref{eq:acceptance_probII} becomes:
\begin{equation}
    \label{eq:acceptance_prob_metropolis}
    a(\theta',\theta) = \min \left(1,
    \frac{\mathbf{P}_X(\theta')}
    {\mathbf{P}_X(\theta)}
    \right)
\end{equation}

This has a nice interpretation: each point's chances of being picked grow in proportion with the probability the target distribution assigns to it. Naturally, if we want to sample from a distribution, we mean to favor precisely the points that it favors, and that too with the insistence that it favors them.

In its earliest version, the algorithm utilized gaussian proposals centered at the starting state, which are obviously symmetric. 
\begin{gather}
    \label{eq:metropolis_proposal}
    \theta' \sim \mathcal{N}(\theta, \sigma) \\
    q(\theta' \mid \theta)  = \mathcal{N}(\theta' \mid \theta, \sigma)
\end{gather}

The standard deviation is often tuned to maintain the acceptance ratio around a pre-defined value, though more sophisticated methods can be employed \cite{Gelman_2007}.

Along with \ref{eq:acceptance_prob_metropolis}, \ref{eq:metropolis_proposal} defines a protocol which we call RWM. It is based on stochastic exploration corrected for probability density. Thinking of the behaviour of such a strategy in response to an increase in dimensions, we can see how it balances the tension between probability density and parameter space volume. On the one hand, the Gaussian generalizes to a multivariate normal distribution, which relays the behaviour of high dimensional spaces through the way the tails take up increasing volume. On the other, the correction of \ref{eq:acceptance_prob_metropolis} always privileges higher probability regions. Together, they will negotiate the intended compromise, focusing on the region that produces significant contributions to the integral.

On the downside, equation \ref{eq:metropolis_proposal} describes a random walk: the direction and speed of exploration are not customized to the target function. Due to this fact, the chain is likely to move in small steps, and further to often retrace them since there is no element of continuity between transitions (which there could still be in a Markovian process: it can by achieved via the target function's geometry rather than past states). 

This tends to worsen with increased dimensions. As the region that yields consequential contributions to the expectations becomes more and more concentrated and the volume outside of it grows, uninformed proposals become each time more inadequate. Either they are bold and likely to fall outside the narrow non-negligible probability region (wherefore they are rejected), or they are conservative and slow down the chain.

For simple problems, RWM may still perform well within a reasonable span of time. Unfortunately, its issues can quickly get out of hand, bringing unacceptable inefficiency. In that case, the differential structure of the posterior distribution may aid the production of more educated guesses. This is a commonly exploited resource in more advanced MCMC methods, namely Hamiltonian Monte Carlo\footnote{Another common approach is Langevin Monte Carlo, also called the Metropolis-adjusted Langevin algorithm, which is a particular case of Hamiltonian Monte Carlo \cite{Barbu_2020}.}.

\subsection{Hamiltonian Monte Carlo}
\label{sub:hmc}

Monte Carlo integration can be performed directly on a target function via importance sampling. A good choice of importance function should resemble the integrand, $f(\theta) \mathbf{P}( \theta \mid \vec{D})$. We can use Markov Chain Monte Carlo to sample from a distribution that are able to evaluate pointwise - section \ref{sec:mcmc}. Alternatively, we can use adaptive processing to gradually zoom in on the interesting regions of space, as in Sequential Monte Carlo (SMC). This is an umbrella term for methods concerned with characterizing the succession of states in a stochastic process, based on consecutive partial observations \cite{Doucet_2001}.

Hamiltonian Monte Carlo (HMC), originally called Hybrid Monte Carlo (also HMC) \cite{Duane_1987}, comes to solve the problems that a naive random walk brings, and reinstate the efficacy of MCMC for high-dimensional spaces. This is approached at length in references \cite{Betancourt_2018}, which contains an intuitive overview on MCMC methods with a focus on Hamiltonian Monte Carlo, and \cite{Neal_2011}, which is a more detailed review of HMC. Finally, the reference manual and source code of the statistical inference platform \textit{Stan} \cite{Gelman_2015,Stan} presents most if not all of the state-of-the-art HMC techniques, along with select bibliography in the first case. 

HMC's name comes from a physical analogy that helps us understand how to use gradient information based on differential geometry, as we we don't simply want to steer towards high probability regions. The target distribution is taken to define a potential energy, and augmented with a fictitious momentum distribution that gives rise to a kinetic energy. Classical Hamiltonian dynamics then guide the exploration. The improvised momentum is used to slide along the potential, as the augmented system explores the extended space (a mechanical phase space) that is built from the original one (the parameter space).

Simulating the evolution prompted by  is conceptually Simple

To generate a proposal for the Markov chain, we must simulate the dynamics described by Hamilton's equations for some evolution time $\Delta t$, using approximate numerical methods. In that case we can regard the last state of the trajectory as an elaborate proposal $\theta' \sim q(\cdot \mid \theta)$, and perform a Metropolis-Hastings step to account for discretization errors. The acceptance rate is as before (defining $v \equiv (\theta,p)$):
\begin{equation}
    \label{eq:acceptance_prob_hmcII}
    a(\theta',\theta) 
    = \min \Big(1, e^{H(v)-H(v')} \Big)
\end{equation}

, where a momentum reflection is addedto ensure symmetry.

The the energy should be approximately preserved, leading to an acceptance rate nearing 100\%. This is how the methodology affords efficient exploration of the space, producing the ideal combination of distant proposals and high acceptance probabilities at the cost of the transition kernel's higher complexity.

It is important that the integrator be a symplectic map, or equivalently a canonical transformation. This means that it should preserve the form of Hamilton's equations, though not necessarily the energy itself. This means that even if it doesn't exactly obey our ideal Hamiltonian, it acts under the action of a \textit{perturbed} one that approximates it. This prevents the trajectory from drifting - it instead oscillates around the correct one, and its fluctuations average out unbiasedly instead of building up for long integration times. 

An example of a symplectic integrator is the leapfrog integrator, which interleaves momentum and position updates to advance the system while evaluating the gradients at midpoints. The algorithm is presented in \ref{alg:leapfrog}.  Note the time reversibility: if we start at the final point and integrate backwards in time, we will retrace our steps all the way back to the initial point.

\begin{algorithm*}[!ht]
\caption{Leapfrog integration for solving Hamilton's equations in the context of Bayesian posterior sampling.}
\textbf{Inputs}: initial position $\theta(0)$ and momentum $p(0)$, integration stepsize $\epsilon$, mass $\mathcal{M}$, integration time $\Delta t$.\\
\textbf{Computes}: sequence of position $\theta(t)$ and momentum
$p(t)$ states for $0<t \lessapprox \Delta t$, where the two instances of $t$ are matched for the last state.\\
\textbf{Assumes}: Ability to evaluate the prior and the likelihoods, and additionally their gradients.\\
\textbf{Assures}: Preserves the form of Hamilton's equations, though not necessarily the Hamiltonian itself.
\begin{enumerate}
    \item Define
    \begin{gather}
        U(\theta) = \mathbf{L}(\theta \mid \vec{D}) \mathbf{P}(\theta)\\
        \theta_{x}, p_{x} = 
        \theta(x \cdot \epsilon), p(x \cdot \epsilon) 
    \end{gather}
    \item Advance the momentum by a half timestep. 
    \begin{equation}
        p_{1/2} = p_0 - \frac{\epsilon}{2} \cdot \nabla U(\theta_0)
    \end{equation}
    \item For $k \in \{1, \dots, L\}$, with $L \equiv \text{round}(\Delta t / \epsilon)$:
    \begin{enumerate}
            \item Advance the position by a timestep.
            \begin{equation}
                \theta_{k} = \theta_{k-1} + \epsilon \cdot \frac{p_{k-1/2}}{\mathcal{M}}
            \end{equation}
            \item If $k \neq L$, advance the momentum by a timestep.
            \begin{equation}
                p_{k+1/2} = p_{k-1/2} - \epsilon \cdot \nabla U(\theta_k)
            \end{equation}
    \end{enumerate}
    \item Advance the momentum by a half timestep.
    \begin{equation}
        p_{L} = p_{L-1/2} - \frac{\epsilon}{2} \cdot \nabla U(\theta_L)
    \end{equation}
\end{enumerate}
\label{alg:leapfrog}
\end{algorithm*}

The leapfrog algorithm can be generalized to higher order, though the increase in accuracy may not justify the cost. Even though this method is very simple, it beats others whose smaller but coherent errors may prove disastrous for the Hamiltonian trajectory.

Another benefit is that even though they are not immune to problems that frustrate exploration (and thus accuracy), these problems are easy to recognize. This is because the same geometrical features - namely high curvature - that bias estimators will affect numerical stability, causing the trajectory to diverge. Such problems can be easily identified so compensatory strategies can be employed. Overall, their simplicity and theoretical grounds facilitate the task of assessing their effectiveness. 

The integrator's performance depends on some hyperparameters, namely the stepsize $\epsilon$ and the path length $L$. Together, they determine the integration time $\epsilon \cdot L$. For a fixed integration time, the stepsize determines how many iterations $L$ must be performed, which defines the cost of the algorithm; in particular, the number of evaluations of the likelihood and its gradient scale inversely with $\epsilon$. Because the accuracy and stability of the integration depends strongly on this same $\epsilon$, which determines the roughness of time discretization, a good balance should be sought when choosing it.

Having discussed the integrator, we can finalize the main details of HMC. The procedure is laid out in algorithm \ref{alg:hmc}. One may note that an important detail is missing in its construction: it doesn't explicitly cater to constraints (parameter space boundaries). They're left to be controlled by the fact that their prior probability will be zero, which will cause a minimum overflow. A way to treat restraints is to consider the effect of attributing an infinite potential energy to out-of-bounds regions, and handle it within the integration \ref{alg:leapfrog}. This leads to the position ($\theta$) advancement step being followed by a \textit{rebounding} motion (the position is mirrored relative to the boundary and the momentum is negated) if any positional boundary has been crossed \cite{Neal_2011,Daviet_2016}.

\begin{algorithm*}[!ht]
\caption{Hamiltonian Monte Carlo algorithm for posterior sampling in Bayesian inference.}
\textbf{Inputs}: starting state $\theta^{(0)}$, mass $\mathcal{M}$, integration time $\Delta t$.\\
\textbf{Computes}: sequence of states
$\mathbf{t}_\text{MC} = \{\theta^{(i)}\}_{i=1}^M$ to be used in \ref{eq:expectation_mcmc}.\\
\textbf{Assumes}: Ability to evaluate the prior and the likelihoods, to sample from gaussian and binomial distributions, and to integrate Hamilton's equations.\\
\textbf{Assures}: The samples are distributed according to the posterior distribution as $M \rightarrow \infty$,.
\begin{enumerate}
    \item Define
        \begin{gather}
            H(\theta,p)
            = -\ln \Big( \mathbf{L}(\theta \mid \vec{D}) 
            \mathbf{P}(\theta) \Big)
            - \frac{p^2}{2\mathcal{M}}\\
            v^{(i)} \equiv (\theta^{(i)},p^{(i)})
        \end{gather}
    
    \item For $i \in \{1, \dots, M\}$:
    
    \begin{enumerate}
        \item Sample the initial momentum to define the initial state.
        \begin{equation}
            p^{(i-1)} \sim \mathcal{N}(\cdot \mid \mu=0, \sigma^2=M)\\
        \end{equation}
        
        \item Simulate Hamilton's equations for $H(\theta^{(i-1)},p^{(i-1)})$ for a time $\Delta t$ and define the proposal according to the final state.
        \begin{equation}
            v^* \equiv (\theta^*,p^*) \equiv \Big(\theta(\Delta t),-p(\Delta t) \Big)
        \end{equation}
        \item Compute the acceptance probability.
        \begin{equation}
            a(v^*,v) = \min 
            \Big(1, \exp \big( H(v^{(i-1)})-H(v^*)\big) \Big)
        \end{equation}
        \item With probability $a(v^*,v^{(i-1)})$, set
        \begin{equation}
            \theta^{(i)} = \theta^*
        \end{equation}
        Otherwise, keep $\theta^{(i)} = \theta^{(i-1)}$. The momentum can be discarded.
    \end{enumerate}
\end{enumerate}
\label{alg:hmc}
\end{algorithm*}

In this HMC implementation of algorithm, the last state obtained by integrating is proposed as sample, and all others ignored.  More advanced strategies may drop the Metropolis correction entirely, ceasing to be a particular case of this algorithm, and instate some other scheme that safeguards correctness.

Dynamic implementations can be used to adaptively adjust the tunable $L$, such as the no U-turn sampler (NUTS) \cite{Betancourt_2018,Hoffman_2011}. The inverse of the mass should resemble the covariance of the target distribution. For a Gaussian distribution, optimal results can be achieved with a Gaussian kinetic energy, but in the general case that is no longer true. As such, a position dependence may be worked in to better treat local features of the distribution's landscape \cite{Girolami_2011}.

As compared to random walk Metropolis, well-tuned HMC only presents random walk behaviour between energy levels. This is induced by the random choice of a momentum variable that lifts the position coordinates up to a double-dimensional (as compared to the original parameter space) phase space, joining the position coordinate to establish the energy\footnote{Even this random walk aspect can be lessened by only partially refreshing the momentum between transitions \cite{Neal_2011}.}. Other than that commencing step, the exploration of the energy level set is deterministic - fixed by the integrator - and informed by the geometry of the target distribution, which brings improved efficacy.

The energy transition distribution is induced by the sampling of the momentum, and should be adapted to the marginal energy distribution of the target distribution. This is similar to how the RWM proposal should be tuned to the distribution itself. On the other hand, the efficacy of the exploration for a fixed energy depends on the integration method and time. As for the integration time, there is a trade-off between completeness of exploration and cost, so the goal would be to keep it \textit{just long enough} to characterize the energy level set before the returns diminish. On the opposite extreme, its being too small may not bring gains as compared to exploration via a random walk. 

Note that HMC is not well suited to multi-modal distributions, and can get trapped in isolated energy minima, especially if the modes are far apart and/or differ in height (since even if the chain gets there the acceptance probability will be low due to the difference in energy). However, not only is this weakness common to most other approaches to MCMC, it also can be mitigated via improvements to the basic HMC algorithm \cite{Betancourt_2015b,Neal_2011,Lan_2014,Gu_2019,Betancourt_2018_talk} or by embedding the chain in an interacting ensemble (as in SMC). 

The circumstances under which this version  of HMC shines the most are high dimensionality, hard-to-resolve (elongated and/or sharp) shapes, and near normality - all of which greatly benefit from the fictitious momentum variable. Due to this enhancement, HMC chains are capable of accomplishing geometric ergodicity for a wider range of functions than RWM or other simplistic approaches, in which case the expectation estimates abide by a central limit theory (CLT) instead of achieving correctness only in the asymptotic limit (as for a usual Markov chain) \cite{Betancourt_2018_talk}. Even if this does not hold, progress tends to occur on a much faster time scale for a geometric method as compared to random walk based exploration. Further, even in a worst case scenario HMC offers an interesting perk: powerful diagnostic tools. This is owing to the fact than when it does fail, it tends to fail quite spectacularly, facilitating fault detection and correction \cite{Betancourt_2018_talk}, as mentioned regarding the construction of the numerical trajectory.

Target distributions that have points of null  probability can be problematic, since the potential energy has discontinuities, being minus infinity at these points. Some ways to deal with discrete or discontinuous likelihoods have been proposed, but this is not straightforward \cite{Nishimura_2020}. This is the case for the sinusoidal example of the main text. This creates potential barriers that are overridden only if the discretized dynamics happens to leap over them by chance. As a consequence, the periodic roots of the precession example then create a minefield of potential wells that trap the particle in.

Even though the particles may be able to escape by resampling induced relocation or luck, if one is to rely on such methods much cheaper alternatives exist than HMC, in particular confessedly luck-based methods (random walks). If there aren't many such asymptotes, a possible way of escaping these regions would be to screen for vanishing acceptance rates and perform a RWM step after HMC if they're found. Such a sequence is posterior-invariant if each step is, and by slightly moving the particle one may hope to introduce some chance that it crosses the potential wall and is positioned more favorably to Hamiltonian dynamics. This is attempted in section \ref{sub:phase_estimation}. However, it may lead to expensive proposals often being constructed just to be (with near unit probability) discarded and replaced by cheaper ones, which is rather unproductive.

\subsection{Markov kernels within SMC}
\label{sub:mcmc_smc}

Despite the alluring simplicity of their underlying principle, MCMC methods have many drawbacks of their own. Coming near their promised correctness presents many hurdles, since theoretical ergodicity often doesn't transfer well to practice. Not only is the stationary regime hard to reach, but also hard to recognize. 

For these reasons and others, they do worse in many aspects as compared to SMC. Namely, they are not suitable for online estimation; tend to require more likelihood evaluations; don't do well under multimodality; do not offer estimates of the model evidence; and can be hard to tune. These drawbacks can be alleviated by combining MCMC with SMC, as the resampling mechanism in \ref{alg:sir}. Their invariance properties make them suitable to refresh the particle cloud while exactly preserving the distribution it represents. For that reason, MCMC kernels are widely used within SMC for particle propagation, \cite{Del_Moral_2006,South_2019,Gunawan_2020,Rios_2013}. After performing the reweighting and multinomial sampling steps as before, the uniformly weighted particles are moved in parallel using one or more MCMC transitions. A RWM kernel is perhaps the most popular approach, but HMC may significantly enhance performance. When using HMC moves, SMC is often termed sequential Hamiltonian Monte Carlo (SHMC).

As compared to LWF, this scheme preserves the entire distribution and not just its first two moments. This automatically solves the issues in capturing multimodality, by addressing the root of the problem. The aid is reciprocal, and the SMC framework also complements the Markov transitions well in several ways. When they're used as variety-introducing mutations, a burn-in period is not required, because the particle cloud is already a Monte Carlo approximation of the desired distribution before they are performed. If it does not approximate it exactly, these moves should aid convergence, since the Markov chain's stationary distribution is precisely the target.

Discarding lag samples is also unnecessary, given that the distribution is characterized by weighted particle density and not successive Markov chain states. However, correlation between samples is still relevant, despite not being strictly necessary for formal correctness in the asymptotic limit (infinite number of particles). It dictates how effectively variety is introduced in the particle positions, which strongly impacts how well covered the parameter space is. Correlation between samples can be used for deciding by how many Markov moves each particle should be displaced, in order to assure an effective exploration \cite{Gunawan_2020}.

The two simplifications mentioned above are direct benefits of shifting responsibility from sequential states to parallel particles, but there are others. For instance, the particle cloud structure can provide an insight into the tuning of the MCMC kernels \cite{Buchholz_2021}, whose efficacy strongly depends on their parameters. This is particularly critical when using Markov transitions which are very sensitive to the parameters and where choosing them is a challenging problem, as with HMC. For instance, the mass matrix should resemble the covariance of the target distribution, which can be approximated by that of the particle cloud.

Finally, the fact that SMC is more supple with respect to the targeted densities brings added flexibility, relaxing the MCMC requirement that the target density be a constant function of the parameters.

Thus constructed SMC samplers enjoy, unlike before, absolute formal correction (asymptotically), in the sense that the resampling step doesn't introduce error. They can also be interpreted as \textit{interacting} Markov chains. This is perhaps the most common implementation of particle filters of this sort, and it is popular across various fields of knowledge. For that reason, it is known by many different names, both more specific and more general: interacting Metropolis algorithms, Feynman-Kac particle models (mostly in measure-theoretic literature), etc. \cite{Del_Moral_2004,Del_Moral_2014}.

It can also be seen as a genetic algorithm which alternates between mutation and selection phases. For the duration of the \textit{mutation} periods, the particles are allowed to evolve independently through Markov transitions. The \textit{selection} stage is performed based on a probabilistic survival of the fittest criterion: the most apt particles \textit{reproduce}, and the least apt ones \textit{die}.

All the advantages of exact \textit{mutations} come at the cost of extra likelihood evaluations. The extra required resources are significant, making the propagation step the most costly part of the algorithm. For the scheme of algorithm \ref{alg:sir}, which is particularly frugal with respect to likelihood evaluations, moving a particle at some step $k$ will demand at least twice as many resources as have been used for all the reweightings of the particle up to that point ($2k$). And this is for RWM mutations - HMC requires on top of that $L\cdot k$ gradient evaluations, $L$ often being of the order of the hundreds or more and the gradients being rather expensive for large datasets and/or complex functions.

The use of Markov kernels also makes the algorithm more memory-intensive, because all the data must be preserved throughout the runtime (to be used at least in the move steps). While in the case of tempered likelihood estimation this was already the case, this still brings some added difficulties, namely if meaning to exploit parallelism \cite{Gunawan_2020}. While parallel Markov chains are in theory \textit{embarrassingly parallel}, large shared datasets make them less so due to the overwhelming memory requirements.

All of this may be dispensable if the mean and variance offer a sufficient description of the distribution, but not for more complex targets or if the accuracy of expectations is of concern. In particular, the model evidence estimator is sensitive to the propagation strategy. Further, not only do Markov steps in themselves improve the accuracy of this estimator, they can also can be structured in a way to enable recycling the proposals for the construction of a different and more robust one. This strategy is proposed in \cite{South_2019}.

Either way, the SMC samplers described here have been employed to deal with complex statistical functions, including tempered inference with HMC move kernels \cite{Daviet_2016,Gunawan_2020} (the most demanding combination of all alternatives reviewed here). Cost doesn't matter \textit{per se}, and for many applications the value of these costly but dependable strategies is completely unmatched by simpler but unavailing ones.

%% file: Apps/subsampling.tex
\section{Data subsampling}
\label{app:subsampling}

Subsampling strategies can help alleviate the cost of inference. However, they can affect the results; for instance, naive subsampling strategies can debase the best features of HMC \cite{Betancourt_2015}. Some solutions have been proposed to remedy this \cite{msc, Betancourt_2015,Chen_2014,Quiroz_2018,Dang_2019,Gunawan_2020}. 

Subsampling strategies attempt to replace theexact likelihood and log-likelihood, which depend on all data,
\begin{gather}
    L(\theta) = \prod_{k=1}^{N} L_k(\theta \mid D_k) \equiv
    \prod_{k=1}^{N} L_k(\theta) \\
    \ell (\theta) = \sum_{k=1}^{N} \ell(\theta \mid D_k)
    \equiv \sum_{k=1}^{N} \ell_k(\theta).
\end{gather}

, by an unbiased estimator based on a subset of the data.   

If $\widehat{\ell}_m(\theta)$ is an unbiased estimator of the log-likelihood based on subsampling $m$ data and $\widehat{\sigma}^2_m(\theta)$ estimates its standard deviation, we can estimate the likelihood by:
\begin{equation}
    \label{eq:likelihood_estimator}
    \widehat{\mathbf{L}}_m(\theta) = 
    \exp \Big( \widehat{\ell}_m(\theta) 
    - \frac{1}{2} \widehat{\sigma}^2_m(\theta)
    \Big)
\end{equation}

The motivation is that, by linearity of expectation, an unbiased estimator can easily be found for the log-likelihood, and $\widehat{\mathbf{L}}(\theta)$ will be unbiased too if $\widehat{\ell}_m(\theta)$ is normal (which is likely for large $N$ as per the CLT) and $\widehat{\sigma}^2_m(\theta)$ is exact. The variance of the estimator corrects for bias; if it is approximate, the estimator is approximately unbiased.

As for the log-likelihood estimator, the starting point is:
\begin{equation}
    \label{eq:naive_subsampling_ll}
    \widehat{\ell}_m(\theta)  = 
    \frac{N}{m} \sum_{j=1}^{m} \ell_{u_j}(\theta) 
\end{equation}

\noindent, where the $u_j$ are subsampling indices sampled at random: $u_j \in \{1,\dots,N\}$. Once we fix the $m$ elements of $\vec{u}$, $\widehat{\ell}_m(\theta)$ is a value that estimates $\widehat{\ell}(\theta)$. Evaluating it has cost $\mathcal{O}(m)$.

Clearly, $\mathbb{E}\big[\widehat{\ell}_m(\theta)\big]=\ell_m(\theta)$ by linearity of expectation. However, subsampling tends to increase the variance in the estimate. A well-known variance reduction technique for Monte Carlo estimates is the use of control variates\footnote{Other variance reduction techniques exist, some of which have been successfully employed in HMC \cite{Li_2019}.}. They exploit the fact that the expected value is unaltered if we add a term and cancel out its expectation:
\begin{equation}
    \widehat{\ell}_m(\theta)  = 
    \frac{N}{m} \sum_{j=1}^{m} \ell_{u_j}(\theta) 
    + C \Big(
    \frac{N}{m}\sum_{j=1}^{m}q_{u_j}(\theta) - \sum_{k=1}^{N}q_k(\theta)
    \Big)
\end{equation}

The $q_k(\theta)$ are auxiliary random variables called \textit{control variates}. Even though the expectation is unchanged, the variance is not, and can be minimized through the choice of the $q_k$ and $C$. If we have some way of approximating $\ell_k(\theta)$, one way of achieving this is by choosing $q_k(\theta) \approx \ell_k(\theta)$ and setting $C=-1$ to get the so-called (for obvious reasons) \textit{difference estimator}:
\begin{equation}
    \label{eq:difference_estimator}
    \widehat{\ell}_m(\theta)  = 
    \sum_{k=1}^{N}q_k(\theta)
    + \frac{N}{m} \sum_{j=1}^{m} 
    \Big( \ell_{u_j}(\theta) 
    - q_{u_j}(\theta) \Big)
\end{equation}

The $q_k(\theta)$ can be Taylor series around some reference value $\theta^*$ for the distribution\footnote{This assumes unimodality (unless provisions are made, such as clustering the data to form  multiple centroids $\theta^*_k$ to serve as references).}. For MCMC within SMC, the reference $\theta^*$ can be the current SMC mean $\bar{\theta}$, and for MCMC it can be determined in a preliminary processing phase. In both cases, the coefficients for the Taylor expansion can be calculated once using the full dataset and are then used for all parallel particle updates/transitions (for SMC/MCMC respectively), signifying $\mathcal{O}(1)$ cost (i.e. no likelihood evaluations, only distances with respect to $\theta^*$). This implies that \ref{eq:difference_estimator} has cost $\mathcal{O}(m)$ just like \ref{eq:naive_subsampling_ll}.

Finally, $\widehat{\sigma}^2_m(\theta)$ too can be obtained from the subsample as the variance of the differences in \ref{eq:difference_estimator} (between parenthesis). This clears the requirements for the computation of the estimator in \ref{eq:likelihood_estimator}, which can be used in any sampling strategy, namely any MCMC and SMC method (for both the weight updates and the particle propagation steps in the latter case). In HMC, the potential energy would become:
\begin{equation}
    \widehat{U} = -\ln \Big( \widehat{\mathbf{L}}_m(\theta) \Big)
\end{equation}

The HMC updates require the log-gradient through $-\nabla \widehat{U}$, which can can be obtained from the sample as well by differentiating \ref{eq:difference_estimator} and its variance. Most of these calculations are laid out in \cite{Quiroz_2018}, which additionally shows that the error in expected values associated with the perturbed posterior decreases as $\mathcal{O}\big(1/(Nm^2)\big)$. In particular, this is the expected convergence rate of the calculated mean and variance for the posterior distribution.  

For instance, letting $m=\mathcal{O}(\sqrt{N})$ means that their error decreases quadratically with the total number of observations, which is favorable. However, this is assuming that $\theta^*$ is the posterior mode based on the full dataset. For SMC, there are convenient statistics are at hand via the particle cloud, whose mean we can use directly; for MCMC, auxiliary methods such as gradient descent become necessary. In that case, \textit{stochastic} gradient descent based on a fixed subsample may be a less purpose-defeating assumption, but it slows down convergence. If the gradient is also based on $m=\mathcal{O}(\sqrt{N})$ observations, the rate becomes $\mathcal{O}(N^{-1/2})$. 

Being in possession of an estimator, we can use it instead of the complete likelihood (or its byproducts) when performing inference. However, this doesn't necessarily guarantee a good cost-to-performance ratio; while improvements in the estimator bring improved results, this is as compared to a crude approximation, not necessarily to the original algorithm.

In the case of HMC, it's particularly tempting to subsample the gradients, which are responsible for the lion's share of the cost. The subsampling indices could be sampled only at the beginning of each trajectory (picking a new $\vec{u}$ for each iteration of algorithm \ref{alg:hmc}), or within the trajectory itself (picking a new $\vec{u}$ for each iteration of algorithm \ref{alg:leapfrog}). 

If we do this but impose the usual Metropolis corrections based on the full dataset, which is in itself costly, we correct the bias that we are introducing. However, we are likely to dramatically lower the acceptance rate, since the energy is no longer stable. If we skip this correction to avoid processing all the data, the trajectory will be biased - just how much depends on the quality of the estimators, the features of the data, and the details of the implementation. Resampling the subsampling indices at each step of the trajectory performs better than not, but still not satisfactorily so \cite{Betancourt_2015}. As such, this approach compromises the scalability that HMC affords. In this context, some revisions of the algorithm have been suggested.

\subsection{Stochastic gradient Hamiltonian Monte Carlo}
\label{sub:sg_hmc}

Reference \cite{Chen_2014} adopts an intuitive view of the dynamics behind HMC in an attempt to remedy the downfalls of subsampling. They point out that using a subsampled gradient amounts to using a noisy gradient, which by the CLT they write as:
\begin{equation}
    \label{eq:stochastic_gradient}
    \Delta \widehat{U}(\theta) \approx 
    \Delta U(\theta) + \mathcal{N}(0,V)
\end{equation}

\noindent, where $V$ is the (co)variance of the stochastic gradient noise and may depend on $\theta$. This defines a diffusion matrix that shows an additional dependence on the stepsize, $B=\epsilon V/2$, which in turn determines the unwanted random force that affects the dynamics.

It is shown that the dynamics associated with \ref{eq:stochastic_gradient} fail to preserve the target distribution, leading to divergent trajectories.To mend this problem, the authors suggest introducing a friction term inspired in the Langevin equation (a stochastic differential equation initially used to describe Brownian motion \cite{Tong_2012}): 
\begin{equation}
    \dfrac{\dd p}{\dd t} = - \dfrac{\partial \widehat{U}}{\partial \theta}
    -\gamma \frac{p}{\mathcal{M}} 
\end{equation}

For $\gamma=B$, equilibrium is restored. In general, the exact noise model won't be known exactly, but rather approximately; the algorithm can be adapted accordingly. The variance $V$ can be estimated using the inverse of the Fisher information matrix based on a data mini-batch. All details can be found in \cite{Chen_2014}. Results of implementing the algorithm are shown in subsection \ref{sub:multi_cos}.

The authors suggest omitting the Metropolis corrections entirely to spare costs, and instead controlling the quality of the sampler (namely the bias) through the discretization of the dynamics (by keeping the integration stepsize small). However, there is then a sensitive trade-off between bias and integration cost for a fixed evolution time. Overall, even though it helps control divergence, this scheme may undermine the efficiency of HMC, especially in high-dimensional scenarios \cite{Betancourt_2015,Dang_2019}.

\subsection{Hamiltonian Monte Carlo with energy conserving subsampling}
\label{sub:shmc_ecs}

A possible pathway to a more correct subsampling HMC strategy is an extended target density. We will consider an approach presented in \cite{Quiroz_2018} for MCMC in general, in \cite{Dang_2019} for HMC in particular, and still more particularly in \cite{Gunawan_2020} for SMC with HMC move steps. The idea is to have the target distribution encompass the subsampling indices, as the random variable that they are:
\begin{equation}
    \label{eq:HMC_ECS_target}
    \mathbf{P}_X(\theta,\vec{u}) 
    = \widehat{\mathbf{L}}_m(\theta,\vec{u})\mathbf{P}(\theta)\mathbf{P}(\vec{u})
\end{equation}

When performing HMC steps, only the posterior is targeted; that is, the Hamiltonian dynamics operate for a constant $\vec{u}$ \footnote{Auxiliary variables have been attributed a momenta of their own in HMC \cite{Alenlov_2019}, but typically they're continuous, as in latent variable models. HMC isn't made for discrete domains like that of $\vec{u}$, though accommodations can be made \cite{Nishimura_2020}}. Accordingly, when we further augment the distribution with a momentum variable, it respects $\theta$ only, and matches it in dimension. Thus, within HMC:
\begin{equation}
    \label{eq:HMC_ECS_target_p}
    \mathbf{P}_X(\theta,\vec{u}) =
    \exp \Big(-\widehat{H}_{\vec{u}}(\theta,\vec{p}) \Big)
    \mathbf{P}(\vec{u})
\end{equation}

This doesn't mean that $\widehat{H}_{\vec{u}}(\theta,\vec{p})$ doesn't depend on $\vec{u}$: it does, because now the potential energy is $\widehat{U}(\theta) = -\ln \widehat{\mathbf{L}}_m(\theta,\vec{u}) -\ln \mathbf{P}(\theta)$. However, this vector is fixed within the Hamiltonian, its variability being relegated to this vector's own distribution $\mathbf{P}(\vec{u})$.

Still, $\vec{u}$ must be updated, or else the extension serves no purpose. That can be done with a simple Metropolis step (using a uniform proposal distribution, which is clearly symmetric too), as long as $\theta$ is held fixed. Updating each parameter by conditioning it on all others is a strategy called \textit{Gibbs sampling}, a special case of the Metropolis-Hastings algorithm where we sample each variable individually while sweeping through the parameter vector \cite{Andrieu2003}.

Here the separation is between $\theta$ and $\vec{u}$. We want to sample $\vec{u}$ according to $\mathbf{P}_X(\theta,\vec{u})$ conditionally on $\theta$; we can do so with a MH step. This strategy is called Metropolis-within-Gibbs, or more descriptively one-variable-at-a-time Metropolis. We fix $\theta$, sample new indices uniformly, and accept them with probability:
\begin{equation}
    \label{eq:pm_mh}
    a(\vec{u}\, ',\vec{u}) = \min \left(1,
    \frac{\widehat{\mathbf{L}}_m(\theta,\vec{u}\, ' )}
    {\widehat{\mathbf{L}}_m(\theta,\vec{u})}
    \right)
\end{equation}

Note that $\theta$ is fixed, corresponding in practice to the latest $\theta$ iterate. This is an instance of \textit{pseudo-marginal} Metropolis-Hastings (PM-MH), a version of MH where the likelihood isn't available analytically and a non-negative unbiased estimator takes its place. This preserves correctness (the chain's invariant distribution is retained), earning marginal methods the title of \textit{exact approximations} \cite{Warne_2020}. 

A common example comes from latent variable models whose integrals can't be solved analytically: importance sampling can be used to get a Monte Carlo estimator of the likelihood (this requires only pointwise evaluations of the integrand, and is for that reason often called a \textit{likelihood-free} method). An alternative would be to target the joint distribution of parameters and latent variables. This is less efficient than the \textit{marginal} methods (such as the ones considered so far), which target the intended distribution directly; as such, pseudo-marginal methods attempt to bring the approach closer to them \cite{Andrieu_2009,Alenlov_2019}. 

In our subsampling case, we introduce not evaluation sites but subsampling indices, since we are truncating a sum and not an integral. Either way, this amounts to extending the target density to encompass some extra random variables, now the auxiliary samples used to estimate the likelihood \cite{Andrieu_2009,Tran_2016,Warne_2020,Quiroz_2018}. They can be refreshed at each iteration; this refreshment is factored in when computing the joint acceptance rate \cite{Andrieu_2009,Tran_2016} (as it is in \ref{eq:pm_mh}). That there is variability in the compound MH step for even fixed $\theta$ - equivalently, that the distribution is extended - slows down convergence, how much depending on the variance of the estimator (hence the attempted reduction through e.g. control variates) \cite{Warne_2020}.

One may wonder whether the samples can't simply be picked anew at each iteration to instance the estimator and then discarded. This is sometimes done, and given the self-explanatory name of Monte Carlo within Metropolis (in the case of a Monte Carlo likelihood estimator). However, in that case, absolute validity is only achieved in the case of perfect \textit{estimation}; otherwise the scheme is rather an \textit{inexact approximation} whose accuracy grows with the number of samples \cite{Andrieu_2009}.

A possible complication of \ref{eq:pm_mh} is index stagnation. Should the estimator significantly overestimate the likelihood associated with some $\vec{u}$, others will suffer from low acceptance, \textit{jamming} the sampler and affecting validity. To solve this, correlation can be introduced between consecutive proposals, by updating only a segment of the previous index set in what is called a block pseudo-marginal strategy \cite{Tran_2016}. This should make the acceptance rate more stable, preventing the monopoly that could happen otherwise.

After the index updating step, a HMC step can be performed while holding $\vec{u}$ fixed (targeting \ref{eq:likelihood_estimator}). The HMC moves preserve the energy because the MH correction step targets precisely the energy according to which the Hamiltonian dynamics are simulated, guaranteeing high acceptance as well as correctness.

In short, energy conserving subsampling is achieved by wrapping the HMC transitions in a Gibbs update, in which a block Metropolis-Hastings step is used to pick the subsampling indices (conditioned on the latest particle location) before the usual HMC step (conditioned on the indices). This amounts to a pseudo-marginal approach where the posterior is augmented to include some supplementary sample identifiers, the joint density being left invariant by the composite - Gibbs - Markov transition comprised of two MH steps (one of which is HMC).  Marginalizing over the indices $\vec{u}$ should yield the intended $\theta$ samples.

The implementation details can be found in \cite{Gunawan_2020}.

%% file: Apps/expsetup.tex
\section{Experimental considerations and set-up}
\label{app:expsetup}

The latest known coherence times are supplied by IBM at any given time. However, variations as high as $50\%$ were observed between consecutive calibrations. As such, curve fits were used as a benchmark to validate and contextualize results. They additionally allow for comparing the achieved uncertainty in view of the data requirements, yielding a more complete performance assessment. The fitting was done using two methods: the SciPy \cite{Scipy} function \texttt{curve\_fit} (non-linear least squares fit) with default parameters, and Qiskit's native coherence fitters (from the \texttt{BaseCoherenceFitter} class).

The former was used for direct comparison, since it can work on the same data (measurement times and pulse schedules) as the inference protocol, whereas the default fitters take only gate sequences. In all cases, the methods were approximately matched for maximum evolution time, and for each of them the measurement times were evenly spaced from some near-zero value to this upper limit. 

The SciPy fitter is the same tool that Qiskit coherence fitters use internally, but Qiskit's error estimates are more reliable. They are adjusted according to the variance of each measurement over all shots, as the default parameters result in an overestimation of the errors. For a fair comparative analysis, the quantitative results of the Qiskit fitters are the ones presented.


In the case of the inference, whose performance is more variable due to using fewer data, these quantitative results were obtained using a number of runs and taking median values for both the parameter estimates and their associated uncertainties. In some cases, the evolution of the uncertainty through the iterations of the SMC algorithm is also presented. Finally, the results will where relevant be represented graphically as the curve associated with the determined parameter(s).

In the last case, the same graphs present data points corresponding to the \textit{(measurement time, probability)} tuples obtained using custom pulse schedules and a curve fitted specifically to them. This is done using external (non-Qiskit) curve fitting. The data used for the points and for these curve fits are exactly the same, whereas the inference results use a lower number of shots. This serves as an assessment of predictive power. Curve fits tend to require higher resolution if reasonable results are to be obtained.

\begin{figure}[!htb]
\captionsetup[subfigure]{width=\textwidth}%
\centering
\includegraphics[width=\linewidth]{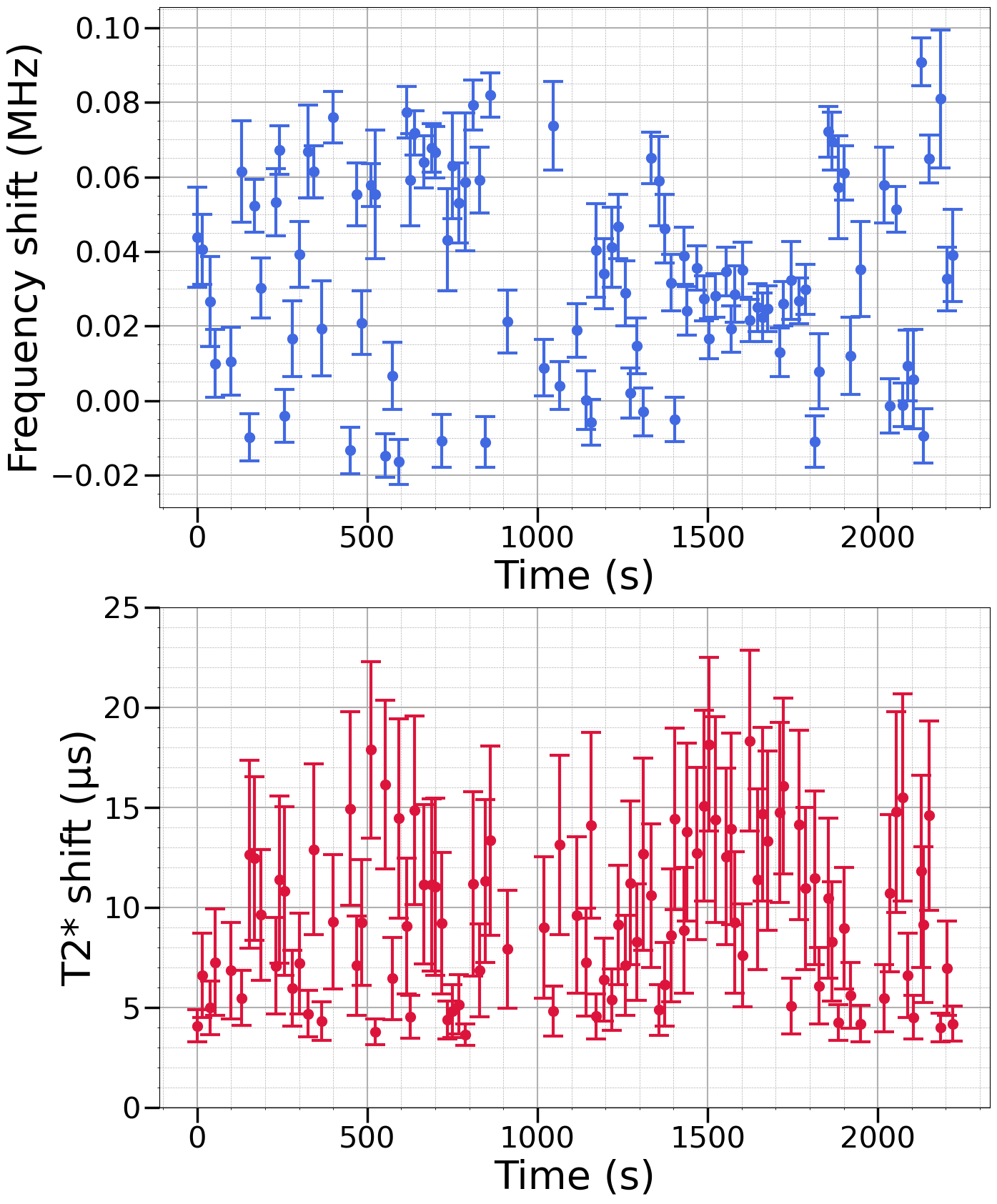}
\caption{Estimated time fluctuations of the resonance frequency - $\mathcal O (\text{GHz})$ - and the unechoed dephasing time for the IBMQ backend \texttt{ibmq\_armonk}, as a function of the starting time of data collection. The error bars represent standard deviations.}
\label{fig:fluctuation}
\end{figure}

In the cases where direct quantitative comparison is intended, the same backend was used, and maximal time proximity for the measurements across several methods was aimed for. Even still, unavoidable factors make it so that gathering datasets in succession where possible does not guarantee that each registered identical behaviour. Figure
\ref{fig:fluctuation} shows the time evolution of the estimated values using inference (whose results can be more local in time due to relying on fewer shots). The delays are due to queuing, execution times and repetition of runs to extract statistics. As such, discrepancies may in some cases be attributable to sources other than error in the estimation.

Unlike curve fits, inference can work with all-different times for a single measurement each. However, Qiskit only allows up to $75$ different pulse schedules per job submitted, each of which can be repeated for up to $8192$ shots. To streamline the data collection process, a maximum of $75$ different times is used for the data vectors, the total desired number of experiments being distributed by these times.

It should be noted that the problem of estimating time constants is fairly simple, and not necessarily a case where inference stands out; oscillations are a more interesting application. Even so, they make an interesting test case, as they elucidate the impact of intrinsic differences between likelihood models (in terms of learning rate, most favorable SMC sequences, particle number, effective sample size, order of magnitude of the evolution times, etc.). They also help understand the difficulties estimating an oscillation frequency along with a damping constant brings, due to conflicts between what is optimal for each.

For the Ramsey experiments, the probabilities of $0$ and $1$ are switched with respect to the usual case. This is because a the half-pi pulses were adapted from a Hadamard gate - using the default instruction \texttt{u2}$(0,\pi)$ -, which is an involution. Because it represents a rotation of $\pi$ around the $x+z$ axis, when applied twice to $\ket{0}$ it brings the state back to $\ket{0}$ rather than to $\ket{1}$. This is without prejudice to the covered angle's being as intended $\pi/2$ on the $x-z$ plane.

The following sections detail the experimental setup for the Hahn-Ramsey experiments. For further details and references on the physical processes, refer to \cite{msc}.

\subsection{Hahn-Ramsey experiment}
\label{app:exp_ramsey_1d}

Ramsey sequences can benefit from error mitigation techniques for improved robustness. 

Some of the sources of the $T_2$ decay are actually reversible, unlike the energy loss component. A common reason for this effect are field inhomogeneities  that lead to low frequency noise. This can occur for an ensemble of physical qubits due to local field variations that change their individual Larmor frequencies, or similarly if there are small temporal fluctuations in the qubit's resonance frequency. 

\begin{figure}[!ht]
    \centering
    \includegraphics[width=\columnwidth]{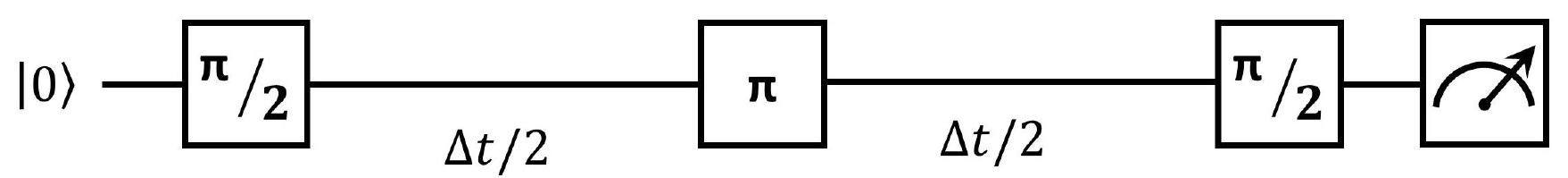}
    \caption{Pulse schedule for measuring $T_2$. This is known as a \textit{Hahn echo} or \textit{spin echo} sequence.}
    \label{fig:hahn_echo_circ}
\end{figure}

This can be exploited using a spin echo or Hahn echo experiment \cite{hahn_1950}, where the noise retraces its steps, canceling itself out for a measurement at $\Delta t$. The result is a longer characteristic time, which we call $T_2$. An echoed adaptation of the circuit of figure \ref{fig:t2_circ} is shown on figure \ref{fig:hahn_echo_circ}. 

\begin{figure}[!ht]
    \centering
    \includegraphics[width=\columnwidth]{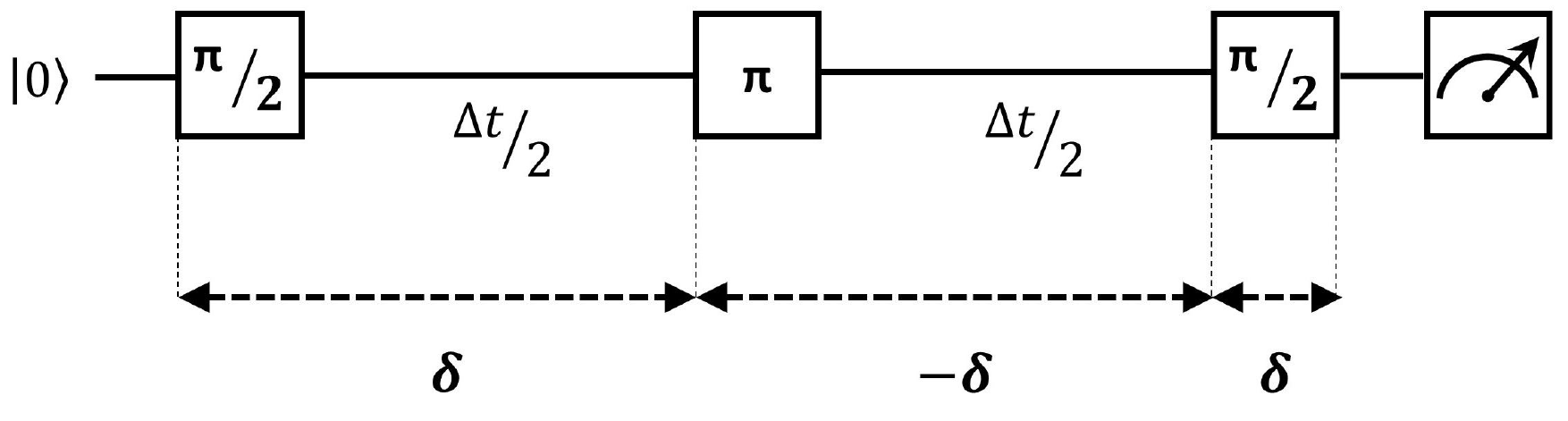}
    \caption{Pulse schedule for measuring $\Omega_0 = \widetilde{\Omega}_0+\delta_\text{meas}$ in an echoed Ramsey experiment (Hahn-Ramsey sequence) \cite{Vitanov_2015}.}
    \label{fig:echoed_ramsey_circ}
\end{figure}

A similar idea can applied to Ramsey oscillations. We refer to \cite{Vitanov_2015} for discussion on how to realize these refinements. Based on the ideas exposed therein, we construct a very simple improved version of the circuit from figure \ref{fig:ramsey_circ}, shown in figure \ref{fig:echoed_ramsey_circ}. 

Within the scope of this work, the purpose of such an arrangement is to make the coherence time large relative to the duration of at least a few oscillation cycles. One may then assume it to be negligible in the scale of the measurement times, and solve the standard undamped oscillatory likelihood problem.

The $\pi/2$ pulses may be constructed from scratch by running a Rabi fit to determine the amplitude given the intended duration, or vice-versa. An alternative is to adapt standard operations, namely rotations around the $x$ or $y$ axis. Since gates are calibrated to be in resonance, the frequency should be altered by a small detuning $\delta$. Changes in frequency necessitate a correction in the pulse duration and/or amplitude (for a fixed angle, e.g. $\pi/2$). Depending on the intended precision, bringing the pulses slightly off-resonance while maintaining the amplitude $\Omega_d$ and the pulse duration may be sufficiently accurate. For a $\delta$ small enough as compared to the resonant frequency $\Omega_0$, the effect in the pulse duration should be negligible.

\begin{figure}[!ht]
\captionsetup[subfigure]{width=.9\textwidth}%
\begin{subfigure}[t]{.5\textwidth}
  \centering
  \includegraphics[width=\linewidth]{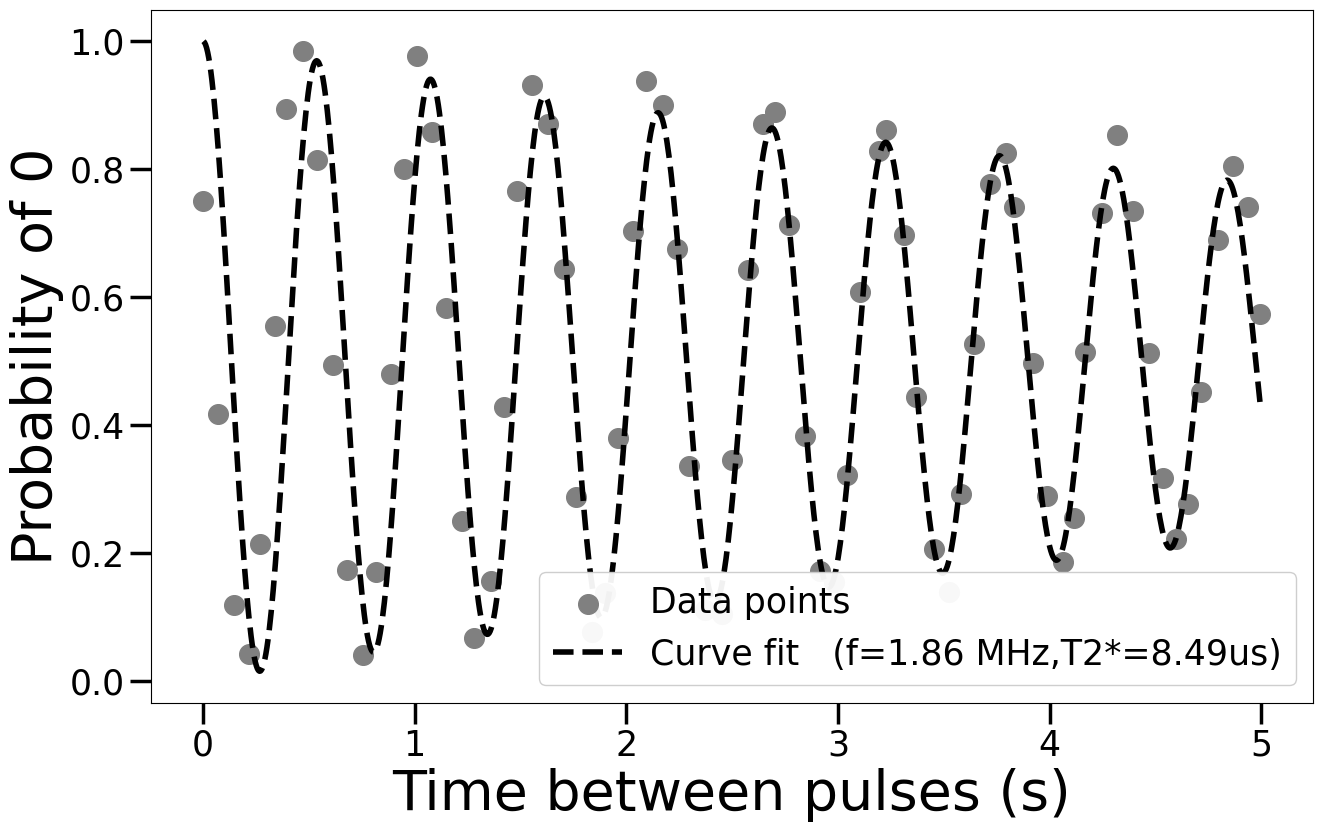}
  \caption{Data points generated using the Ramsey sequence of figure \ref{fig:ramsey_circ}.}
  \label{fig:unechoed}
\end{subfigure}
\begin{subfigure}[t]{.5\textwidth}
  \centering
  \includegraphics[width=\textwidth]{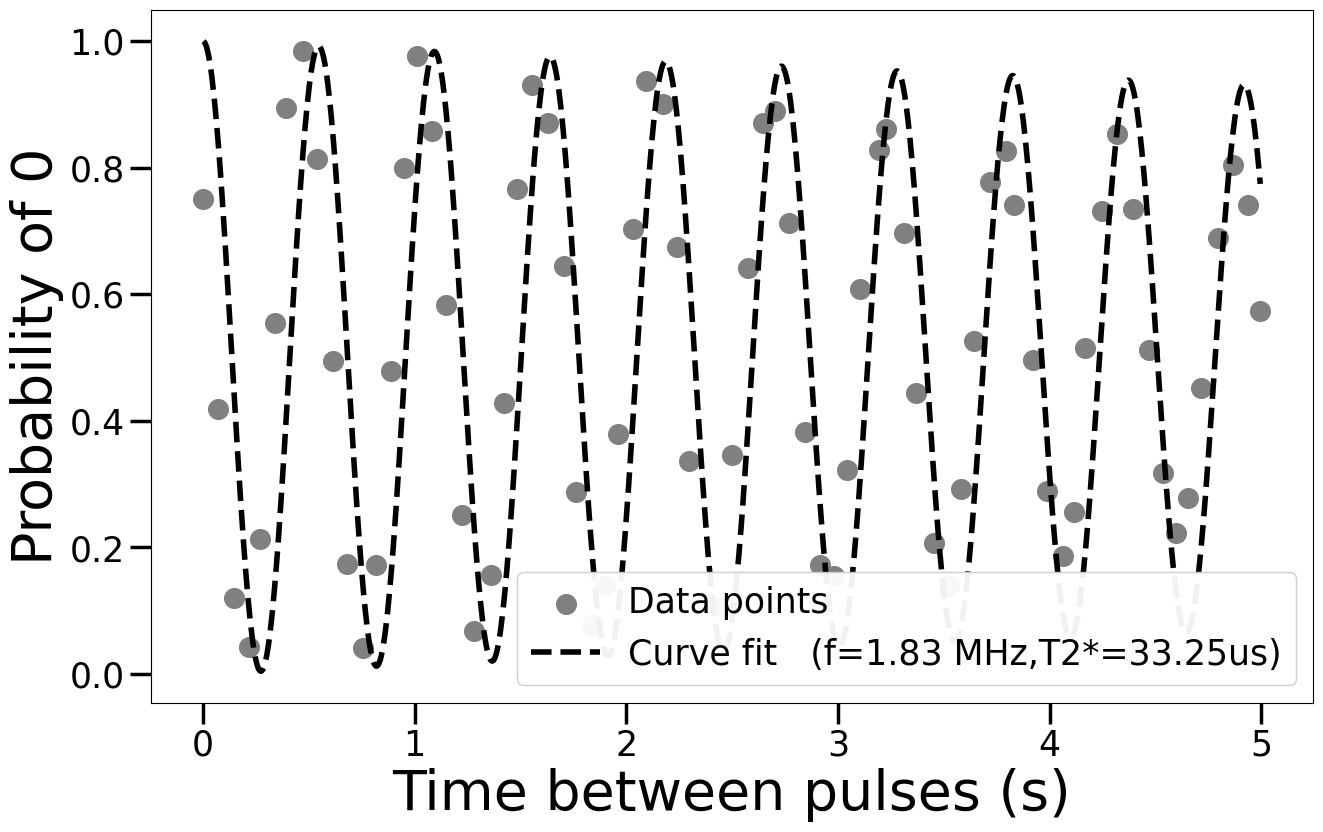}
  \caption{Data points generated using the echoed Ramsey sequence of figure \ref{fig:echoed_ramsey_circ}.}
  \label{fig:echoed}
\end{subfigure}
\caption{Effect of inserting a refocusing pulse in a Ramsey sequence for \texttt{ibmq\_armonk}. }
\label{fig:echoed_vs_not}
\end{figure}

The impact of the pi-pulse is illustrated in figure  \ref{fig:echoed_vs_not}. When comparing figure \ref{fig:echoed} to the original case of figure \ref{fig:unechoed}, we can see that the decoherence in the represented interval (up to $5\mu s$) is slight. For measurements within that interval, we can approximate the amplitude as $1$ while still obtaining reasonable results. The coherence time was improved by roughly $400\%$.

